\documentclass[useAMS,usenatbib,twocolumn]{mn2e}
\usepackage{amsmath}
\usepackage{amssymb}
\usepackage{graphicx}
\usepackage{color}
\usepackage{amsmath,amsbsy}

\title[]{Synthetic observations of molecular clouds in a galactic center environment: I. Studying maps of column density and integrated intensity}
\author[Bertram et al.]{Erik~Bertram$^1$, Simon~C.~O.~Glover$^1$, Paul~C.~Clark$^2$, Sarah~E.~Ragan$^3$ \& Ralf~S.~Klessen$^{1}$\\
$^1$Universit\"at Heidelberg, Zentrum f\"ur Astronomie, Institut f\"ur Theoretische  Astrophysik, Albert-Ueberle-Str.~2, 69120 Heidelberg, Germany \\
$^2$School of Physics and Astronomy, Cardiff University, CF24 3AA, UK \\
$^3$School of Physics and Astronomy, University of Leeds, LS2 9JT, England \\
}

\begin{document}

\maketitle

\abstract
We run numerical simulations of molecular clouds (MCs), adopting properties similar to those found in the Central Molecular Zone (CMZ) of the Milky Way. For this, we employ the moving mesh code A{\sc repo} and perform simulations which account for a simplified treatment of time-dependent chemistry and the non-isothermal nature of gas and dust. We perform simulations using an initial density of $n_0 = 10^3\,$cm$^{-3}$ and a mass of $1.3 \times 10^5\,$M$_{\odot}$. Furthermore, we vary the virial parameter, defined as the ratio of kinetic and potential energy, $\alpha = E_{\text{kin}} / |E_{\text{pot}}|$, by adjusting the velocity dispersion. We set it to $\alpha = 0.5, 2.0$ and $8.0$, in order to analyze the impact of the kinetic energy on our results. We account for the extreme conditions in the CMZ and increase both the interstellar radiation field (ISRF) and the cosmic-ray flux (CRF) by a factor of 1000 compared to the values found in the solar neighbourhood. We use the radiative transfer code R{\sc admc}-3{\sc d} to compute synthetic images in various diagnostic lines. These are {[C{\sc ii}]} at 158\,$\mu$m, [O{\sc i}] (145\,$\mu$m), [O{\sc i}] (63\,$\mu$m), $^{12}$CO ($J=1 \rightarrow 0$) and $^{13}$CO ($J=1 \rightarrow 0$) at 2600\,$\mu$m and 2720\,$\mu$m, respectively. When $\alpha$ is large, the turbulence disperses much of the gas in the cloud, reducing its mean density and allowing the ISRF to penetrate more deeply into the cloud's interior. This significantly alters the chemical composition of the cloud, leading to the dissociation of a significant amount of the molecular gas. On the other hand, when $\alpha$ is small, the cloud remains compact, allowing more of the molecular gas to survive. We show that in each case the atomic tracers accurately reflect most of the physical properties of both the H$_2$ and the total gas of the cloud and that they provide a useful alternative to molecular lines when studying the ISM in the CMZ.
\endabstract

\begin{keywords}
Galaxy: centre -- galaxies: ISM -- ISM: clouds -- stars: formation
\end{keywords}

\section{Introduction}
\label{sec:introduction}

The inner few hundred parsecs of our Milky Way, known as the Central Molecular Zone (CMZ), are rich in molecular and dense gas and account for about 5\% of the molecular gas content of the Galaxy \citep{MorrisAndSerabyn1996}. As shown by \citet{LongmoreEtAl2013}, the CMZ contains gas with densities of the order of several $10^3\,$cm$^{-3}$ and a total mass of $\sim10^7\,$M$_{\odot}$. Furthermore, it is highly turbulent \citep{ShettyEtAl2012} and illuminated by a strong interstellar radiation field (ISRF) combined with a high cosmic-ray flux (CRF) \citep{Yusef-ZadehEtAl2007,ClarkEtAl2013}. It thus provides an excellent laboratory for studying the physics of the interstellar medium (ISM) under extreme conditions (such as those that also occur in distant starburst galaxies) with very high resolution.

Previous observations suggested the existence of copious CO gas in the CMZ near the Galactic Center (GC) \citep{Bania1977,BurtonEtAl1978,LisztAndBurton1978,MorrisAndSerabyn1996,BitranEtAl1997,OkaEtAl1998,MartinEtAl2004}, which might be used to trace the density and velocity structure of the underlying molecular component of the ISM. However, the effect of the extreme physical conditions in the CMZ on the CO distribution within the gas and the degree to which CO is a biased tracer of the underlying cloud properties remain relatively unexplored issues. Previous numerical models of clouds in the CMZ \citep{ClarkEtAl2013} or in other harsh environments \citep{ClarkAndGlover2015} have shown that the gas is likely to be highly chemically inhomogeneous. This means that molecular tracers might tell us only little about the kinematics, the temperature and density distributions of the cloud and so on. Hence, in order to learn more about the internal physics (e.g. heating, cooling and the chemistry) of the warm gas, which makes up a large fraction of the total mass, we have to use other probes.

Recently, \citet{ClarkEtAl2013} have shown that the [O{\sc i}] 63\,$\mu$m and the {[C{\sc ii}]} 158\,$\mu$m fine structure lines dominate the cooling in CMZ clouds over a wide range in densities. Moreover, \citet{Rodriguez-Fernandez2004} present observations of these fine structure lines in the CMZ, carried out with the ISO satellite. They show that both lines are very strong in this region and can be used to infer important information about the physics of the gas in the CMZ. Hence, using such atomic tracers in order to study the physics of MCs in the CMZ seems to be a promising approach.

In this paper we follow up on this idea and investigate the applicability of atomic tracers to study the properties of the ISM in the CMZ. We aim to provide synthetic observations, readily comparable to observations of fine structure lines. To do so, we compute synthetic observations of MCs in a CMZ-like environment using the radiative transfer code R{\sc admc}-3{\sc d} \citep{Dullemond2012} for various diagnostic lines, specifically for {[C{\sc ii}]} (158\,$\mu$m), [O{\sc i}] (145\,$\mu$m), [O{\sc i}] (63\,$\mu$m), $^{12}$CO (2600\,$\mu$m) and $^{13}$CO (2720\,$\mu$m). We perform simulations of various clouds with the moving mesh code A{\sc repo} \citep{Springel2010} using environmental properties similar to those experienced by a typical CMZ cloud. In particular, we adopt values for the ISRF strength and the CRF comparable to those inferred for the dense CMZ cloud G0.253+0.016, also known as ``The Brick'' \citep[see e.g.][]{GuestenEtAl1981,LisEtAl1994,LisAndMenten1998,LisEtAl2001,MolinariEtAl2011,ImmerEtAl2012,LongmoreEtAl2012,KauffmannEtAl2013,ClarkEtAl2013,JohnstonEtAl2014,RathborneEtAl2014,RathborneEtAl2015}. We use values for the ISRF and the CRF that mimic the harsh conditions assumed to be found in the CMZ, which are a factor of $\sim1000$ larger than the values measured in the solar neighbourhood \citep{Habing1968,Draine1978,MathisEtAl1983}. We simulated clouds with an initial number density of $n_0 = 10^3\,$cm$^{-3}$ and studied the impact of different virial $\alpha$ parameters, $\alpha = 0.5, 2.0$ and $8.0$, on the physical properties. More information about the simulations can be found in \citet{BertramEtAl2015}. Furthermore, we note that this study is the first of two papers which analyze the properties of CMZ-like clouds. In this paper, we focus on maps of column density and integrated intensity, while in Paper II (Bertram et~al., in prep.), we study the kinematic properties of our model clouds.

This paper is structured as follows. In section \ref{sec:methods} we present our numerical simulations and the radiative transfer post-processing tool. In section \ref{sec:results} we show and discuss the results of our studies. We present a summary and our conclusions in section \ref{sec:summary}.

\section{Methods}
\label{sec:methods}

\begin{table*}
\begin{tabular}{l|c|c|c|c|c|c}
\hline\hline
Model name & Virial $\alpha$ & Initial number density $n_0$ & 3D velocity dispersion $\sigma_v$ & Radius $R$ & Free-fall time $t_{\text{ff}}$ & Simulation end $t_{\text{end}}$ \\
 & & [cm$^{-3}$] & [km/s] & [pc] & [Myr] & [Myr] \\
\hline
GC-0.5-1000 & 0.5 & 1000 & 5.4 & 8.9 & 1.40 & 1.00 \\
GC-2.0-1000 & 2.0 & 1000 & 10.8 & 8.9 & 1.40 & 1.40 \\
GC-8.0-1000 & 8.0 & 1000 & 21.6 & 8.9 & 1.40 & 1.40 \\
\hline
\end{tabular}
\caption{Overview of the initial conditions for our different cloud models.}
\label{tab:setup}
\end{table*}

The numerical simulations studied in this paper are described in more detail in \citet{BertramEtAl2015}. However, we summarize the most important aspects of the simulations here and introduce the radiative transfer post-processing, which we use for computing synthetic observational maps of the clouds in various diagnostic lines.

\subsection{Hydrodynamical and chemical model}
\label{subsec:hydromodel}

We use the moving mesh code A{\sc repo} \citep{Springel2010} in order to run numerical simulations of various model clouds. We have added to A{\sc repo} a detailed atomic and molecular cooling function, described in detail in \citet{GloverEtAl2010} and \citet{GloverAndClark2012}, and a simplified treatment of the basic chemistry of the gas. The chemical network is based on the work of \citet{NelsonAndLanger1997} and \citet{GloverAndMacLow2007}, and allows us to follow the formation and destruction of H$_{2}$ and CO self-consistently within our simulations. The network tracks the abundances of 6 species and follows 14 chemical reactions. The simplified \citet{NelsonAndLanger1997} network is known to somewhat overestimate the rate at which CO forms, owing to its neglect of atomic carbon \citep[see the detailed discussion in][]{GloverAndClark2012}. It also neglects the reaction
\begin{equation}
\label{eq:chem}
\text{CO} + \text{He}^+ \rightarrow \text{C}^+ + \text{O} + \text{He},
\end{equation}
which plays an important role in regulating the CO abundance in gas exposed to a high CRF \citep[see, e.g.][]{ClarkAndGlover2015,BisbasEtAl2015}. Our simulations thereby overestimate the CO abundance in the simulated clouds, and hence potentially underestimate the C$^+$ and O abundance in regions where the CO abundance is large. Therefore, in order to establish the effect that this has on the synthetic maps, we make two versions of each map: one in which we use values for the fractional abundances of C$^+$ and O taken from the simulations and a second version in which we assume that all of the carbon is in the form of C$^+$ and all of the oxygen is present as O. This can be achieved by assuming zero CO abundances in our simulations. As discussed in more detail in Appendix \ref{sec:noCO}, we find only minor differences in the [O{\sc i}] (145\,$\mu$m) emission PDF and negligible differences in the [O{\sc i}] (63\,$\mu$m) and [C{\sc ii}] (158\,$\mu$m) PDFs. We therefore conclude that the known weaknesses of our simplified chemical model have little influence on the results we obtain for [C{\sc ii}] and [O{\sc i}]. Full details of the chemical model with a description of how the chemistry interacts with the ISRF via the T{\sc ree}C{\sc ol} algorithm can be found in \citet{ClarkEtAl2012}. Further examples of the use of our chemical model with the A{\sc repo} code can be found in \citeauthor{SmithEtAl2014a}~(2014a) and \citeauthor{SmithEtAl2014b}~(2014b).

We assume that the gas has a uniform solar metallicity and adopt the standard ratio of helium to hydrogen. The abundances of carbon and oxygen are taken from \citet{SembachEtAl2000}. We use $x_{\text{C}} = 1.4 \times 10^{-4}$ and $x_{\text{O}} = 3.2 \times 10^{-4}$, where $x_{\text{C}}$ and $x_{\text{O}}$ are the fractional abundances by number of carbon and oxygen relative to hydrogen. We note that the CMZ has super-solar metallicity. Nevertheless, we use a uniform solar value in order to be conservative regarding the cooling rates in our runs. When we start the simulations, hydrogen, helium and oxygen are in atomic form, while carbon is assumed to be in singly ionized form, as C$^{+}$. We adopt the standard local value for the dust-to-gas ratio of 1:100 \citep[for further discussion see, e.g.][]{GloverEtAl2010}. We set the cosmic ray ionization rate (CRIR) of atomic hydrogen to $\zeta = 3 \times 10^{-14} \: {\rm s^{-1}}$ \citep{ClarkEtAl2013}. This value is a factor $\sim1000$ higher than the value measured in dense clouds in the solar neighbourhood \citep{vanderTak2000} and is comparable to the high value that has previously been inferred in the CMZ \citep{Yusef-ZadehEtAl2007,ClarkEtAl2013}. Note, however, that it is only around a factor of 100 higher than the value derived in more diffuse gas in the solar neighbourhood \citep{IndrioloEtAl2015}. For the interstellar radiation field, we adopt the same spectral shape as given in \citet{Draine1978}. We denote the strength of the Draine ISRF as $G_0 = 1$ and perform simulations with a field strength $G_0 = 1000$ \citep{ClarkEtAl2013}. This corresponds to an integral flux in the energy range $6-13.6\,$eV of $2.7 \times 10^{-3}\,$erg\,cm$^{-2}$s$^{-1}$.

\subsection{Model parameters}
\label{subsec:parameters}

The clouds that we model are initially spherical and located at the center of a large box of low-density gas. For the box, we use periodic boundary conditions. However, this is simply a convenient choice, since the boundaries do not influence the dense cloud evolution, because the size of the box is taken to be much larger than the size of the cloud. The cloud has a uniform initial hydrogen nuclei number density, which we set to $10^3\,$cm$^{-3}$, and a total mass of $M_{\text{tot}} = 1.3 \times 10^5\,$M$_{\odot}$. The initial cloud radius is $R \approx 8.9\,$pc. In all of our simulations, the density of the gas surrounding the cloud is $\approx1$\,cm$^{-3}$, but we note that our results are insensitive to this value provided that it is much smaller than the mean cloud density. The cubic side length of the total simulation domain is set to $5 \times$ the individual cloud radius of 8.9\,pc, i.e. 44.5\,pc in total. All clouds have zero bulk velocity and are placed in the center of the box. The Voronoi cells initially have approximately constant volumes. We initially start with $2 \times 10^6$ cells in total. The initial cell mass within the cloud corresponds to $\approx2\,$M$_{\odot}$.

We make use of a Jeans refinement criterion in order to accurately refine dense and collapsed gas regions in the box over the whole simulation time. We use a constant number of 8 cells per Jeans length, which is sufficient to avoid artificial fragmentation (see, e.g. \citeauthor{TrueloveEtAl1998}~1998, \citeauthor{GreifEtAl2011}~2011 as well as \citeauthor{FederrathEtAl2010b}~2010b). Our code also includes a sink particle formation algorithm \citep{BateEtAl1995,JappsenEtAl2005,GreifEtAl2011} to properly track the formation of stars during each run. Although our numerical models show active star formation by the end of the simulations, we ignore the sink particles in this study, since we only want to focus on the physical properties of the different gas phases. A more detailed analysis of the star formation history of our runs and the sink particle algorithm that we use can be found in \citet{BertramEtAl2015}.

We assume that the gas has a turbulent velocity field with an initial power spectrum $P(k) \propto k^{-4}$ and consists of a natural mixture of solenoidal and compressive modes (see, e.g. \citeauthor{FederrathEtAl2010a}~2010a). The strength of the turbulence is set by our choice of the initial virial $\alpha$ parameter. This is defined as $\alpha = E_ {\text{kin}} / |E_ {\text{pot}}|$, i.e. the kinetic energy $E_{\text{kin}} = 1/2M_{\text{tot}}\sigma_v^2$ divided by the potential energy $E_{\text{pot}} = -3GM_{\text{tot}}^2/(5R)$ measured at the beginning of each run. The quantity $\sigma_v$ denotes the 3D rms velocity dispersion in the cloud. We note that the virial parameter is also often calculated as $\alpha = 2E_{\text{kin}} / |E_{\text{pot}}|$ in the literature, which is different from the notation used in this paper \citep[see, e.g.][]{FederrathEtAl2012}. With our definition, a value of $\alpha = 0.5$ corresponds to a cloud in virial equilibrium and $\alpha > 1.0$ describes clouds that are gravitationally unbound. In order to span a large range in the virial parameter space, we analyze models with $\alpha = 0.5, 2.0\,$ and $8.0$. Note that although it is convenient to parameterize the models in terms of $\alpha$, the underlying physical quantity that changes as we change $\alpha$ is the turbulent velocity dispersion of the gas, since we keep $M_{\rm tot}$ and the initial radius $R$ fixed in all three models. It is therefore the changes to the velocity field, and the consequent changes to the density distribution of the gas, that dictate the observed trends seen in our simulations. However, for simplicity and clarity, we will simply refer to the different $\alpha$ models in our discussions in the following sections.

The turbulence is not driven and hence decays throughout the simulation \citep{MacLowEtAl1998}. The timesteps between the individual snapshots are $\Delta t_{\text{snap}} \approx 7\,$kyr, corresponding to $\sim200$ snapshots per simulation in total. Table \ref{tab:setup} summarizes the initial conditions of our clouds. We note that due to the high computational cost of our $\alpha = 0.5$ model, we had to stop our run already at $t \approx 0.7\,t_{\text{ff}}$, while the other runs end at one free-fall time. This is due to an intense refinement of dense gas regions within the cloud. Furthermore, regarding the densities in the CMZ, we have to keep in mind that a significant number of CMZ clouds have densities higher than those modelled in our simulations, which would lead to even larger turbulent velocities in our simulations for the same constant total mass.

\subsection{Radiative transfer}
\label{subsec:RADMC}

We post-process our data in order to generate synthetic line emission maps for the atomic and molecular tracers listed in Table \ref{tab:lines}, using the radiative transfer code R{\sc admc}-3{\sc d}\footnote{www.ita.uni-heidelberg.de/$\sim$dullemond/software/radmc-3d/} \citep{Dullemond2012}. We use the Large Velocity Gradient (LVG) approximation \citep{Sobolev1957} to compute the C$^+$, O and CO level populations. The LVG implementation in R{\sc admc}-3{\sc d} is described in \citeauthor{ShettyEtAl2011a}~(2011a). We use the rate coefficients for collisional excitation and de-excitation by atomic and molecular hydrogen tabulated in the Leiden database \citep{SchoierEtAl2005}.

At the moment, R{\sc admc}-3{\sc d} cannot deal with A{\sc repo} data directly. Thus, we have to map the simulation output onto a cubic grid. For that, we adopt a resolution of 512 cells in all three spatial dimensions, corresponding to a cell size of $\Delta x = \Delta y = \Delta z = 0.087$\,pc. Moreover, we also use 512 channels in velocity space for our radiative transfer post-processing, corresponding to a spectral resolution of 0.057~km~s$^{-1}$, 0.066~km~s$^{-1}$ and 0.102~km~s$^{-1}$ for our models with $\alpha = 0.5, 2.0$ and 8.0. Our choice of spatial resolution is determined by technical limitations within R{\sc admc}-3{\sc d}: as it is a serial code, our grid size is limited by the requirement that the data should all fit within the memory of the computer that we use to carry our the post-processing. However, for most of our tracers, we find only minor differences in the emission as we increase the grid resolution. In particular, we find that a resolution of $512^3$ grid cells is enough in order to properly recover the emission of the cloud for all of the tracers apart from the [O{\sc i}] 63$\mu$m line (see Appendix \ref{sec:resolution}). Hence, our main conclusions are not strongly affected by our choice of spatial resolution.

Our simulations do not explicitly track the abundance of $^{13}$CO and so we need a procedure to relate the $^{13}$CO number density to that of $^{12}$CO. A common assumption is that the ratio of $^{12}$CO to $^{13}$CO is identical to the elemental abundance ratio of $^{12}$C to $^{13}$C \citep[see, e.g.][]{RomanDuvalEtAl2010}. We make the same assumption and set the $^{12}$CO to $^{13}$CO ratio to a constant value, $R_{12/13} = 60$. In reality, physical effects like chemical fractionation \citep{WatsonEtAl1976} and photodissociation \citep{VisserAndDishoeckAndBlack2009} will alter the abundances of $^{13}$CO relative to $^{12}$CO, leading to a variable value of $R_{12/13}$ within the cloud. However, \citet{SzucsEtAl2014} have shown that the resulting $^{13}$CO maps do not differ greatly from those that we would obtain by assuming a constant ratio of $^{12}$CO to $^{13}$CO, and so we expect this approximation to be adequate.

The radiative transfer calculation yields position-position-velocity (PPV) cubes of brightness temperatures $T_B$, which are related to the intensity via the Rayleigh-Jeans approximation,
\begin{equation}
\label{eq:RQ}
T_B(\nu) = \left ( \frac{c}{\nu} \right )^2 \frac{I_{\nu}}{2k_B},
\end{equation}
where $I_{\nu}$ is the specific intensity at frequency $\nu$ and $k_B$ the Boltzmann constant. For each run, we carry out the post-processing for each of the tracers listed in Table \ref{tab:lines} along an arbitrary line of sight (LoS), using the last snapshot of each model. Since we use isotropic turbulence without magnetic fields, we do not expect our results to significantly depend on the specific choice of the LoS and hence we only focus on emission observed along the $z$-direction (see also Appendix \ref{sec:xyzPDFs}).

\begin{table}
\begin{tabular}{l|c|c|c|c}
\hline\hline
Tracer & Type & Transition & $\lambda$ & $\nu$ \\
 & & & [$\mu$m] & [GHz] \\
\hline
$^{12}$CO & Molecular & $J=1 \rightarrow 0$ & 2600 & 115 \\
$^{13}$CO & Molecular & $J=1 \rightarrow 0$ & 2720 & 110 \\
{[C{\sc ii}]} & Atomic & $^2$$P$$_{3/2} \rightarrow ^2$$P$$_{1/2}$ & 158 & 1900 \\
{[O{\sc i}]} & Atomic & $^3$$P$$_1 \rightarrow ^3$$P$$_2$ & 63 & 4744 \\
{[O{\sc i}]} & Atomic & $^3$$P$$_0 \rightarrow ^3$$P$$_1$ & 145 & 2060 \\
\hline
\end{tabular}
\caption{Tracer name and type, quantum mechanical transition, wavelength $\lambda$ and frequency $\nu$ of various fine structure lines, which we model with the radiative transfer code R{\sc admc}-3{\sc d}.}
\label{tab:lines}
\end{table}

\section{Results}
\label{sec:results}

In this section, we analyze various physical cloud parameters using the latest time snapshot (see Table \ref{tab:setup}). We start with the thermal state of the MCs (Section \ref{subsec:state}) and investigate which chemical components best trace the total shape of the cloud (Section \ref{subsec:shape}). Afterwards, we estimate the effective cloud radii by using synthetic observations in various diagnostic lines (Section \ref{subsec:cloudradii}). We then explore the ability of the lines listed in Table \ref{tab:lines} to trace dense MC regions (Section \ref{subsec:denseregions_tot}) and continue with a quantitative estimate of the mass fraction of the cloud that is traced by the line emission (Section \ref{subsec:massfraction}). At the end, we discuss the $X_{\text{CO}}$-factor of the MCs.

\subsection{Thermal state of the clouds}
\label{subsec:state}

\begin{figure*}
\centerline{
\includegraphics[width=0.245\linewidth]{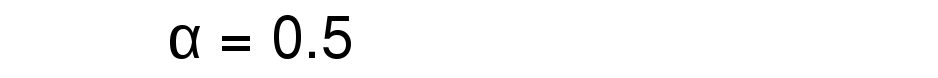}
\includegraphics[width=0.245\linewidth]{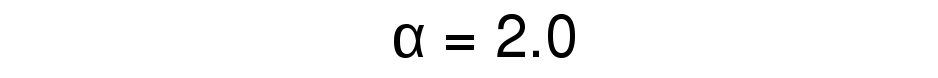}
\includegraphics[width=0.245\linewidth]{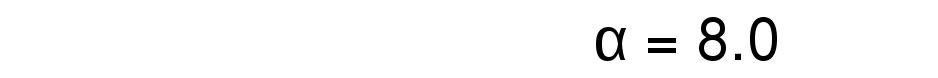}
}
\centerline{
\includegraphics[height=0.245\linewidth]{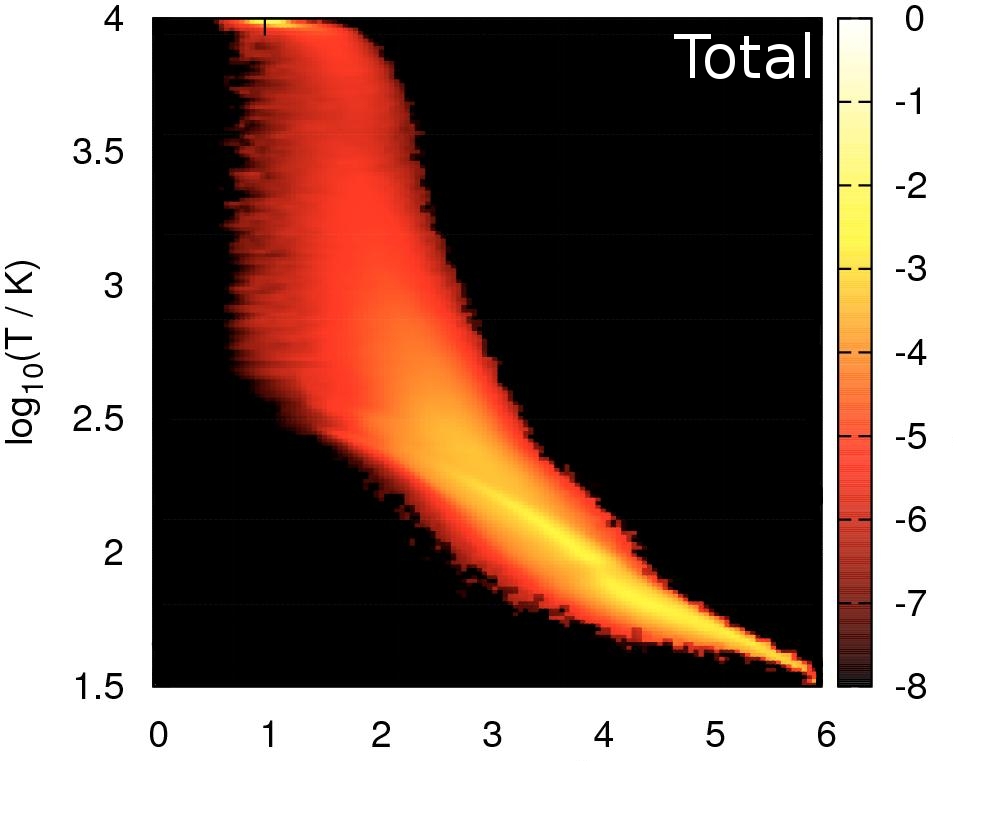}
\includegraphics[height=0.245\linewidth]{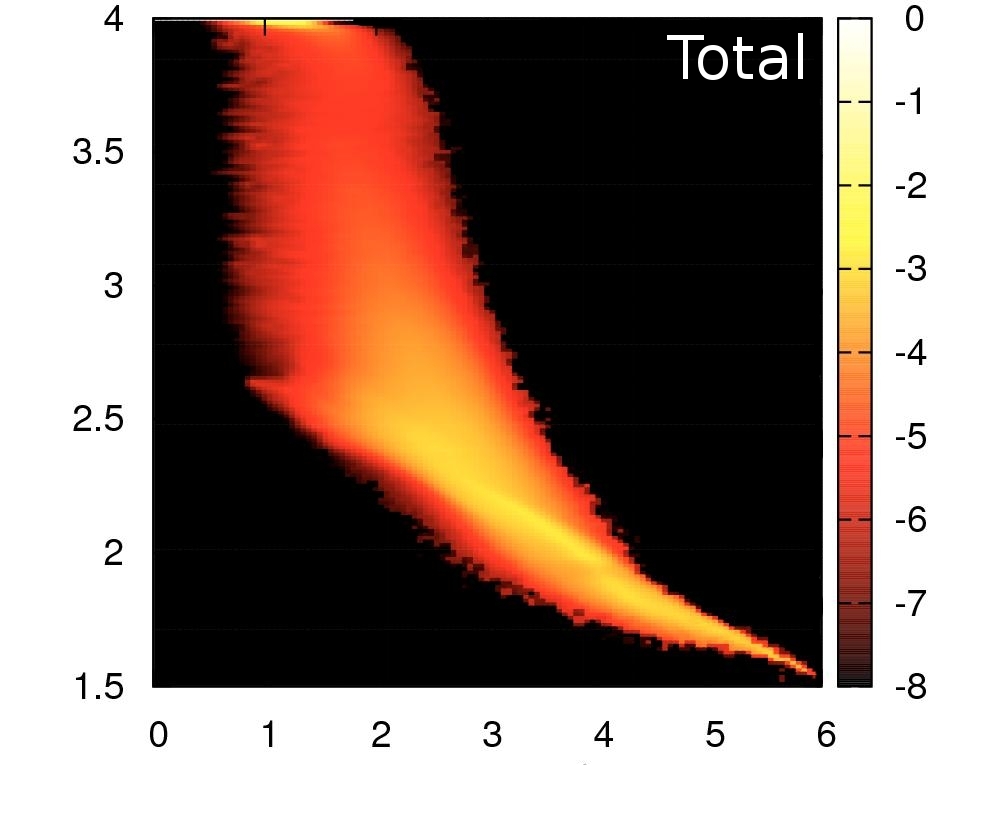}
\includegraphics[height=0.245\linewidth]{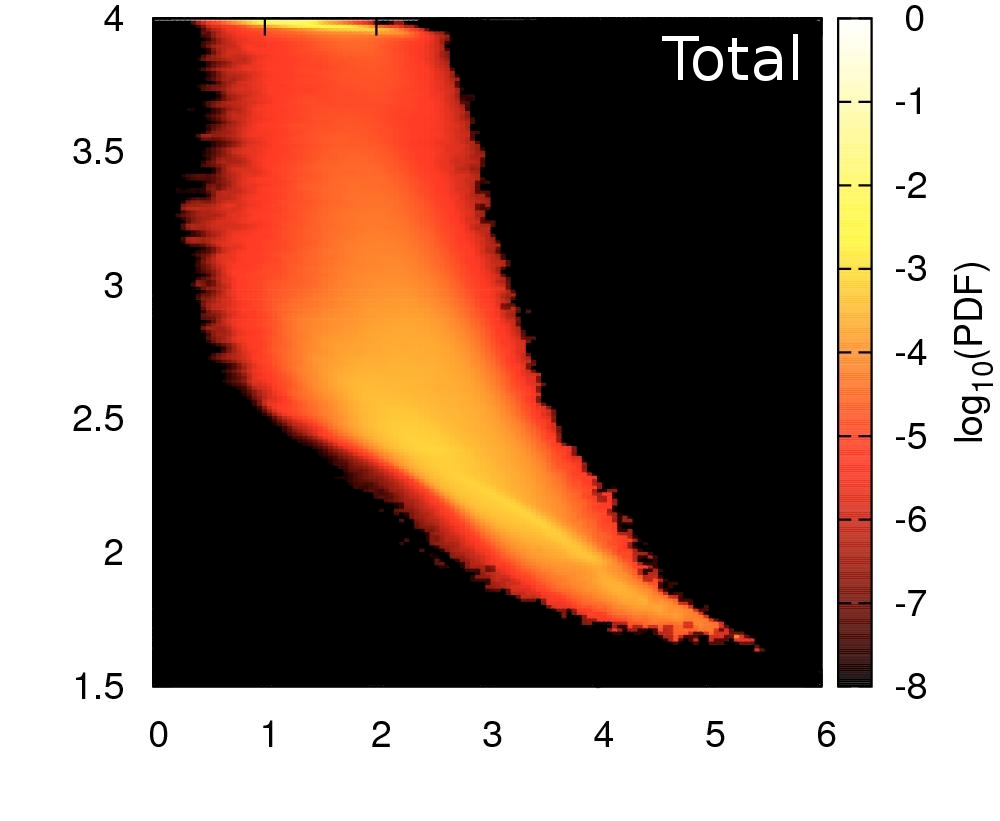}
}
\centerline{
\includegraphics[height=0.245\linewidth]{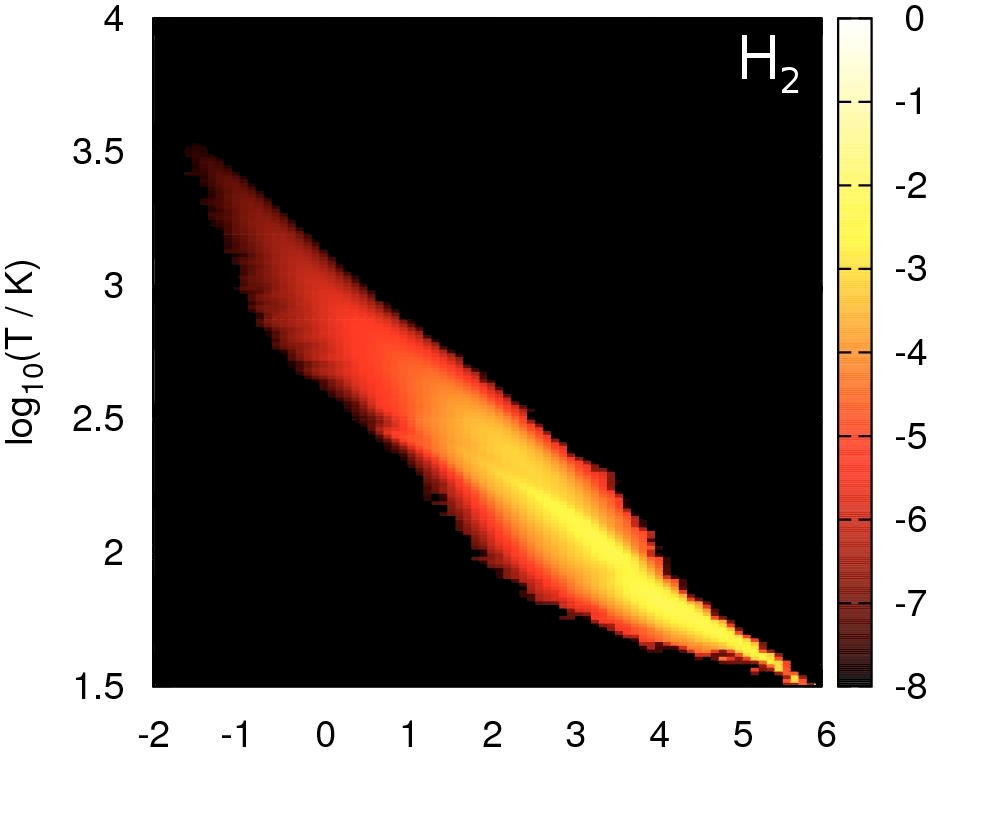}
\includegraphics[height=0.245\linewidth]{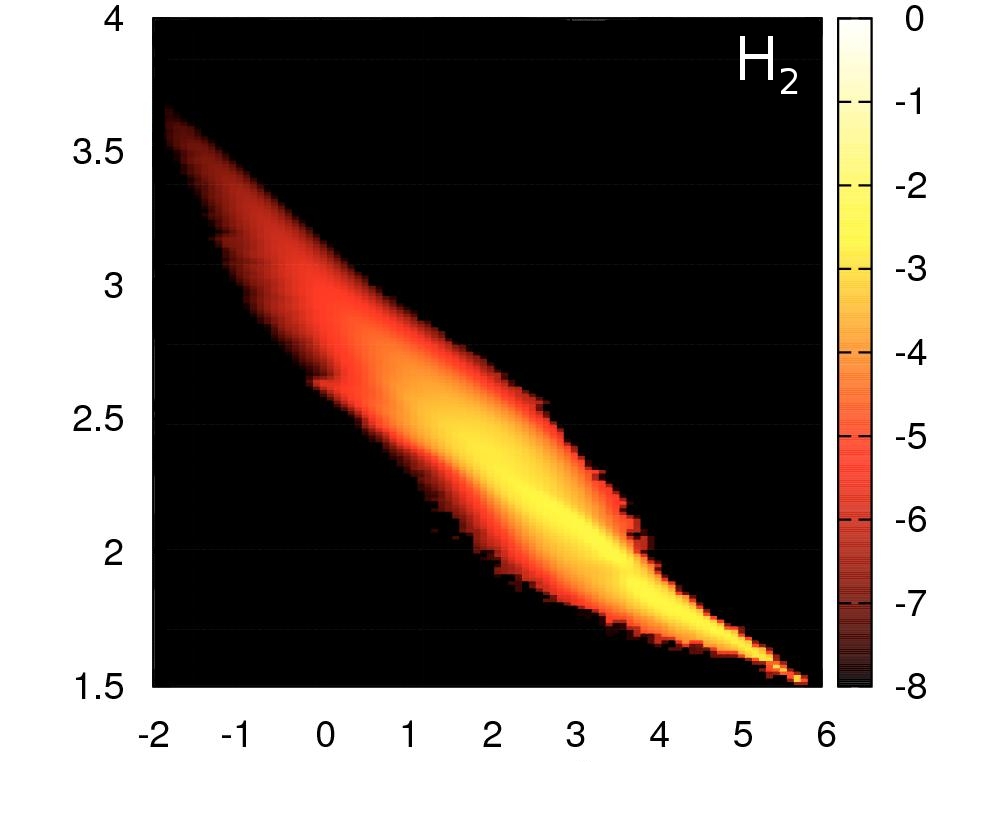}
\includegraphics[height=0.245\linewidth]{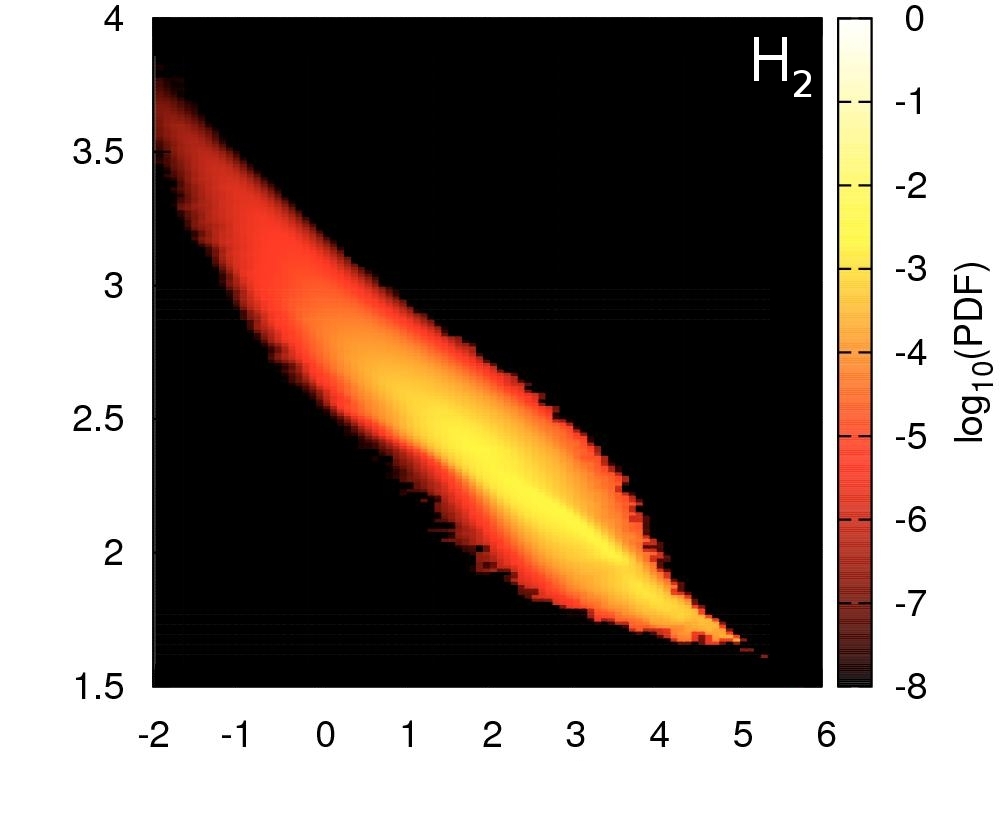}
}
\centerline{
\includegraphics[height=0.245\linewidth]{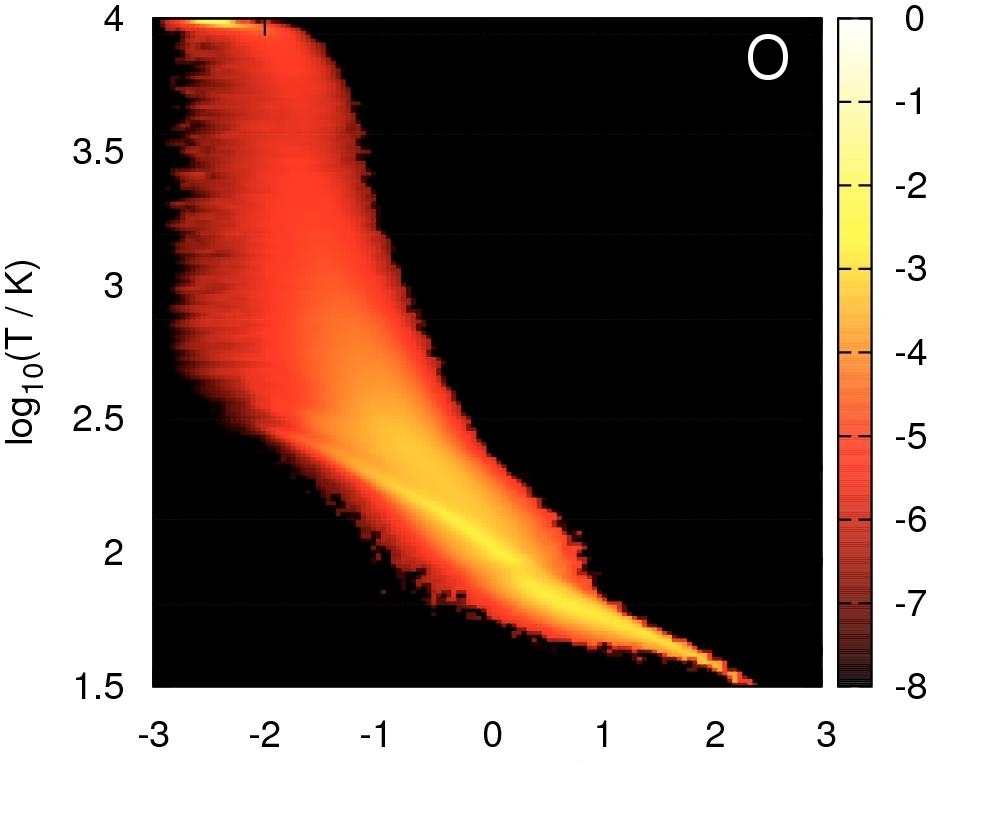}
\includegraphics[height=0.245\linewidth]{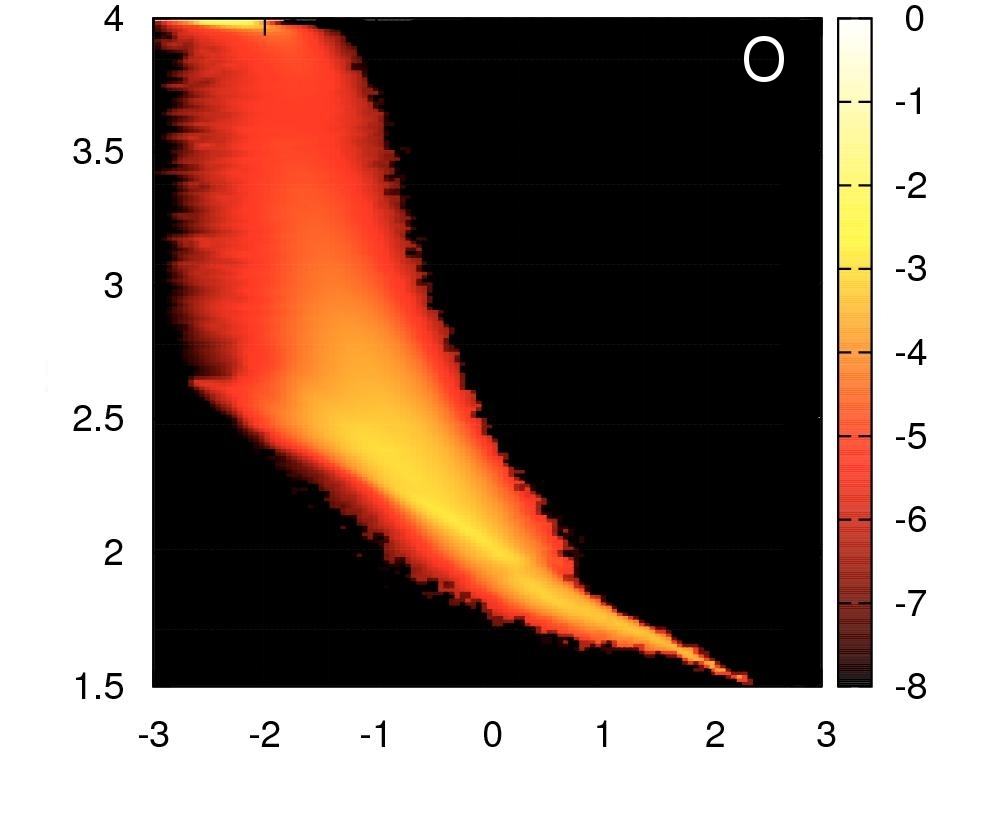}
\includegraphics[height=0.245\linewidth]{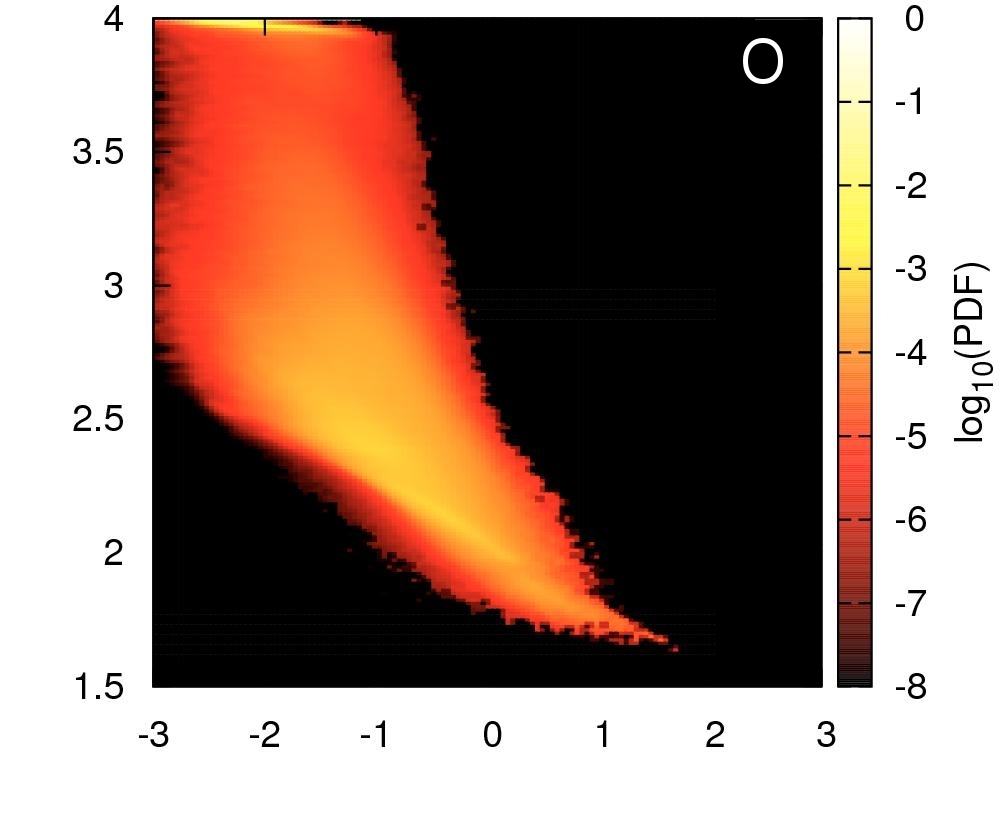}
}
\centerline{
\includegraphics[height=0.245\linewidth]{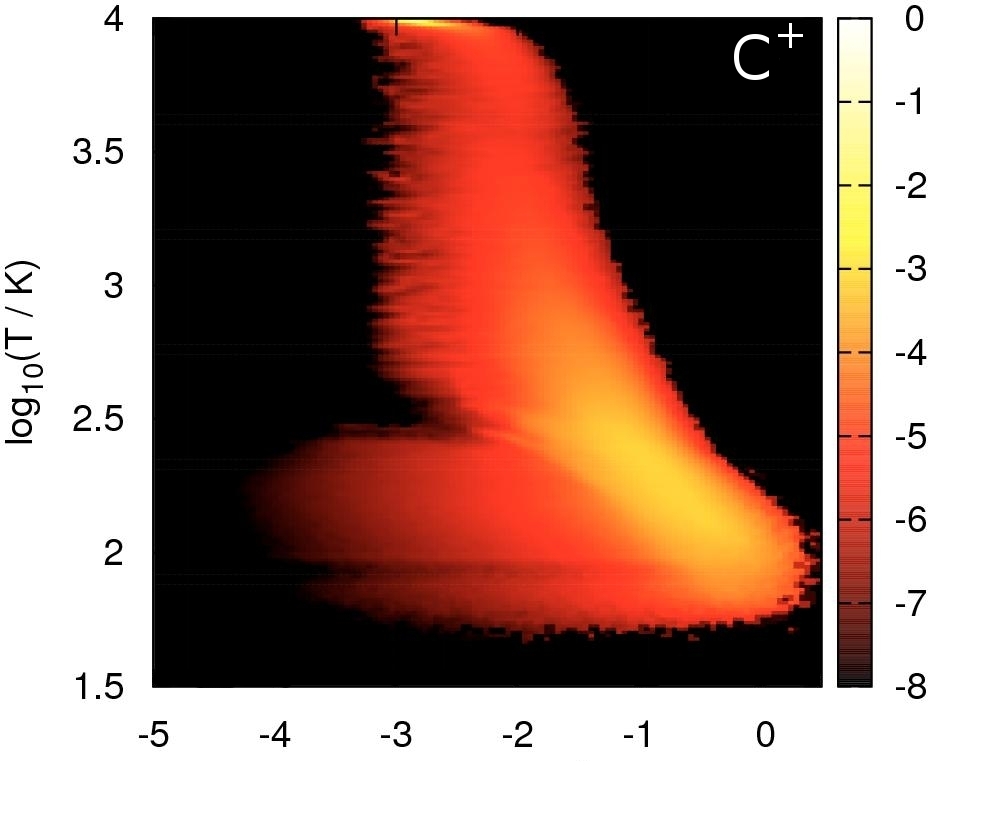}
\includegraphics[height=0.245\linewidth]{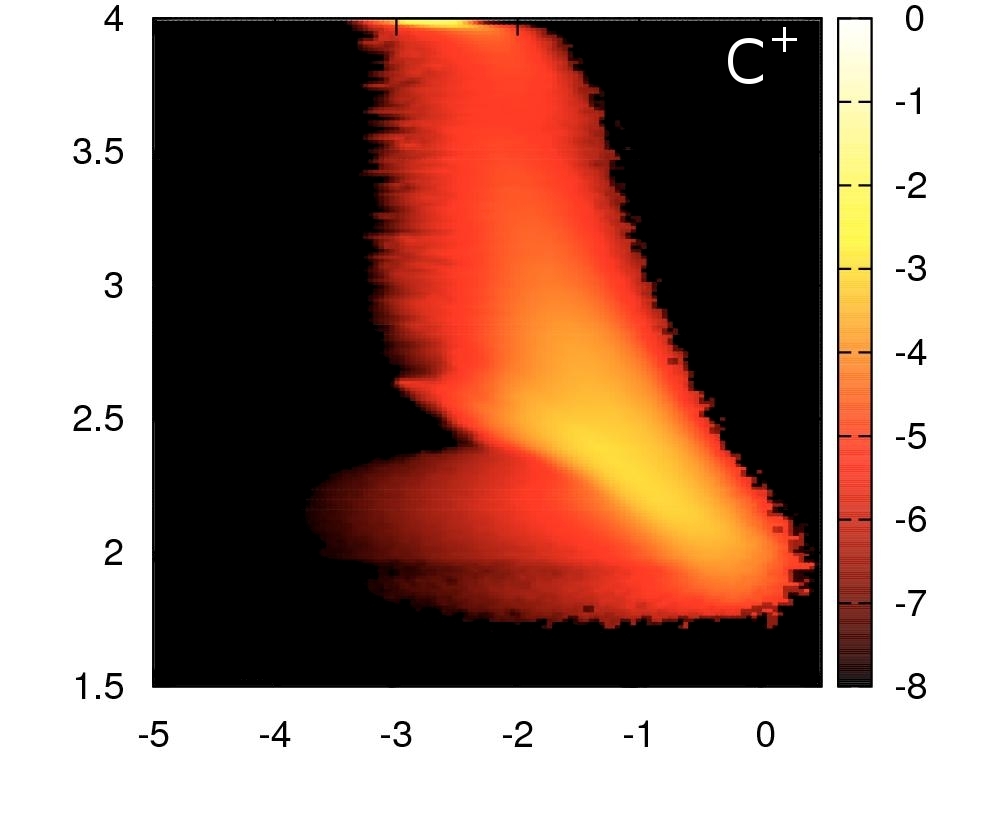}
\includegraphics[height=0.245\linewidth]{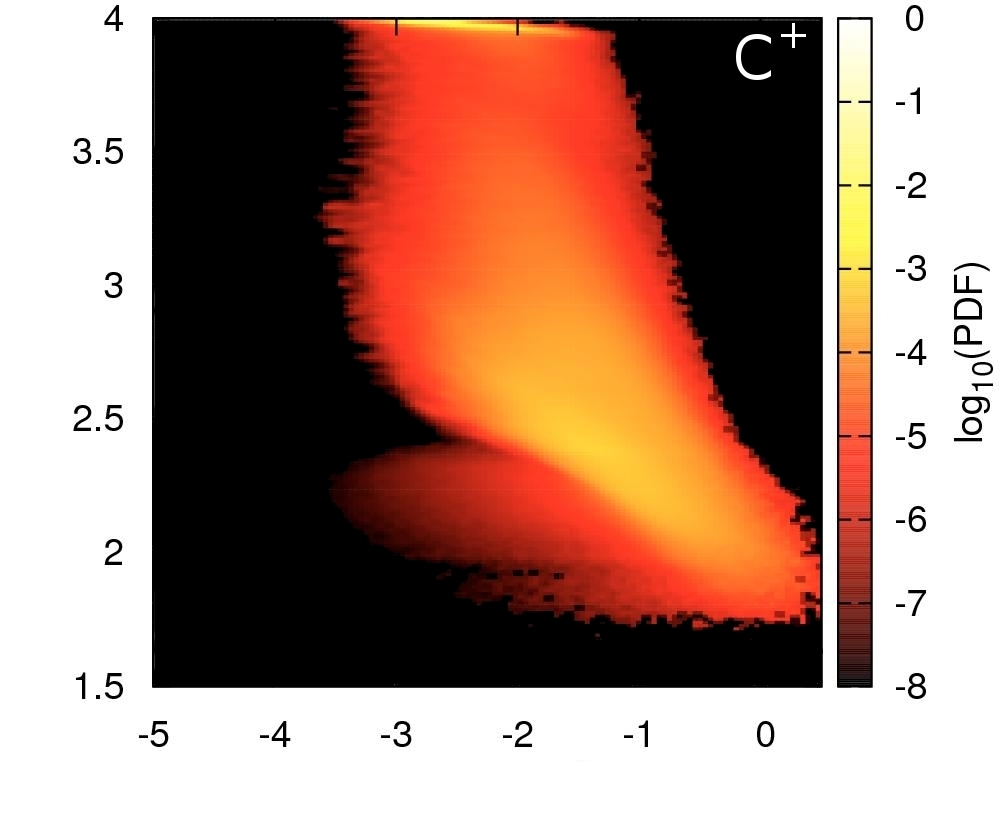}
}
\centerline{
\includegraphics[height=0.245\linewidth]{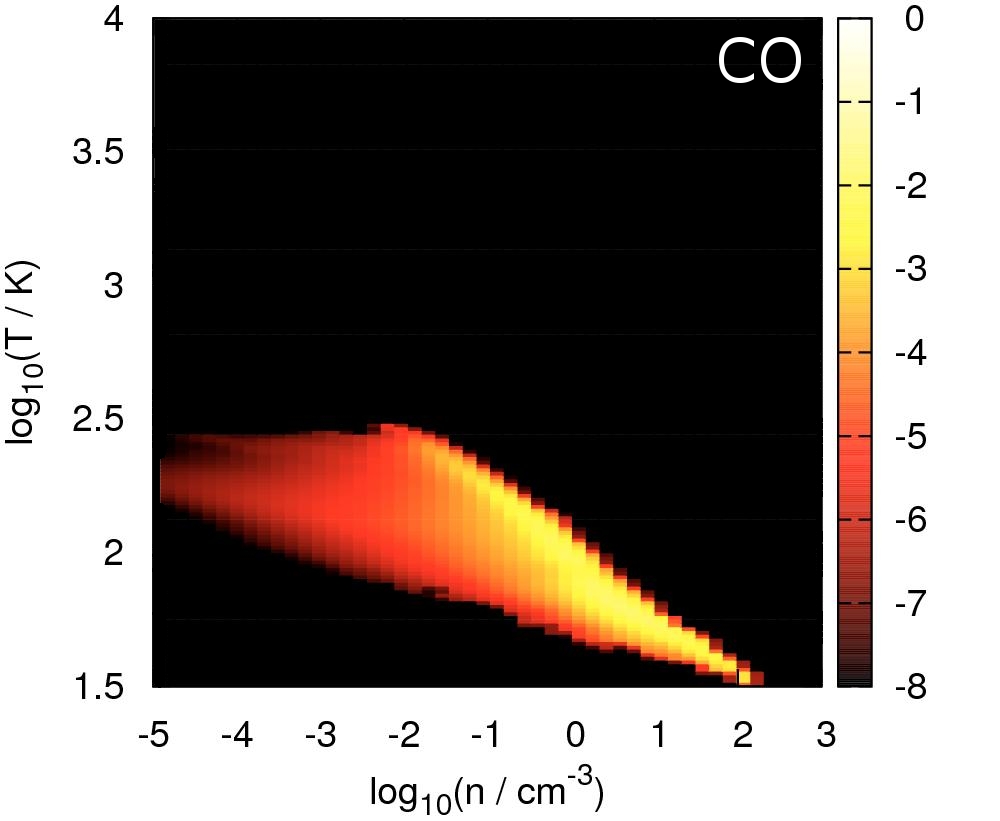}
\includegraphics[height=0.245\linewidth]{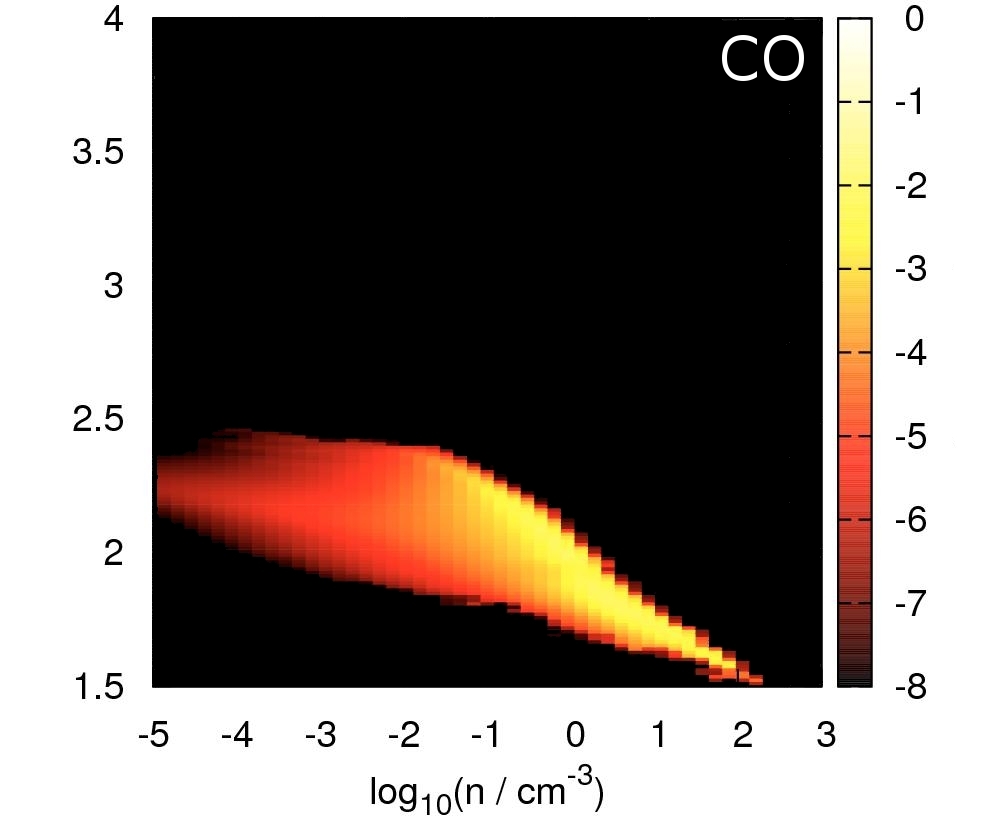}
\includegraphics[height=0.245\linewidth]{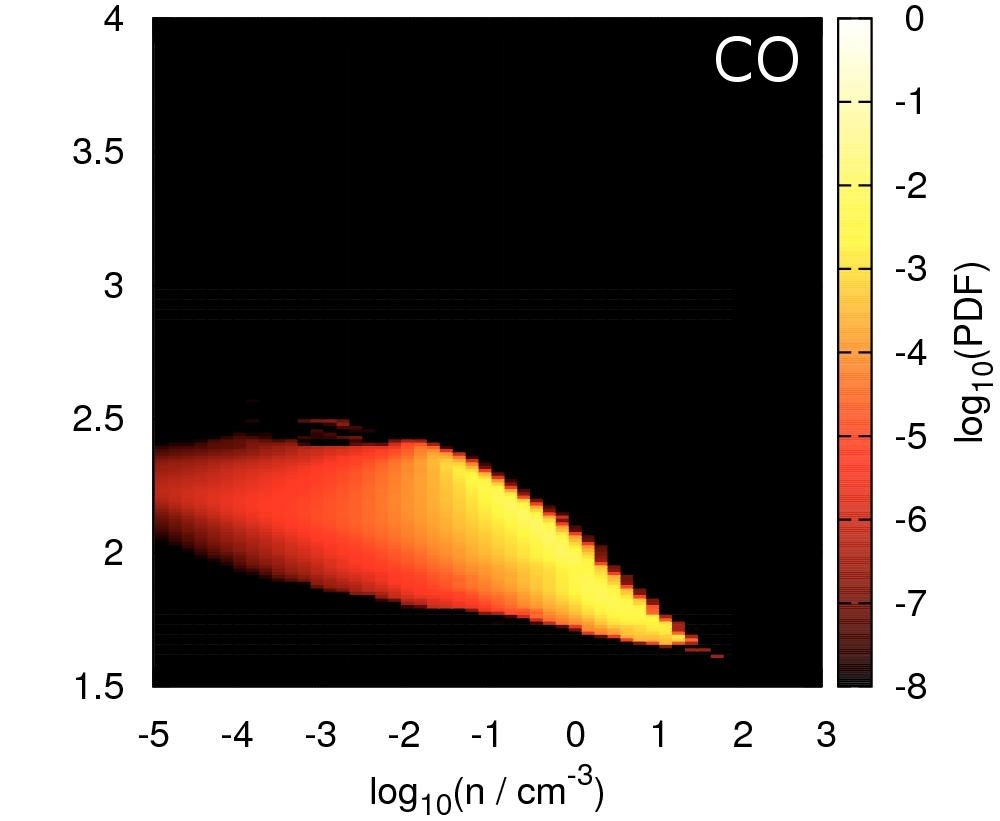}
}
\caption{Physical state of the clouds at the end of our simulations. We show two-dimensional PDFs of gas temperature and number density for various chemical components for runs with $\alpha = 0.5$ (left column), 2.0 (middle column), and 8.0 (right column). In the upper row, the number densities shown are the total particle number density. In the remaining rows, on the other hand, they are the number density of the chemical species indicated in the panel (e.g.\ $n_{\rm H_{2}}$, $n_{\rm O}$, etc.). Note that the scaling of the density axis therefore differs from panel to panel. The color coding shows the fraction of the total gas mass in each logarithmic density and temperature bin.}
\label{fig:state}
\end{figure*}

In Fig. \ref{fig:state} we show the two-dimensional PDFs of gas temperatures and densities within the clouds for the different virial parameters and for all chemical components: total mass, molecular hydrogen, atomic oxygen, ionized carbon and carbon monoxide. These PDFs are mass-weighted, with the colour-coding indicating the fraction of the total mass of the cloud (in the upper row of panels) or of the species of interest (in the remaining panels) located at each point in density-temperature space. The number densities shown in the upper row of panels are the total number density of particles; in the remaining panels, we show instead the number density of the species of interest ($n_{\rm H_{2}}$, $n_{\rm O}$, etc.). Figure \ref{fig:state} demonstrates that the thermal state of the clouds for all of the components only weakly depends on the virial parameter. However, we find strong differences in the PDFs between the various tracers.

We find a significant amount of warm gas at densities $n \lesssim 1000\,$cm$^{-3}$ and temperatures between $T \approx 300-10^4\,$K. We also note that there is a substantial scatter in the temperature at every value of the density in this regime. This gas corresponds to the warm and tenuous envelope in the outer regions of the clouds with low extinction. It is heated by the strong external ISRF via the photoelectric effect. For densities higher than $n \gtrsim 1000\,$cm$^{-3}$, we find an inverse relationship between the temperature and the density. The temperature decreases with increasing density, owing to the efficient self-shielding of the dense gas from the external radiation field. Such high densities only occur in the inner regions of the clouds with temperatures of $T < 100\,$K, which are cold enough to result in active star formation (for further information about the star formation history of our numerical simulations, we refer the reader to \citeauthor{BertramEtAl2015}~2015). These are also the parts of the cloud in which we find molecular H$_2$ and CO gas in our simulations.

If we compare the thermal state of the total mass to the thermal state of the other tracers, we find that only the atomic oxygen and ionized carbon show a similar behavior over the whole range of temperatures. This is because these components exist in the diffuse cloud regions as well as in the denser, more shielded gas. We also see that molecular hydrogen is strongly dissociated in the diffuse, warm envelope of the cloud. It only reaches maximal temperatures of $\sim4000\,$K, which is lower than the warmest regions observed for the total gas. In the case of CO, the range of temperatures ($30-300\,$K) is even smaller. This suggests that CO traces a denser, colder part of the cloud than {[C{\sc ii}]} or {[O{\sc i}]}. We will examine the consequences of this in more detail in the following sections.

\subsection{Tracing the shape of the MC}
\label{subsec:shape}

\begin{figure*}
\centerline{
\includegraphics[width=0.27\linewidth]{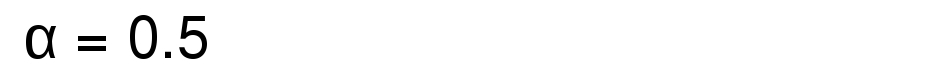}
\includegraphics[width=0.27\linewidth]{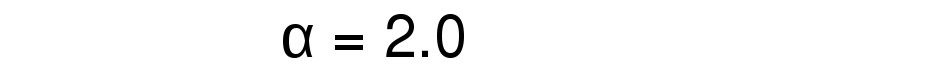}
\includegraphics[width=0.27\linewidth]{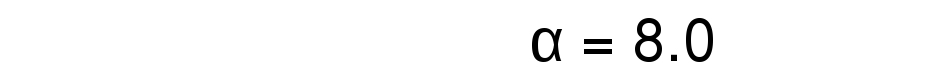}
}
\centerline{
\includegraphics[height=0.26\linewidth]{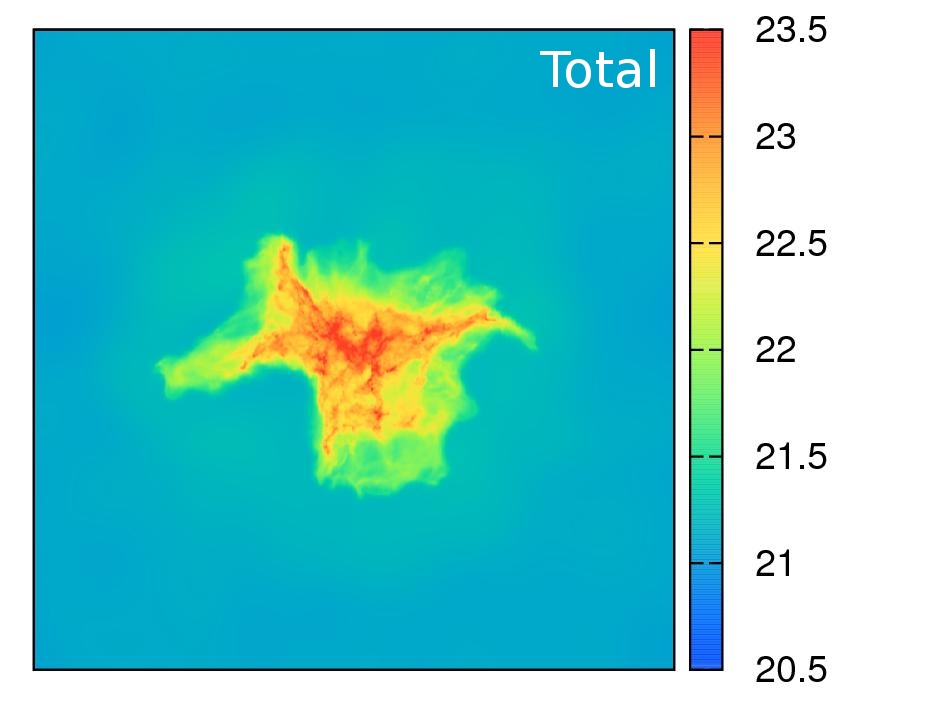}
\includegraphics[height=0.26\linewidth]{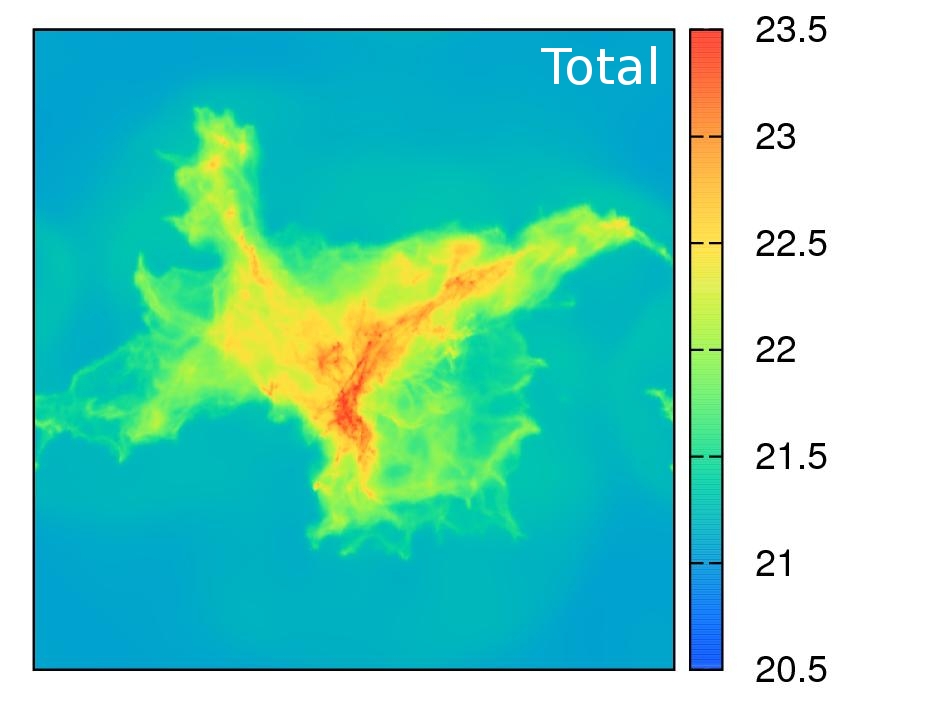}
\includegraphics[height=0.26\linewidth]{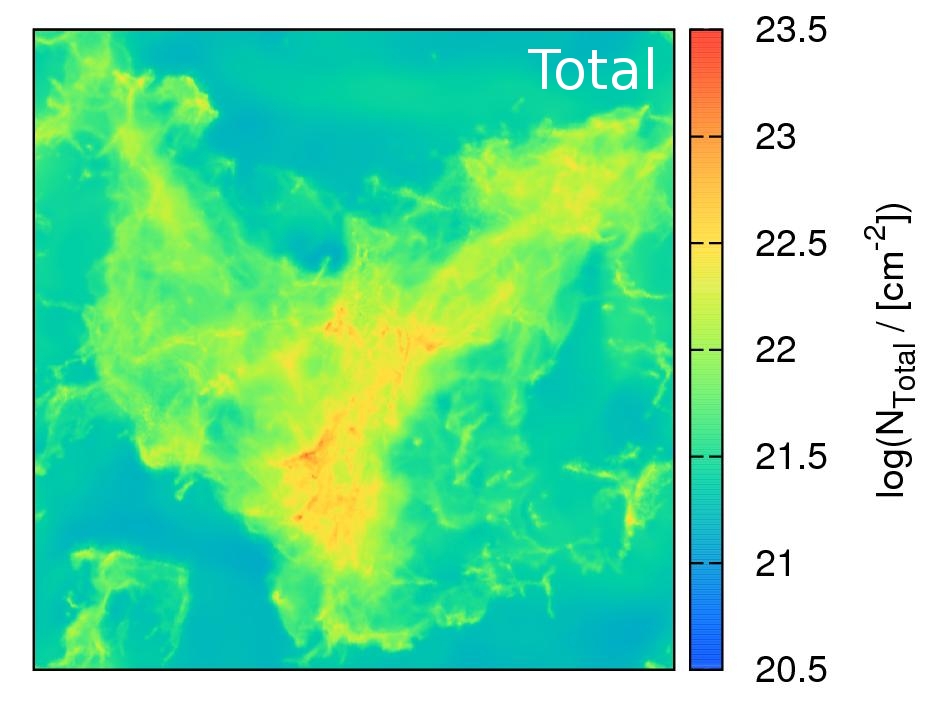}
}
\centerline{
\includegraphics[height=0.26\linewidth]{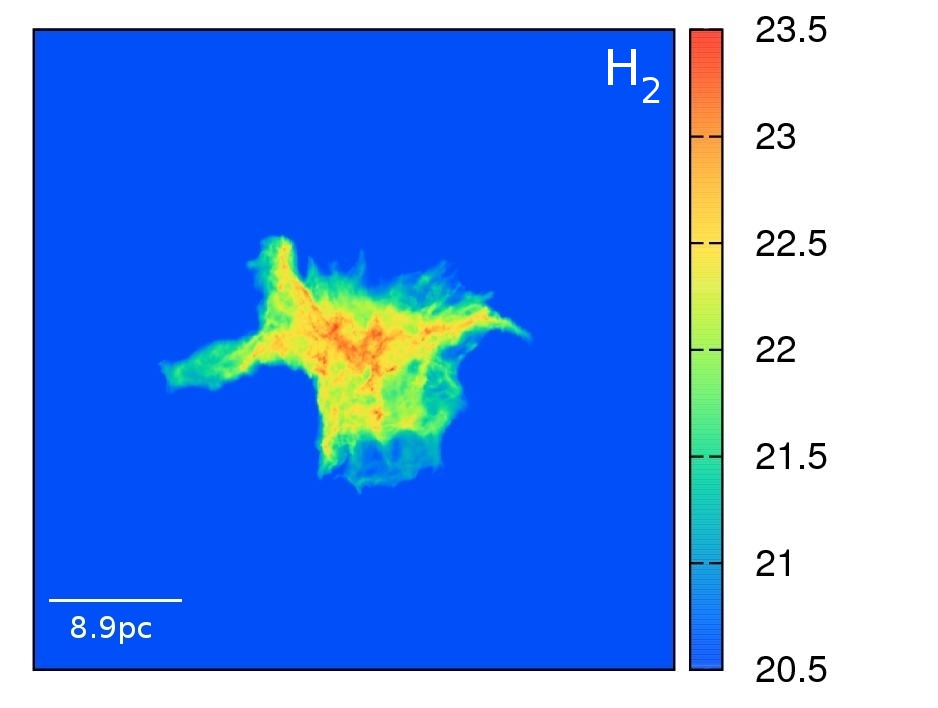}
\includegraphics[height=0.26\linewidth]{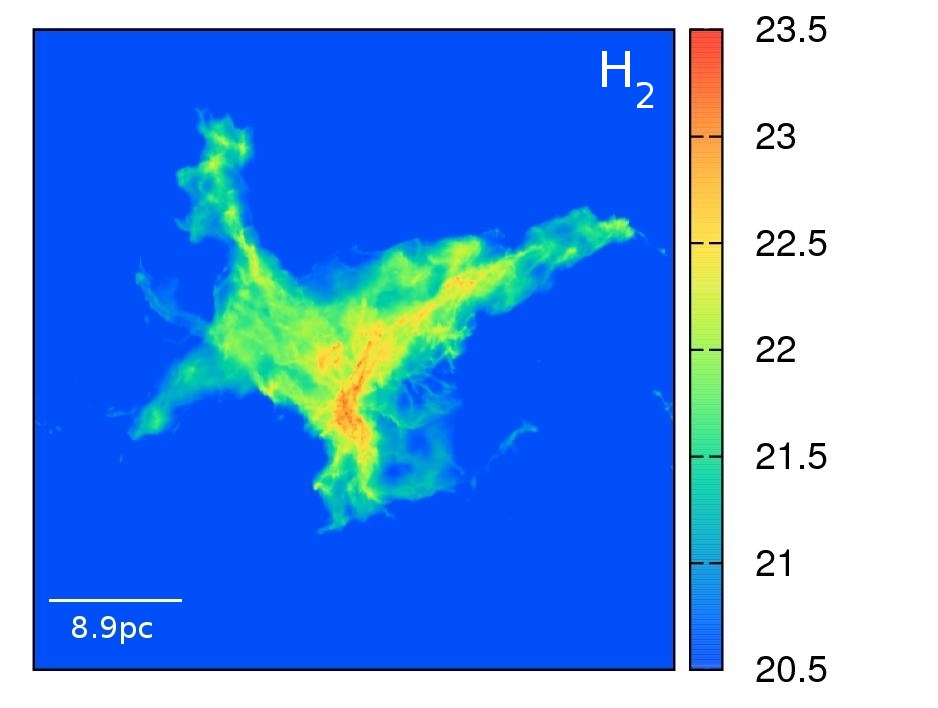}
\includegraphics[height=0.26\linewidth]{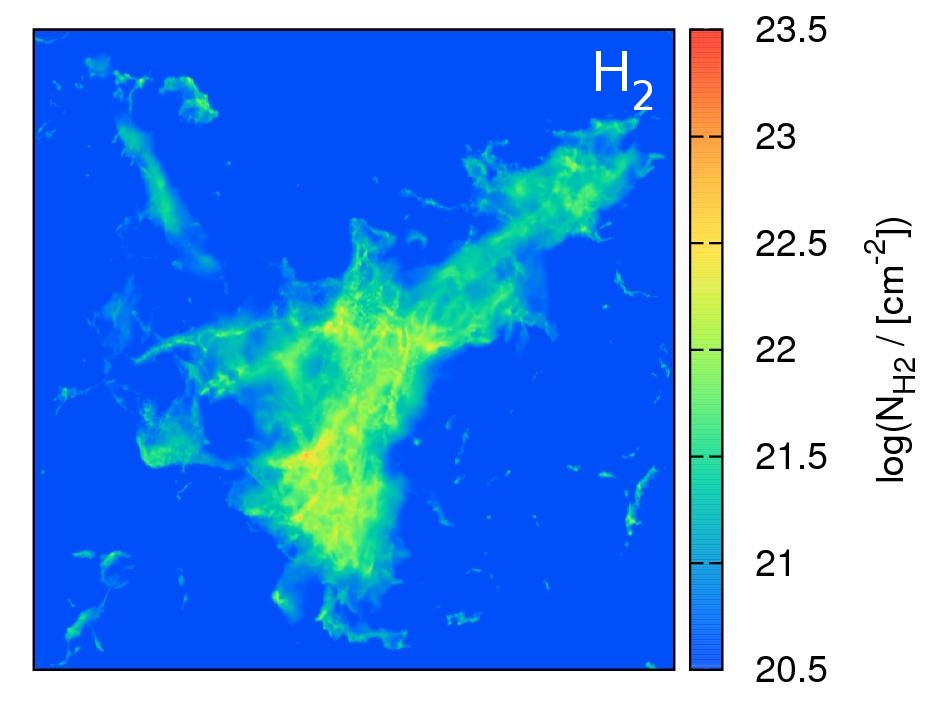}
}
\caption{Logarithmic column density maps computed along the LoS in the $z$-direction for the total density (top row) and the H$_2$ density (bottom row) for different virial parameters: $\alpha = 0.5$ (left column), 2.0 (middle column) and 8.0 (right column). Each side has a length of $44.5\,$pc. The initial cloud radius of 8.9\,pc is indicated in the bottom panels.}
\label{fig:totdensh2}
\end{figure*}

\begin{figure}
\centerline{
\includegraphics[height=0.53\linewidth]{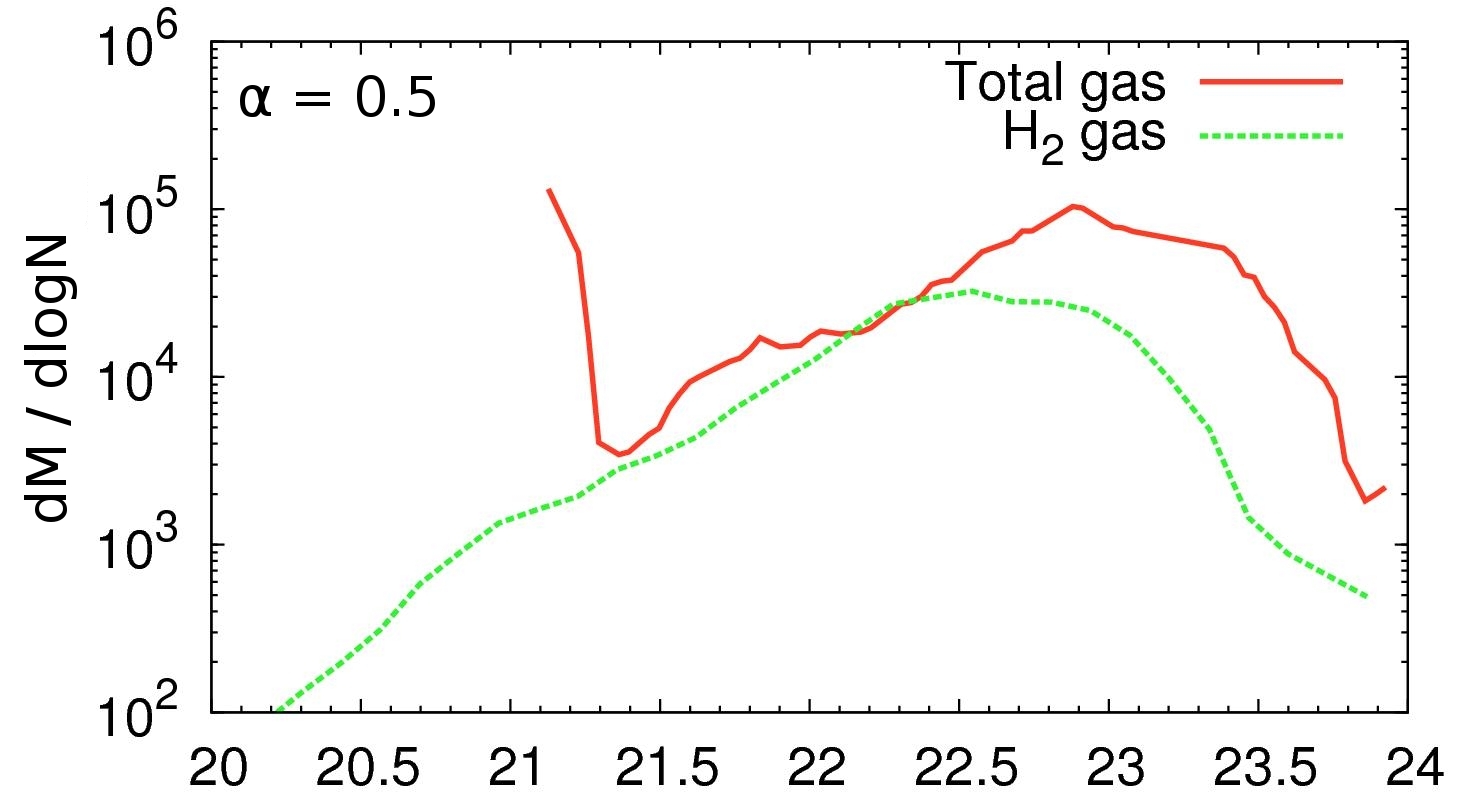}
}
\centerline{
\includegraphics[height=0.53\linewidth]{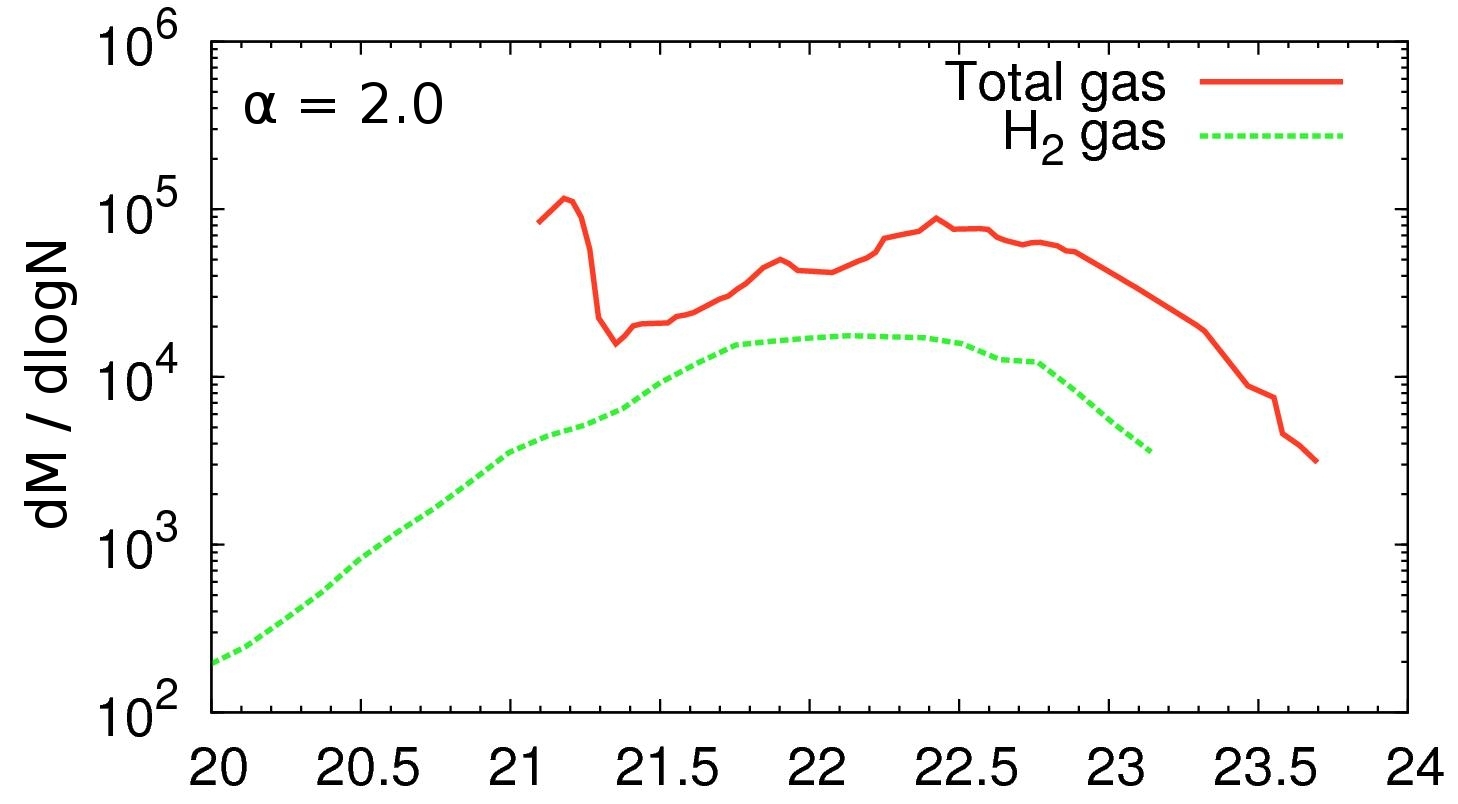}
}
\centerline{
\includegraphics[height=0.59\linewidth]{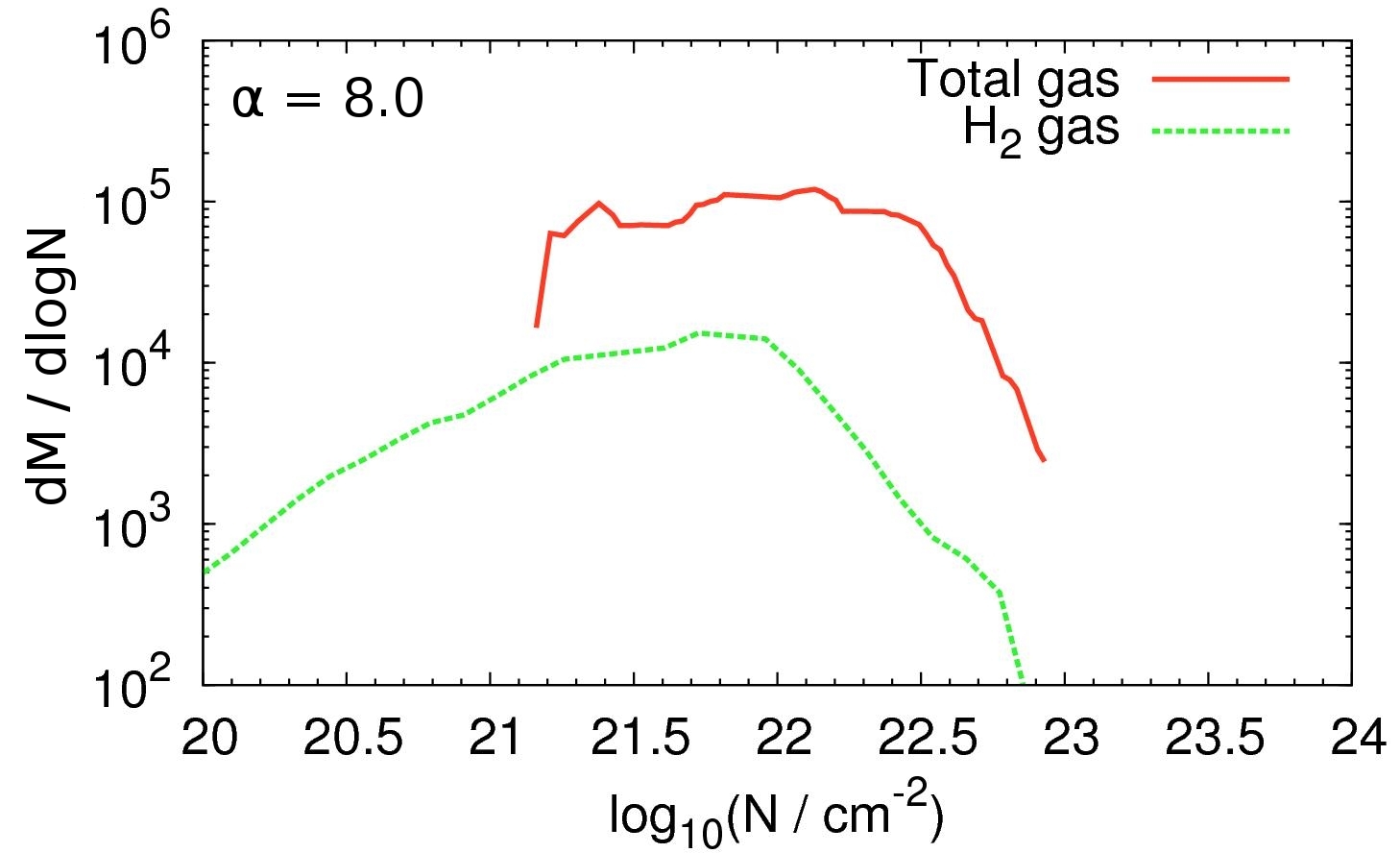}
}
\caption{Total mass per unit logarithmic column density for the total and the H$_2$ gas as shown in Fig. \ref{fig:totdensh2} for our different virial parameters: $\alpha = 0.5$ (top), 2.0 (middle) and 8.0 (bottom).}
\label{fig:cdpdf}
\end{figure}

\begin{table*}
\begin{tabular}{l|c|c|c|c|c}
\hline\hline
Model & {[C{\sc ii}]} (158\,$\mu$m) & [O{\sc i}] (145\,$\mu$m) & [O{\sc i}] (63\,$\mu$m) & $^{12}$CO (2600\,$\mu$m) & $^{13}$CO (2720\,$\mu$m) \\
 & [K\,km\,s$^{-1}$] & [K\,km\,s$^{-1}$] & [K\,km\,s$^{-1}$] & [K\,km\,s$^{-1}$] & [K\,km\,s$^{-1}$] \\
\hline
GC-0.5-1000 & 3.0 & 1.0 & 0.6 & 1.0 & 1.0 \\
GC-2.0-1000 & 3.0 & 1.0 & 0.6 & 0.3 & 0.3 \\
GC-8.0-1000 & 3.0 & 0.8 & 0.5 & 0.1 & 0.1 \\
\hline
\end{tabular}
\caption{Threshold values $W_{\text{thresh}}$ used for the integrated intensity maps in Fig. \ref{fig:tracers} in order to estimate the effective cloud radius $R_{\text{eff}}$. The values are chosen such that the denser cloud regions can be clearly separated from the diffuse gas phase around it, leading to a reasonable estimate for the effective cloud radii presented in Table \ref{tab:radii}.}
\label{tab:thresh}
\end{table*}

\begin{table*}
\begin{tabular}{l|c|c|c|c|c|c}
\hline\hline
Model & 80\% of total mass & {[C{\sc ii}]} (158\,$\mu$m) & [O{\sc i}] (145\,$\mu$m) & [O{\sc i}] (63\,$\mu$m) & $^{12}$CO (2600\,$\mu$m) & $^{13}$CO (2720\,$\mu$m) \\
 & [pc] & [pc] & [pc] & [pc] & [pc] & [pc] \\
\hline
GC-0.5-1000 & 11.0 & 9.0 & 8.7 & 8.1 & 6.6 & 6.2 \\
GC-2.0-1000 & 13.1 & 13.8 & 13.0 & 12.2 & 7.6 & 6.8 \\
GC-8.0-1000 & 17.5 & 20.1 & 19.9 & 19.7 & 8.4 & 6.9 \\
\hline
\end{tabular}
\caption{Effective radii of the clouds shown in Fig. \ref{fig:tracers}, as inferred from equation \ref{eq:Reff}, based on the threshold intensities given in Table \ref{tab:thresh}. For comparison, we also give the effective radii of the clouds that incorporate $\sim80$\% of the total gas mass in the second column.}
\label{tab:radii}
\end{table*}

Fig.~\ref{fig:totdensh2} shows logarithmic column density maps for the total density and the H$_2$ density computed along the LoS in the $z$-direction for virial parameters $\alpha = 0.5, 2.0$ and 8.0 for the last time snapshots (see Table \ref{tab:setup}). We see that as we increase $\alpha$, and hence the velocity dispersion of the gas, the structure of the cloud changes significantly. When $\alpha = 0.5$ and the cloud is gravitationally bound, it remains relatively compact. It is therefore able to shield itself relatively well from the effects of the high external ISRF, with the result that the H$_{2}$ column density traces the total column density fairly well. In the runs with $\alpha = 2$ and $\alpha = 8$, however, the cloud is not gravitationally bound and hence starts to expand. In addition, the strong turbulence opens up large channels into the cloud by lowering the column density locally, allowing radiation to penetrate deeper into the MC, leading to enhanced photodissociation of the molecular hydrogen. Hence, we do not find a good correlation between the total and the H$_2$ column density in the $\alpha = 2.0$ and $\alpha = 8.0$ runs. Molecular hydrogen can only exist in the most dense regions of the cloud, where it is best able to self-shield from the strong external radiation field.

This is also illustrated in Fig. \ref{fig:cdpdf}, which shows the corresponding column density PDFs for three virial parameter models. If we compare the H$_2$ column density PDFs with each other, we find that the peak column density in the $\alpha = 0.5$ model is shifted to larger values compared to the $\alpha = 8.0$ model by about one order of magnitude. This is because the turbulent kinetic energy in the virialized $\alpha = 0.5$ run is much smaller than in the $\alpha = 8.0$ run, leading to denser cloud regions and thus to larger average column densities. Moreover, the fraction of H$_2$ gas in the total box decreases with increasing virial parameter, owing to the high external ISRF, which photodissociates those molecular components more effectively as the radiation can penetrate deeper into the cloud's interior. Furthermore, we see that the H$_2$ gas generally extends to much lower column densities than the total gas. This is because the total gas consists of both atomic and molecular hydrogen and thus does not fall below lower values than $N_{\text{tot}} \approx 10^{21}\,$cm$^{-2}$ due to the H{\sc i} background, while the fraction of H$_2$ gas can continuously decrease in the outer regions due to photodissociation.

In analogy to Fig. \ref{fig:totdensh2}, Fig. \ref{fig:tracers} shows velocity-integrated intensity maps computed along the LoS in the $z$-direction for the same virial parameters and the different tracers presented in Table \ref{tab:lines}. We again find that the molecular tracers $^{12}$CO and $^{13}$CO are photodissociated at the edges of the MC by the strong external ISRF. This effect is even stronger if the amount of turbulent kinetic energy in the box is high. Comparing the different tracers to the total gas column in Fig. \ref{fig:totdensh2}, we find that the cloud's shape is well reproduced by all atomic tracers, although the integrated brightness temperatures of these lines are significantly smaller than those of the $^{12}$CO and $^{13}$CO lines. This holds for all three virial parameter models. Conversely, in the case of CO, we find that neither of its isotopologues describes the total gas distribution of the cloud. However, we observe that they do remain reasonably good tracers of the dense H$_2$, owing to the effect of self-shielding. We will quantify this finding in Section \ref{subsec:massfraction}.

\subsection{Estimating the effective cloud radii}
\label{subsec:cloudradii}

In the previous section we established that carbon monoxide is not a good tracer for the total column density, while the atomic species better reflect the distribution of the total gas mass. In this context, CO tends to significantly underestimate the radius of the total cloud in such an extreme environment. This can compromise any observational estimates of the virial parameter $\alpha$, which requires an accurate estimate for the cloud's radius (see also our follow-up study in Paper II). However, we note that it is generally complicated to define a proper cloud radius. Molecular clouds are complex hierarchical systems and estimates of cloud radii always come with some assumptions of how to define outer cloud boundaries. Hence, if we refer to the effective cloud radius in the following, we always mean the radius of a spherical cloud with the same surface area.

\begin{figure*}
\centerline{
\includegraphics[width=0.26\linewidth]{images/alpha05_map.jpg}
\includegraphics[width=0.26\linewidth]{images/alpha2_map.jpg}
\includegraphics[width=0.26\linewidth]{images/alpha8_map.jpg}
}
\centerline{
\includegraphics[height=0.245\linewidth]{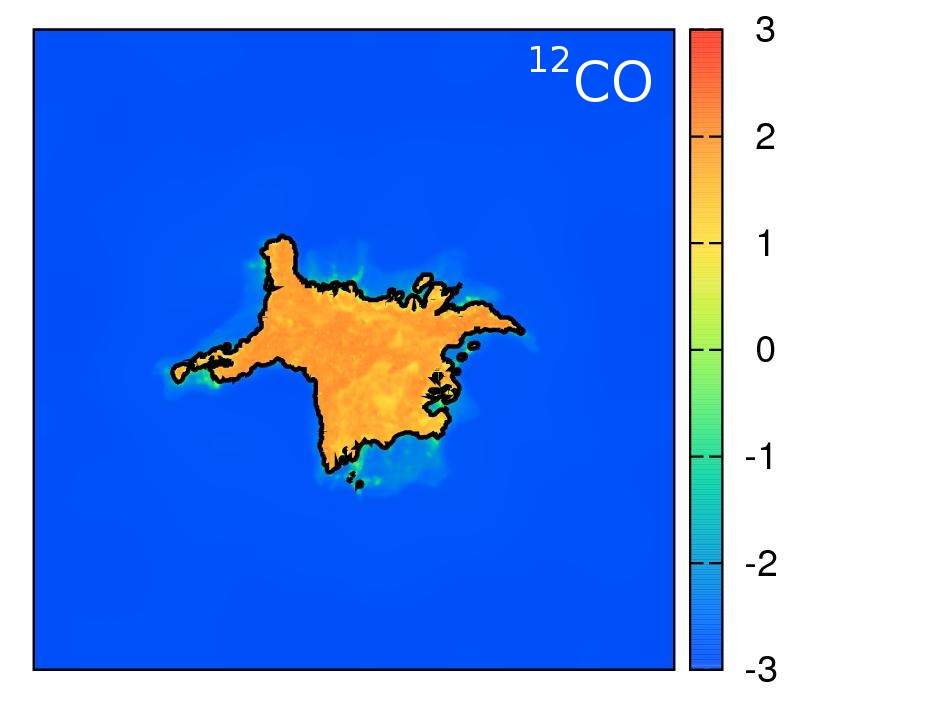}
\includegraphics[height=0.245\linewidth]{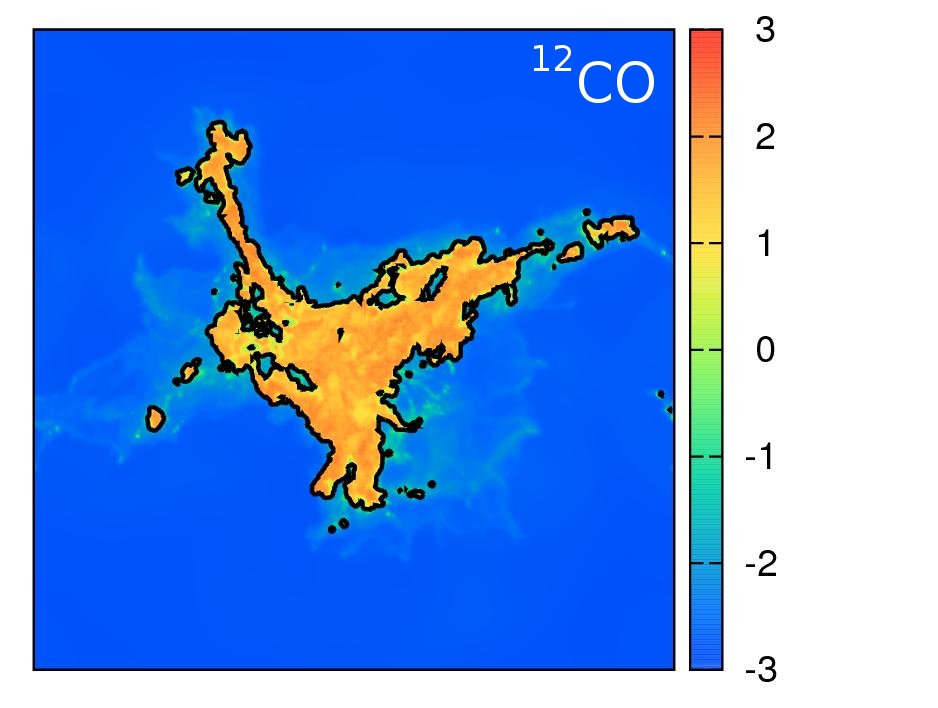}
\includegraphics[height=0.245\linewidth]{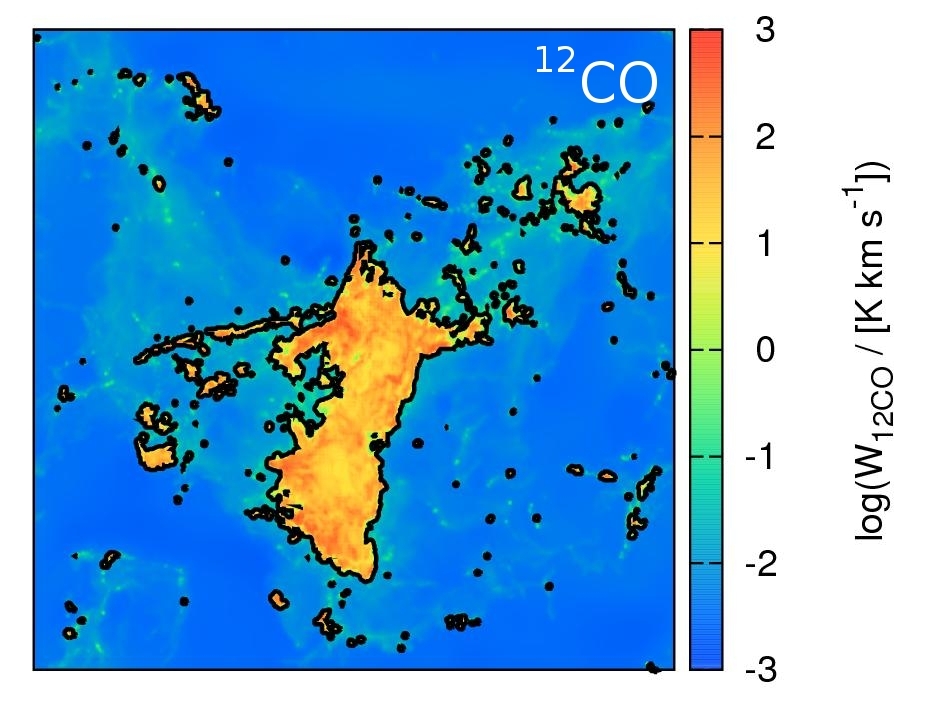}
}
\centerline{
\includegraphics[height=0.245\linewidth]{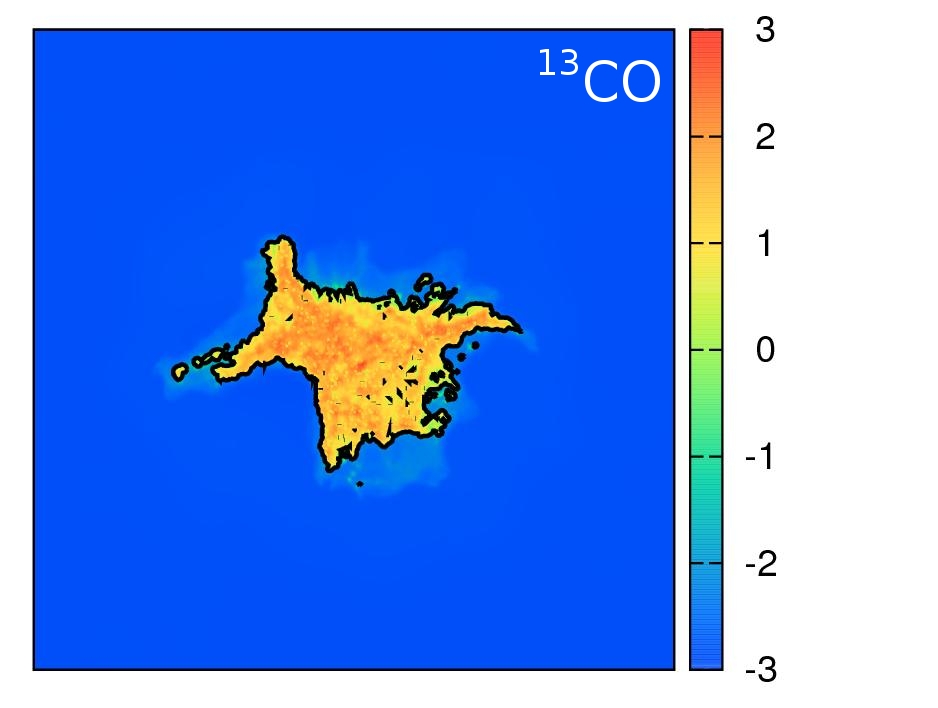}
\includegraphics[height=0.245\linewidth]{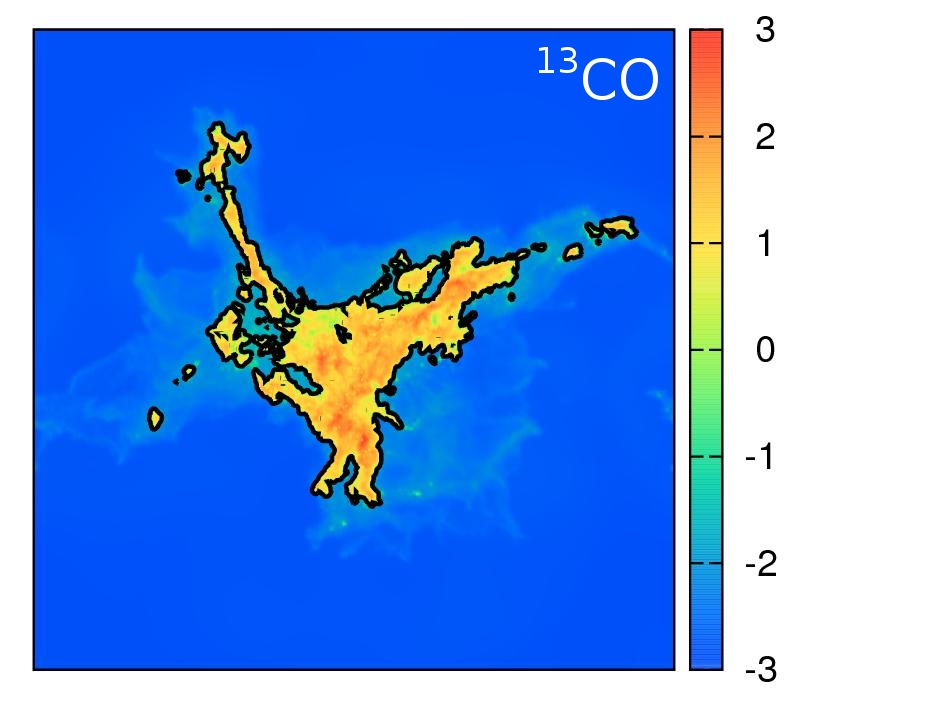}
\includegraphics[height=0.245\linewidth]{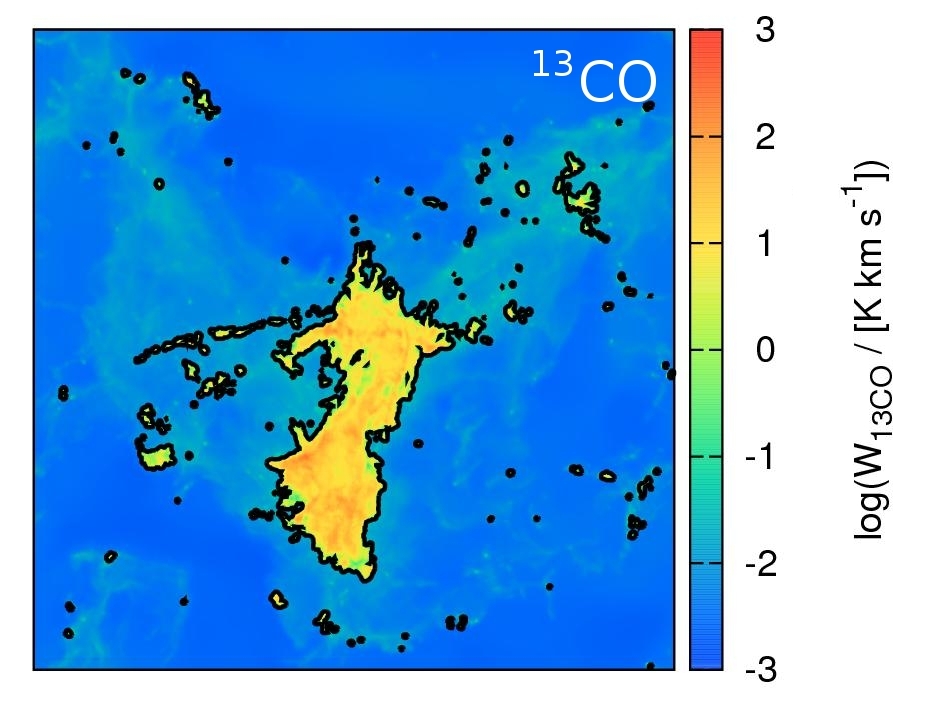}
}
\centerline{
\includegraphics[height=0.245\linewidth]{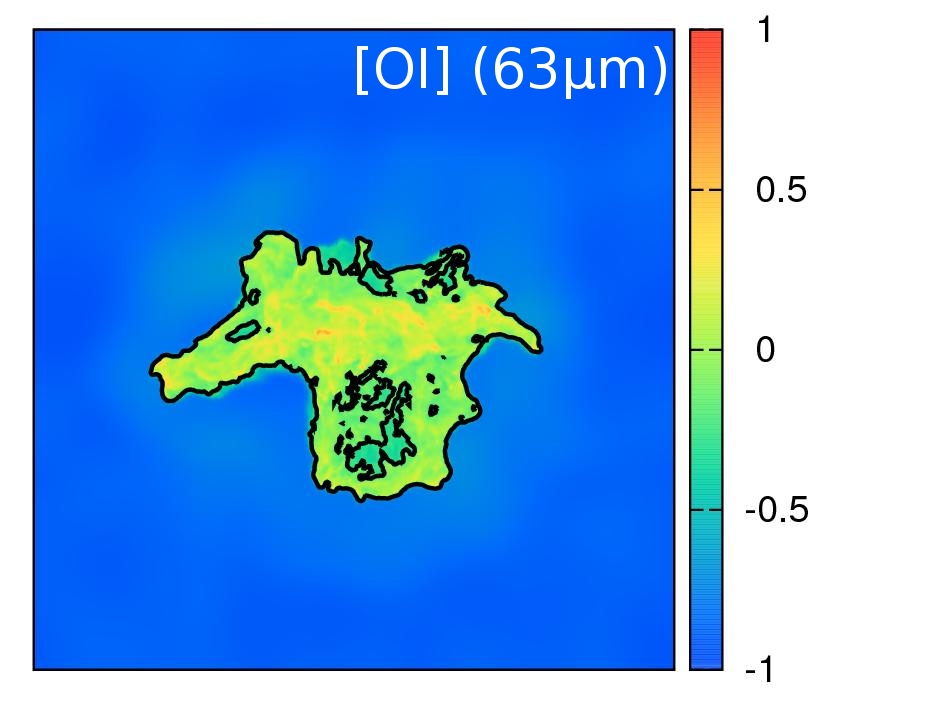}
\includegraphics[height=0.245\linewidth]{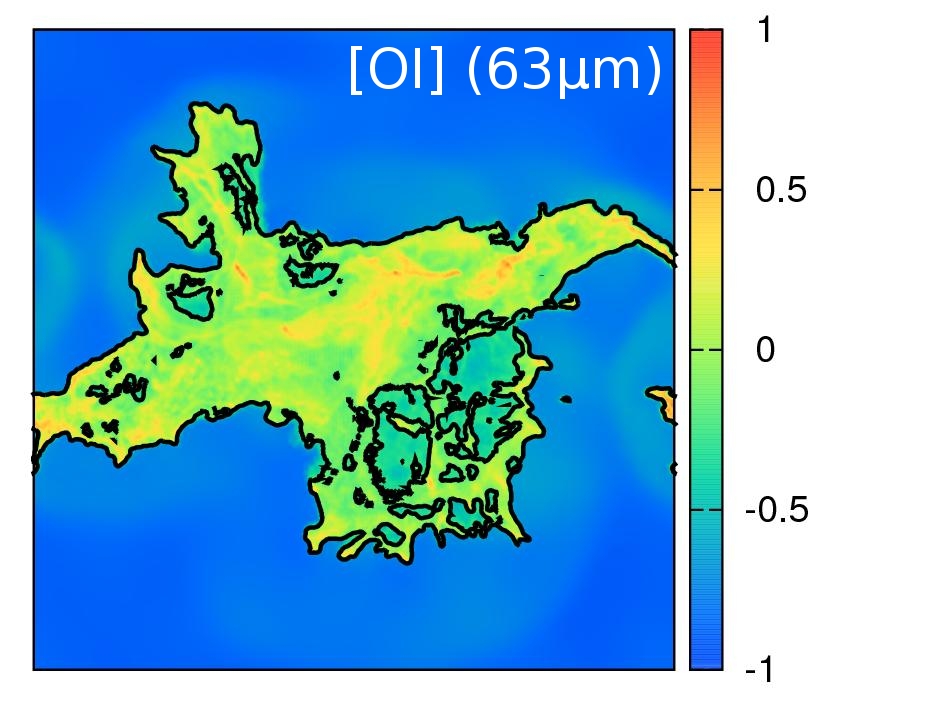}
\includegraphics[height=0.245\linewidth]{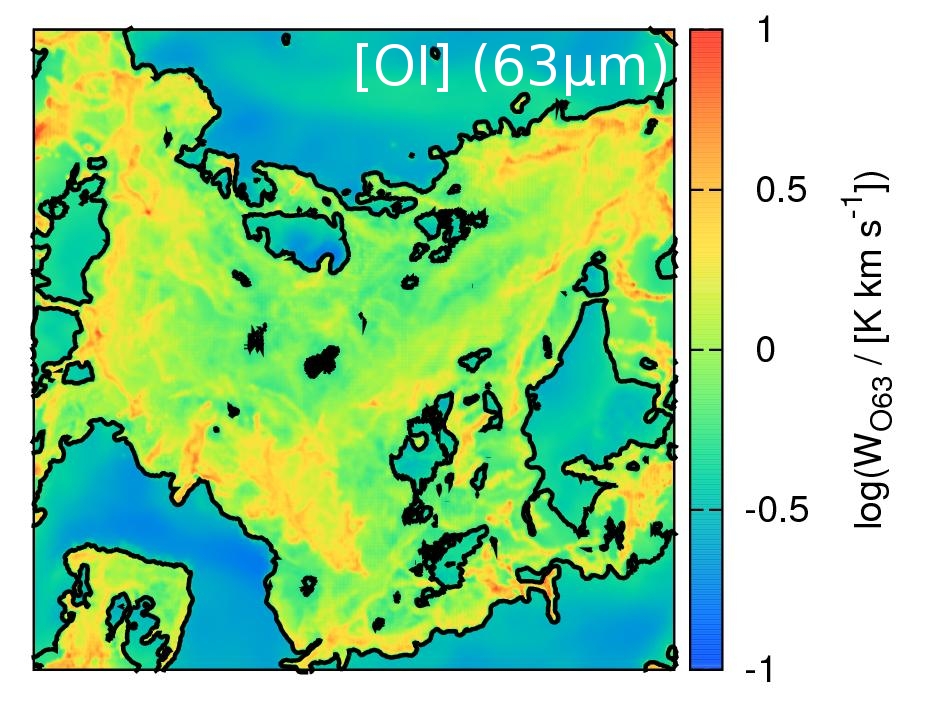}
}
\centerline{
\includegraphics[height=0.245\linewidth]{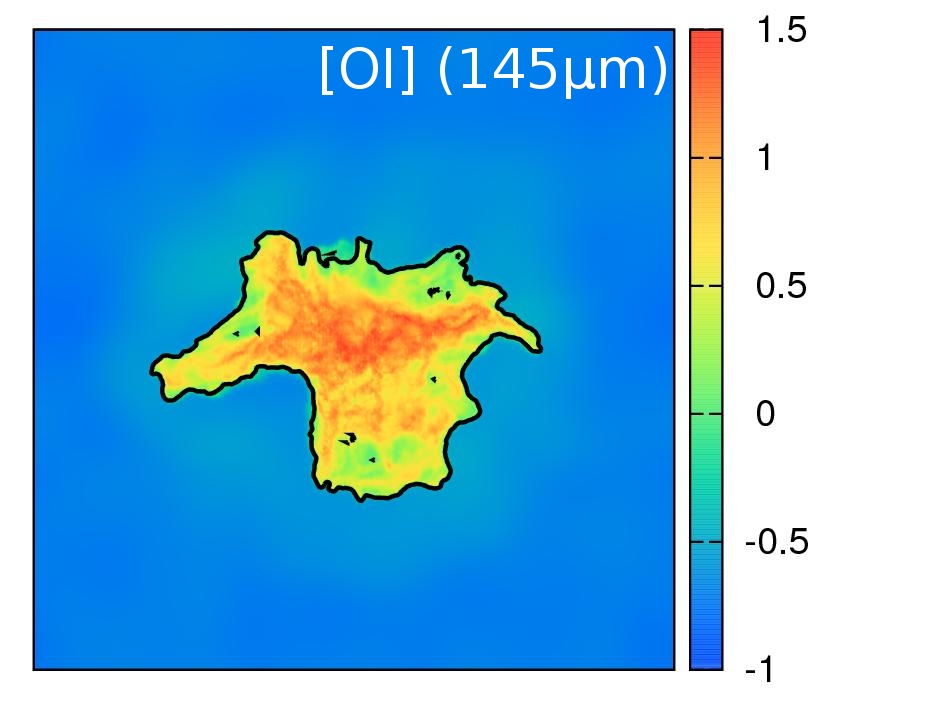}
\includegraphics[height=0.245\linewidth]{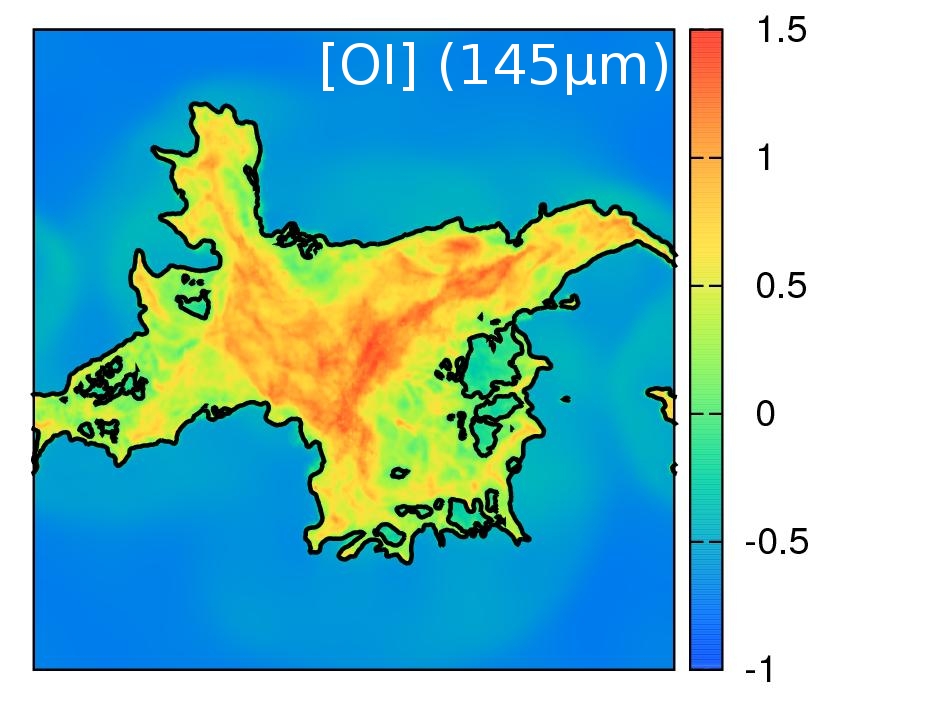}
\includegraphics[height=0.245\linewidth]{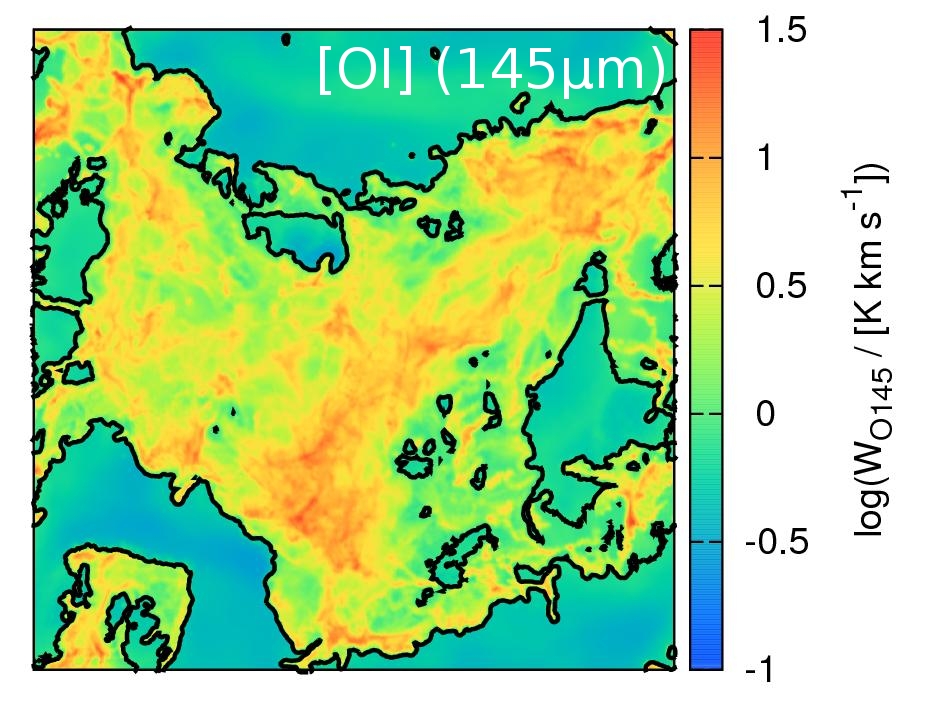}
}
\centerline{
\includegraphics[height=0.245\linewidth]{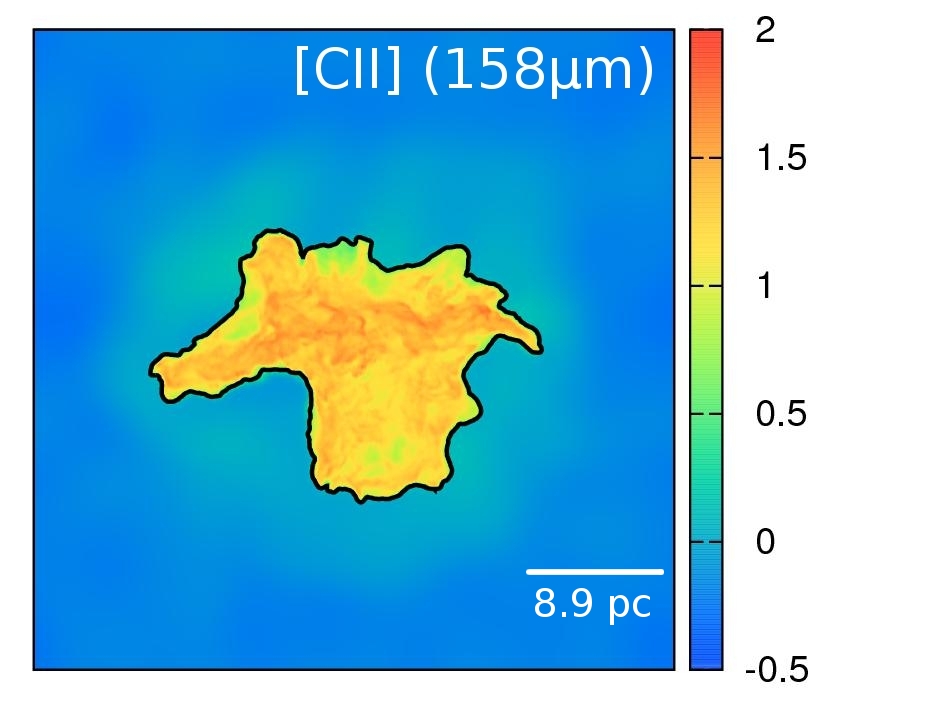}
\includegraphics[height=0.245\linewidth]{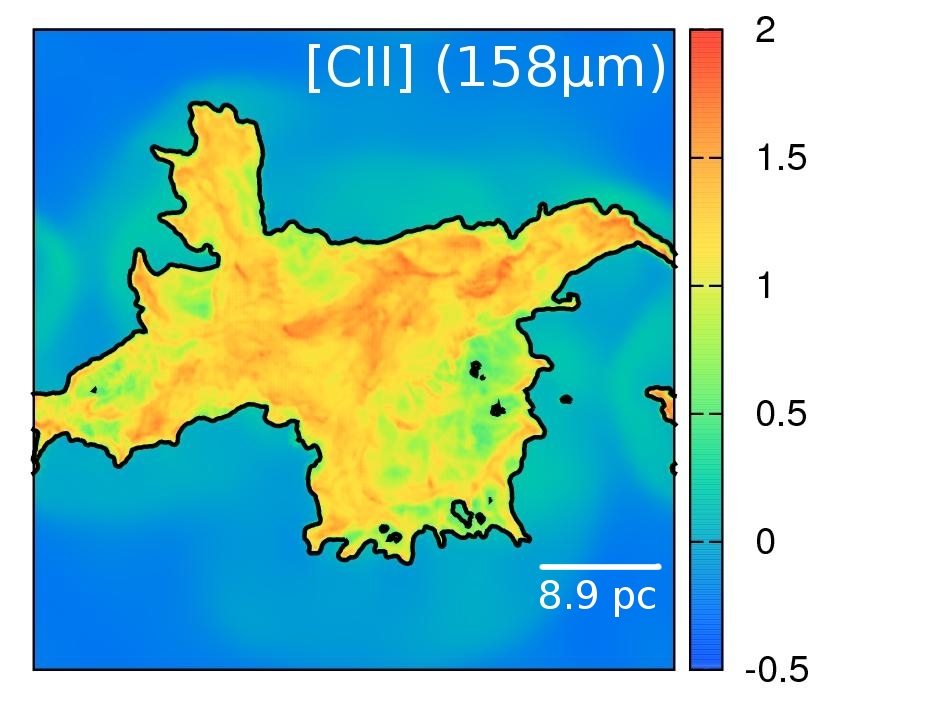}
\includegraphics[height=0.245\linewidth]{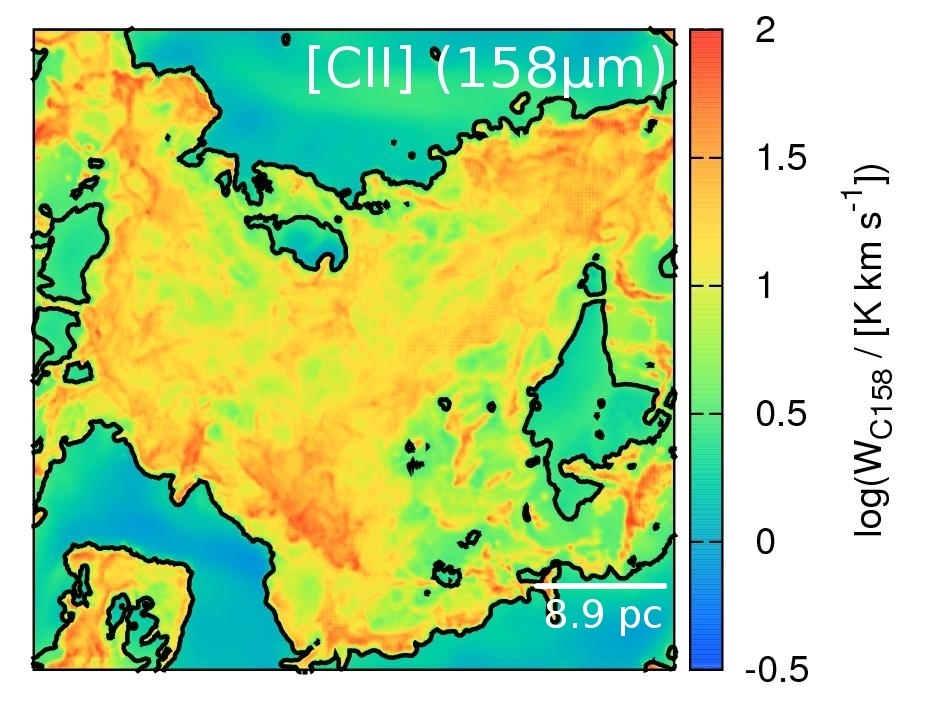}
}
\caption{Velocity-integrated intensity maps computed along the LoS in the $z$-direction for different virial parameters: $\alpha = 0.5$ (left column), 2.0 (middle column) and 8.0 (right column). From top to bottom: integrated intensity maps for $^{12}$CO ($J=1 \rightarrow 0$), $^{13}$CO ($J=1 \rightarrow 0$), [O{\sc i}] (63\,$\mu$m), [O{\sc i}] (145\,$\mu$m) and {[C{\sc ii}]} (158\,$\mu$m). Note the different scaling in the colorbars at the right hand side. Each side has a length of $44.5\,$pc. The initial cloud radius of 8.9\,pc is indicated in the bottom panels. The contour lines show the threshold values given in Table \ref{tab:thresh}.}
\label{fig:tracers}
\end{figure*}

Below, we try to estimate the effective cloud radius $R_{\text{eff}}$ in a similar fashion to what would be applied to observational data sets. At first sight, this seems to be complicated, since all clouds shown in Fig. \ref{fig:tracers} are far away from a spherical geometry. Nevertheless, we can estimate an effective radius by adopting an integrated intensity threshold $W_{\text{thresh}}$ for each map and counting the number of cells $N$, which fulfill $W_{\text{cloud}} > W_{\text{thresh}}$. The values for $W_{\text{thresh}}$ are listed in Table \ref{tab:thresh} and are chosen so that there is a clear separation between emission from the cloud and from the diffuse gas in which it is embedded. Our results are not particularly sensitive to the exact value chosen for $W_{\text{thresh}}$, as long as it is large enough to not be contaminated by the diffuse emission, but not so large that we miss significant emission from the cloud itself. For example, in the case of model GC-0.5-1000,  values of $W_{\text{thresh}}$ anywhere in the range $2 < W_{\text{thresh}} < 10\,$K\,km\,s$^{-1}$ give essentially the same results for $R_{\text{eff}}$. For comparison, we also give the effective radii of the clouds that incorporate $\sim80$\% of the total gas mass.

We have also verified that our adapted thresholds are comparable or greater than the practical thresholds that we would obtain using modern instruments to measure the line emission. For example, in a single pointing, the GREAT instrument \citep[German REceiver for Astronomy at Terahertz Frequencies,][]{HeyminckEtAl2012} on-board the Stratospheric Observatory for Infrared Astronomy (SOFIA) can measure integrated intensities of the {[C{\sc ii}]} line down to around $\sim0.06\,$K\,km\,s$^{-1}$ in around one hour of integration time\footnote{https://great.sofia.usra.edu/cgi-bin/great/great.cgi}, although multipixel array detectors are available today that become more and more efficient. Mapping an entire cloud using multiple pointings obviously allows one to spend much less time per pointing, yielding a higher threshold, but it remains plausible to map extended regions with a sensitivity comparable to or better than the adopted threshold.

We compute the effective cloud radius $R_{\text{eff}}$ by assuming an effective cloud area $A_{\text{eff}} = \pi R_{\text{eff}}^2$, where $A_{\text{eff}} = N \Delta x^2$ and $\Delta x$ is the cell size introduced in Section \ref{subsec:RADMC}. Thus, we get
\begin{equation}
\label{eq:Reff}
R_{\text{eff}} = \sqrt{\frac{A_{\text{eff}}}{\pi}} = \Delta x \sqrt{\frac{N}{\pi}}.
\end{equation}
Table \ref{tab:radii} shows the different effective radii of the various clouds presented in Fig. \ref{fig:tracers}, computed via equation (\ref{eq:Reff}). If we compare the different values listed in Table \ref{tab:radii}, we find that the cloud radii measured with the integrated intensities of $^{12}$CO and $^{13}$CO are significantly smaller by a factor of $1.5-2.5$ compared to the values measured with the integrated intensities of the atomic tracers. This is because the strong ISRF and high CRF destroy the molecular tracers in the more diffuse regions, leading to smaller effective radii. Conversely, the radii measured with the atomic components roughly agree with each other for one specific virial parameter model and give the best radius estimate for the total cloud that incorporates $\sim80$\% of all gas. These values confirm our assumption that the atomic species accurately reflect the spatial distribution of the total gas column and that they yield a robust estimate of the effective radius of the cloud.

\subsection{Tracing the total and H$_2$ column density}
\label{subsec:denseregions_tot}

\begin{figure*}
\centerline{
\includegraphics[width=0.27\linewidth]{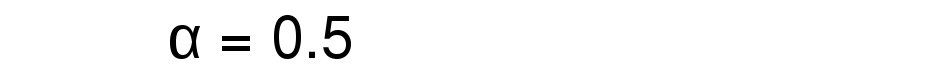}
\includegraphics[width=0.27\linewidth]{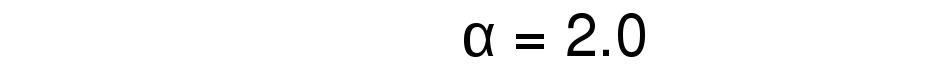}
\includegraphics[width=0.27\linewidth]{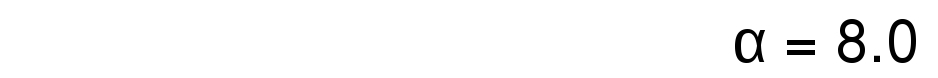}
}
\centerline{
\includegraphics[height=0.23\linewidth]{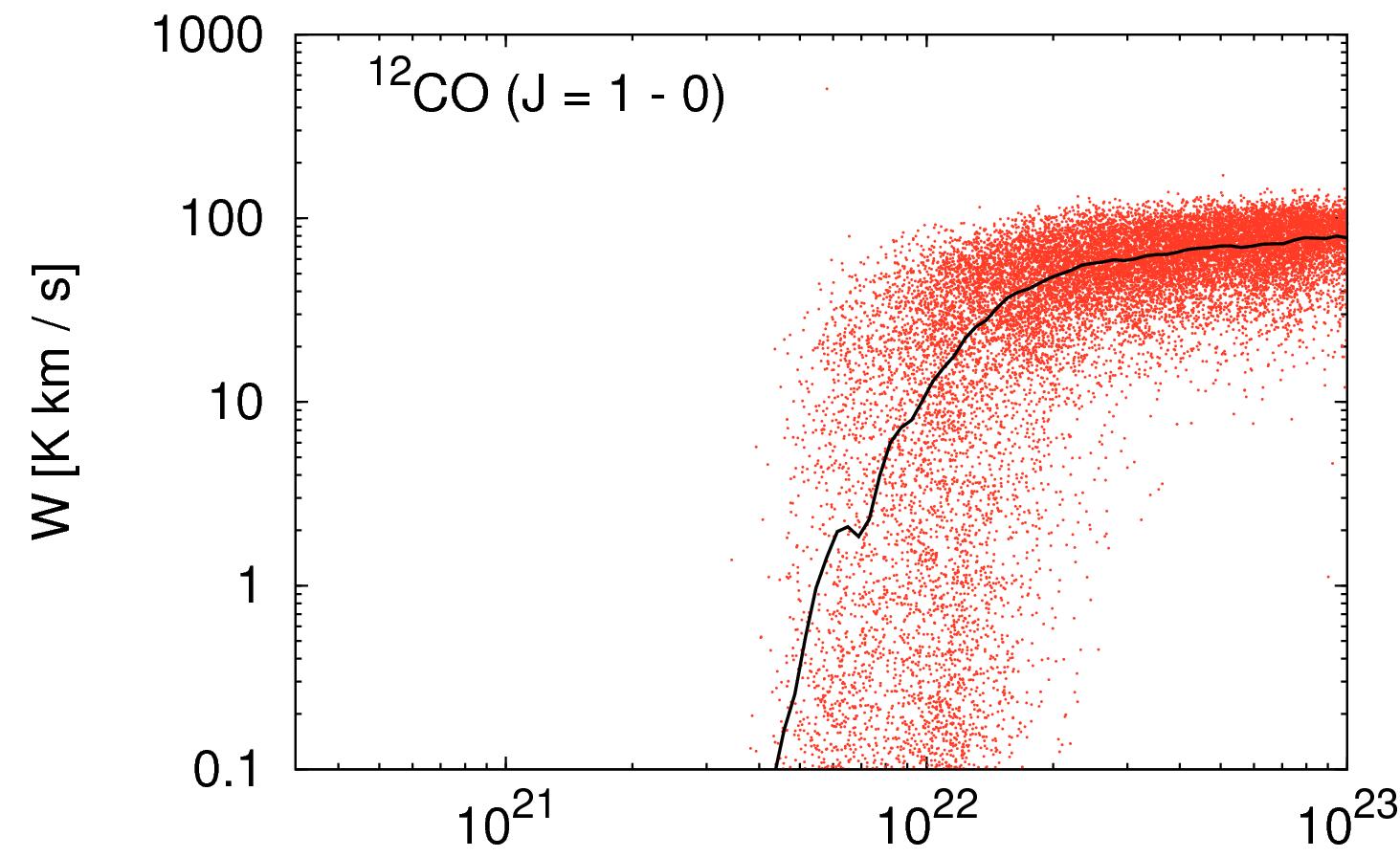}
\includegraphics[height=0.23\linewidth]{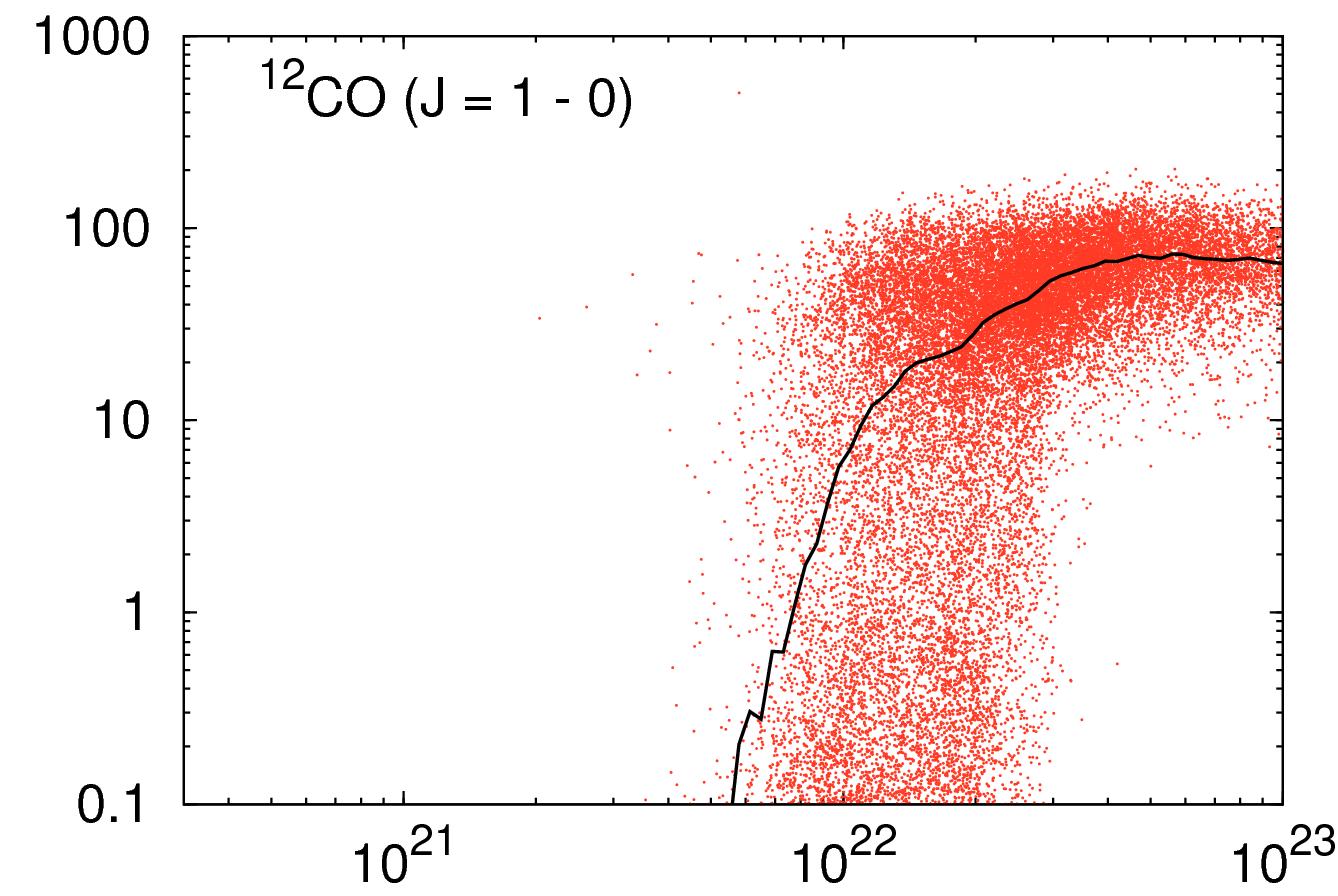}
\includegraphics[height=0.23\linewidth]{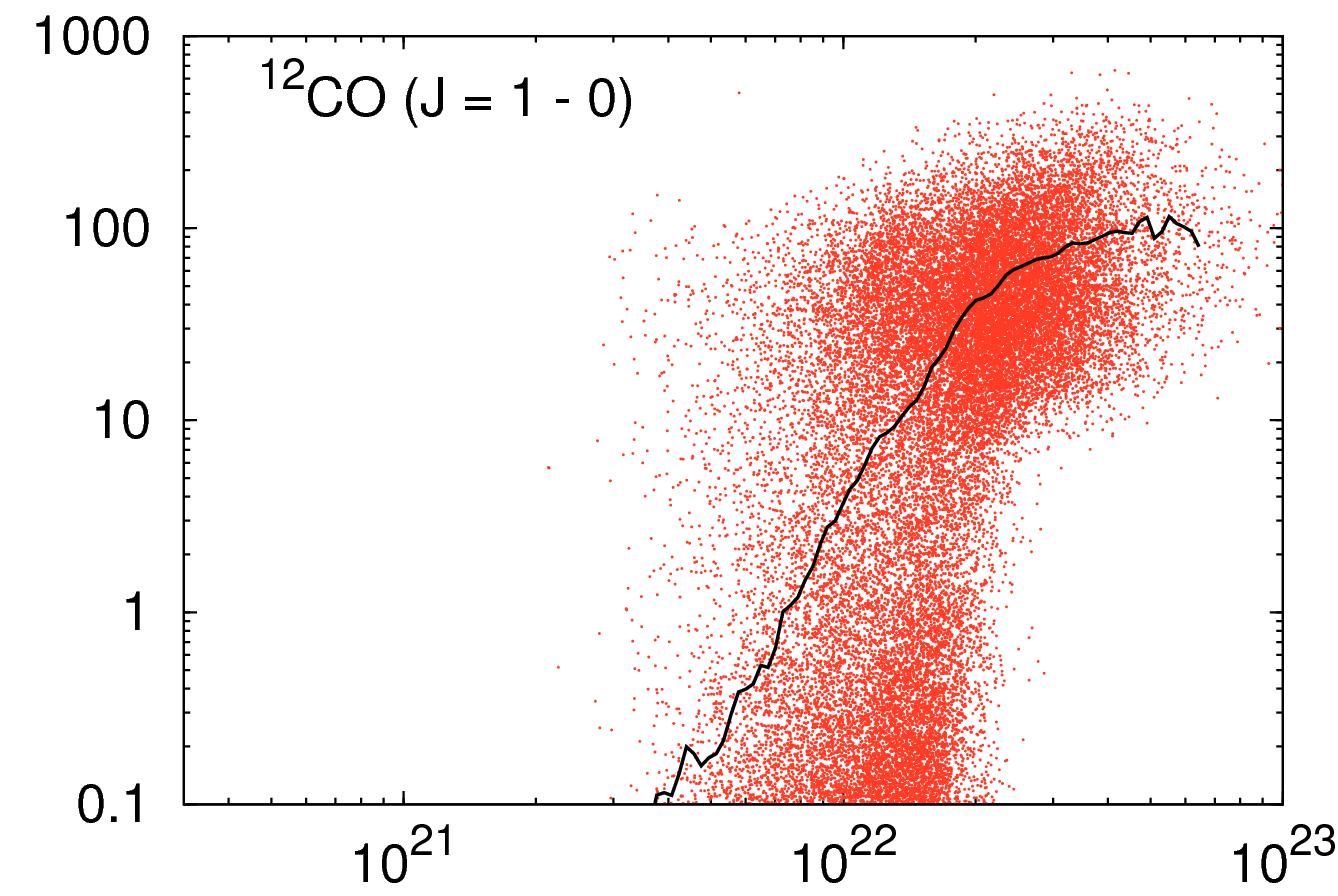}
}
\centerline{
\includegraphics[height=0.23\linewidth]{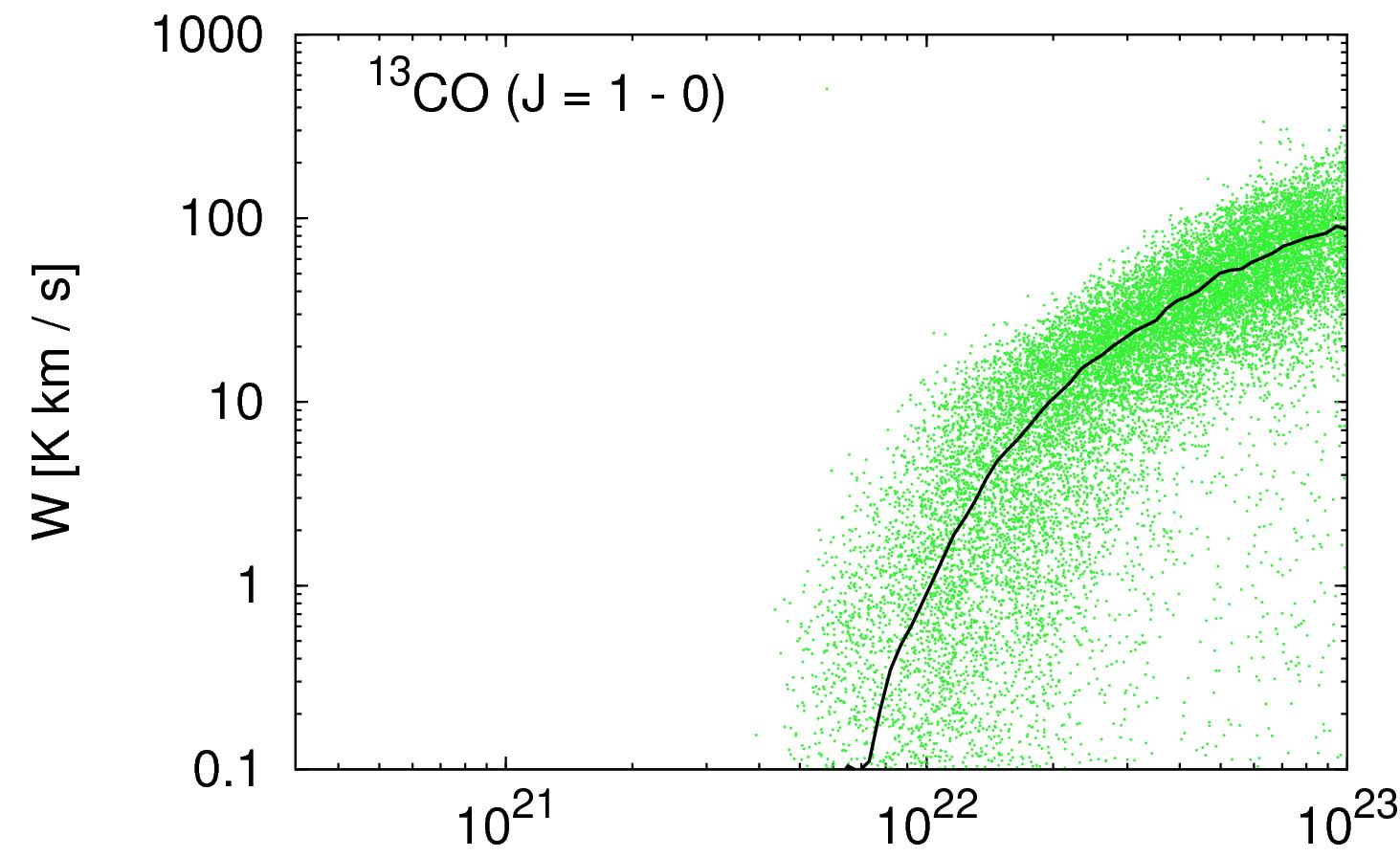}
\includegraphics[height=0.23\linewidth]{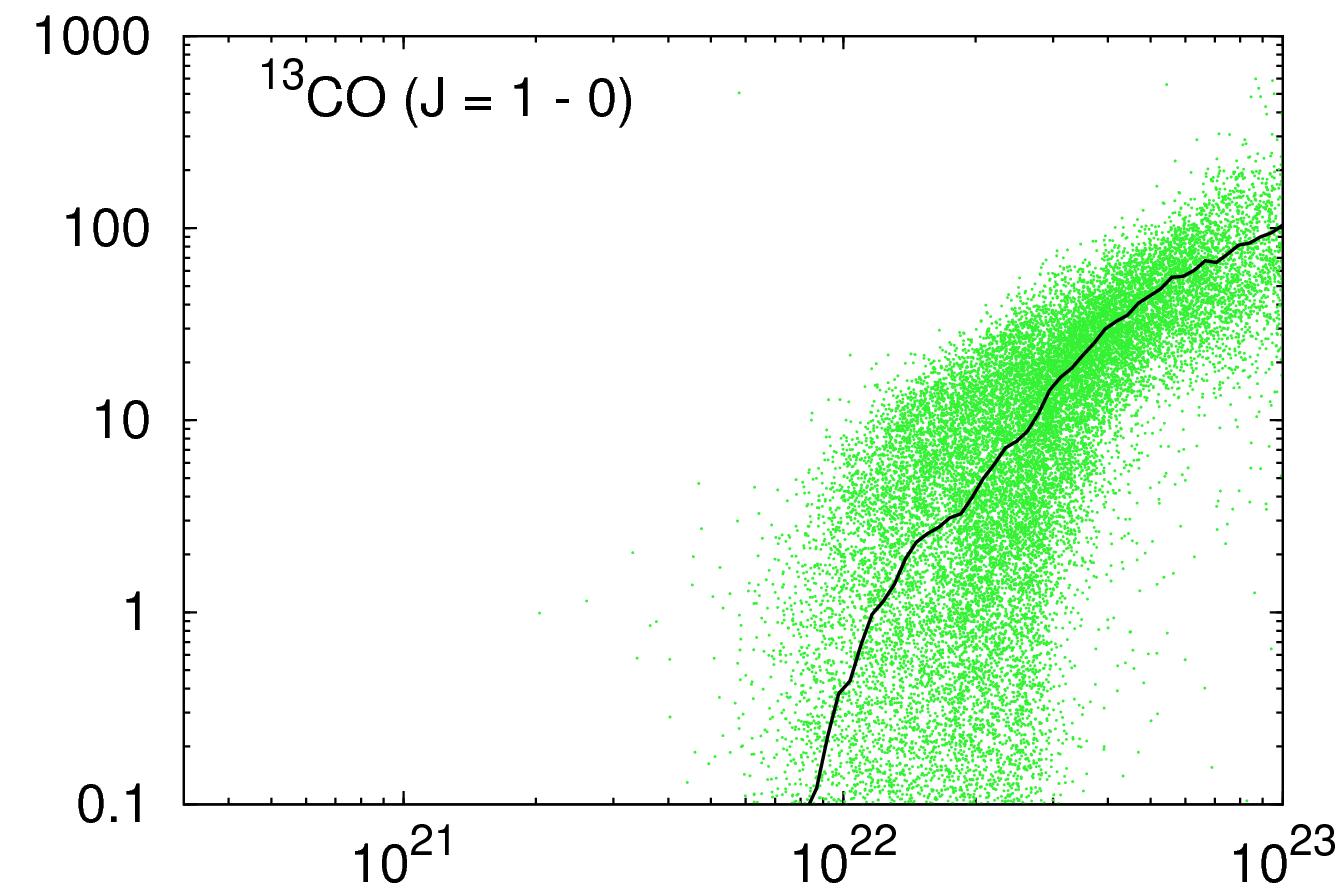}
\includegraphics[height=0.23\linewidth]{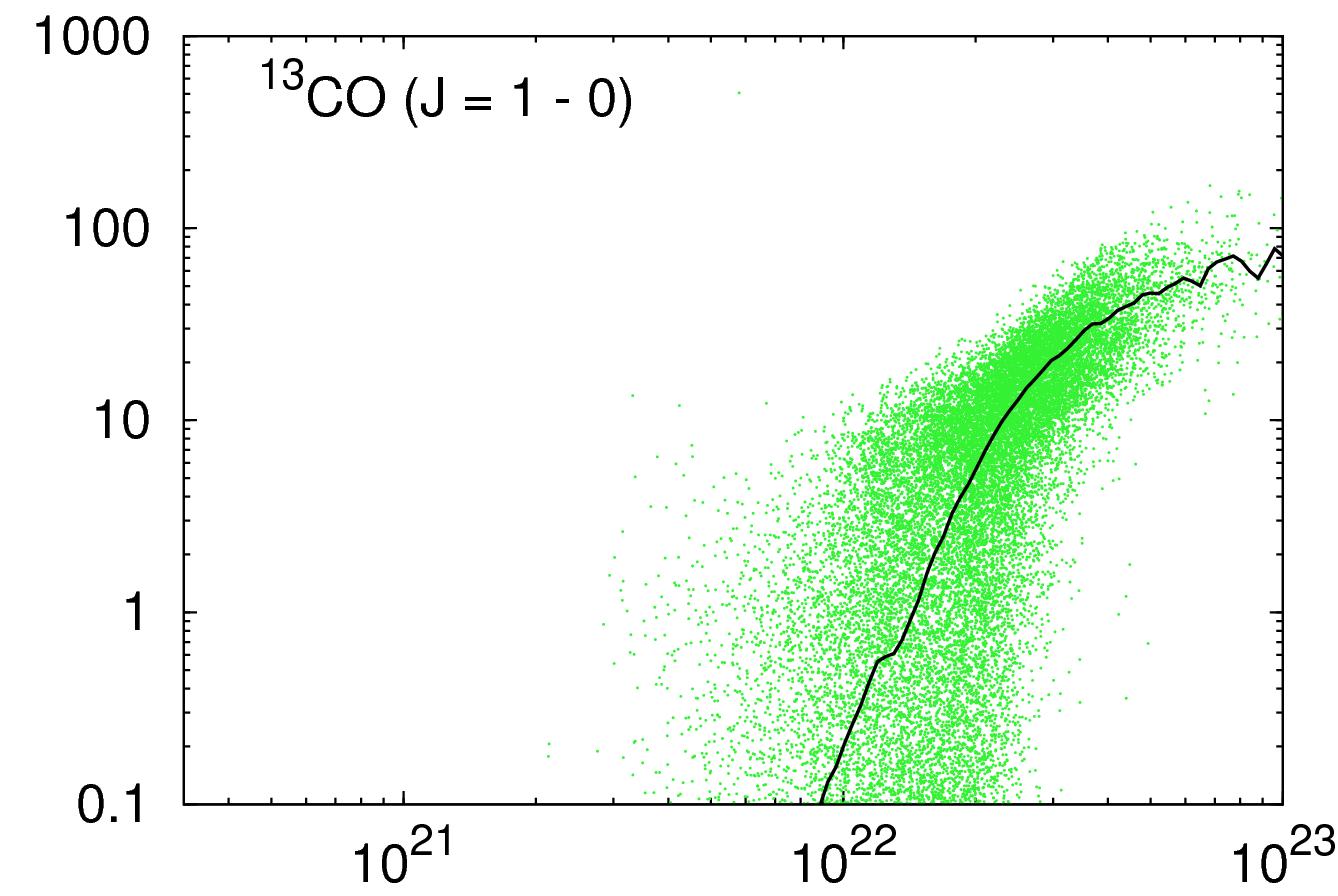}
}
\centerline{
\includegraphics[height=0.23\linewidth]{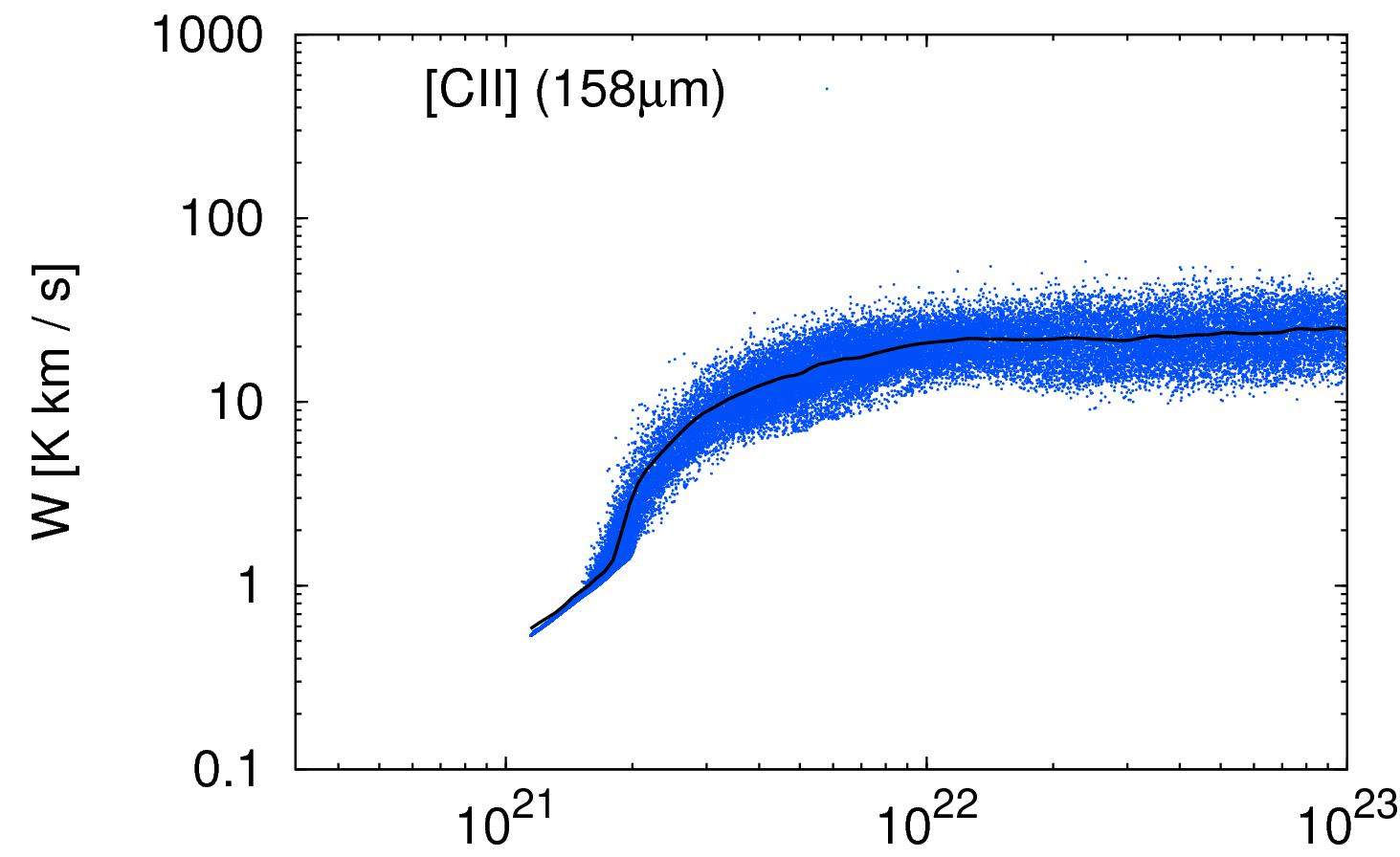}
\includegraphics[height=0.23\linewidth]{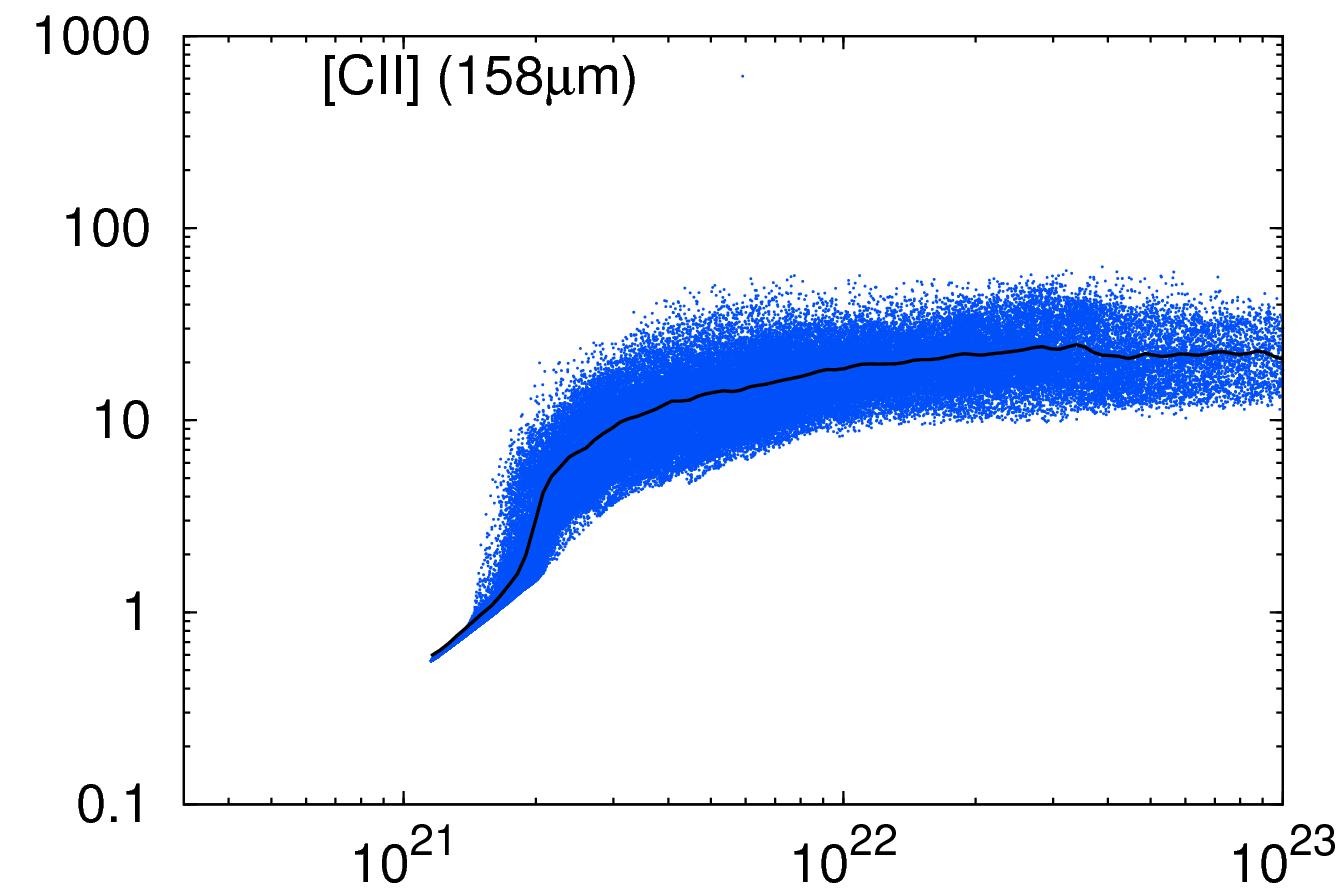}
\includegraphics[height=0.23\linewidth]{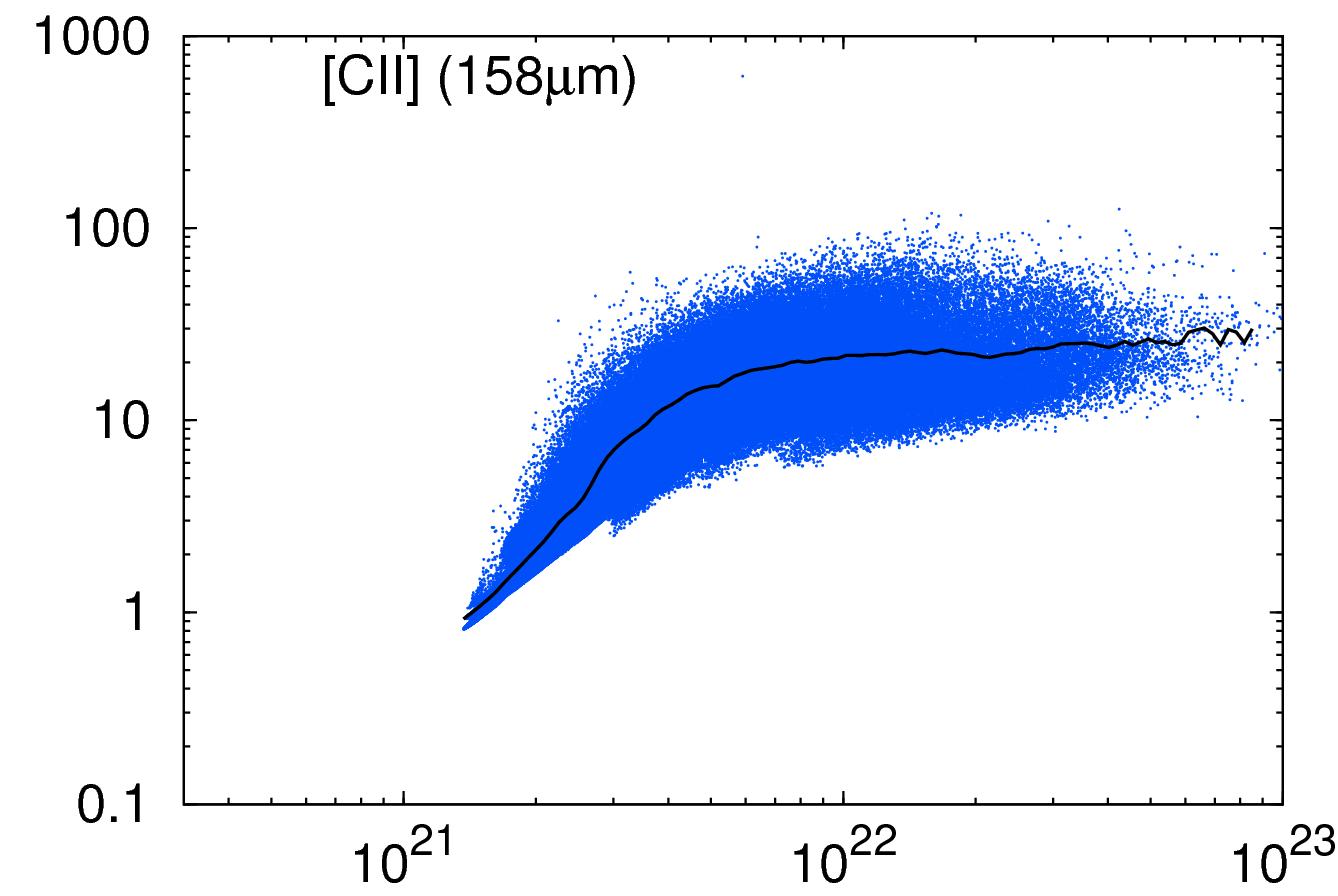}
}
\centerline{
\includegraphics[height=0.23\linewidth]{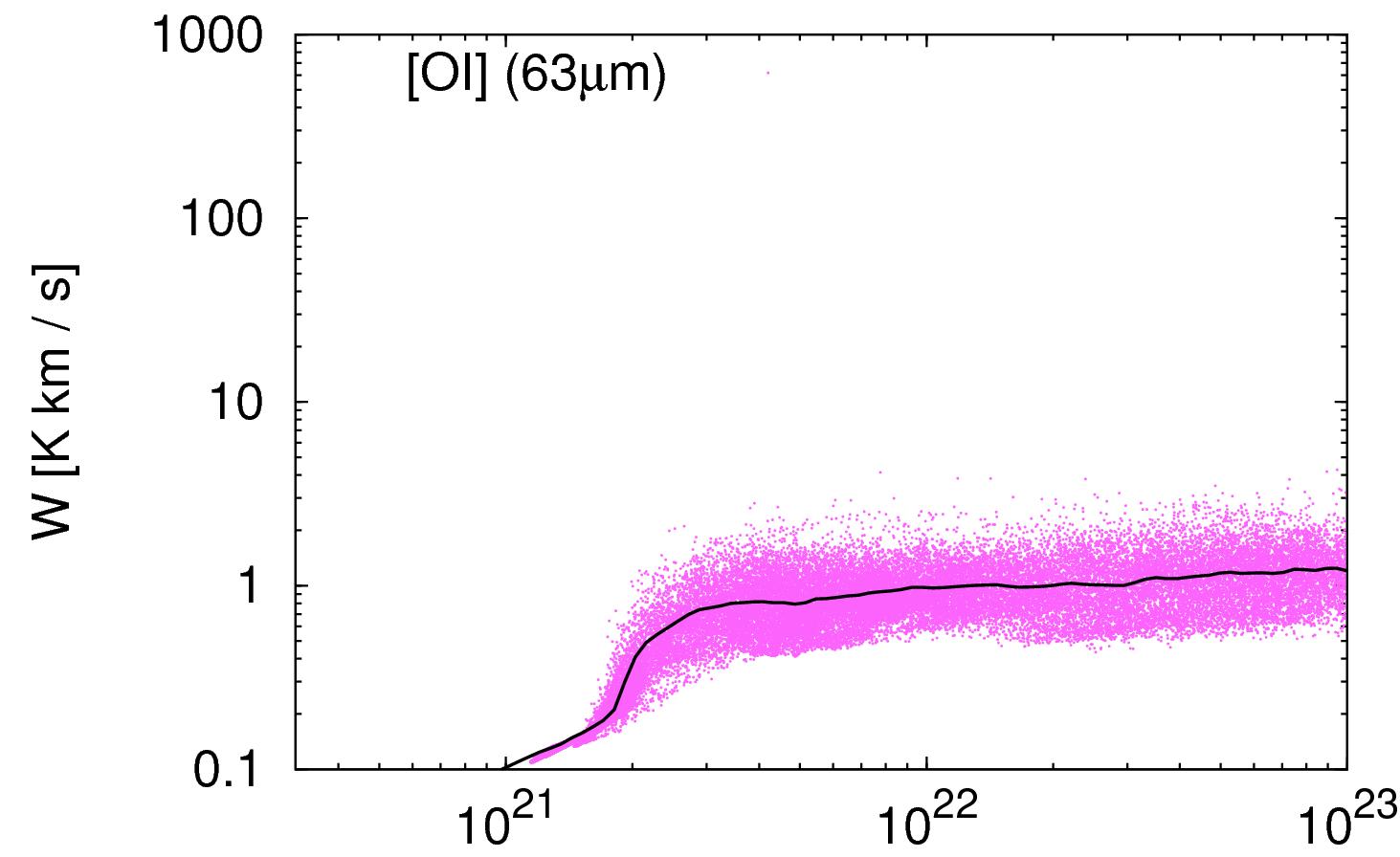}
\includegraphics[height=0.23\linewidth]{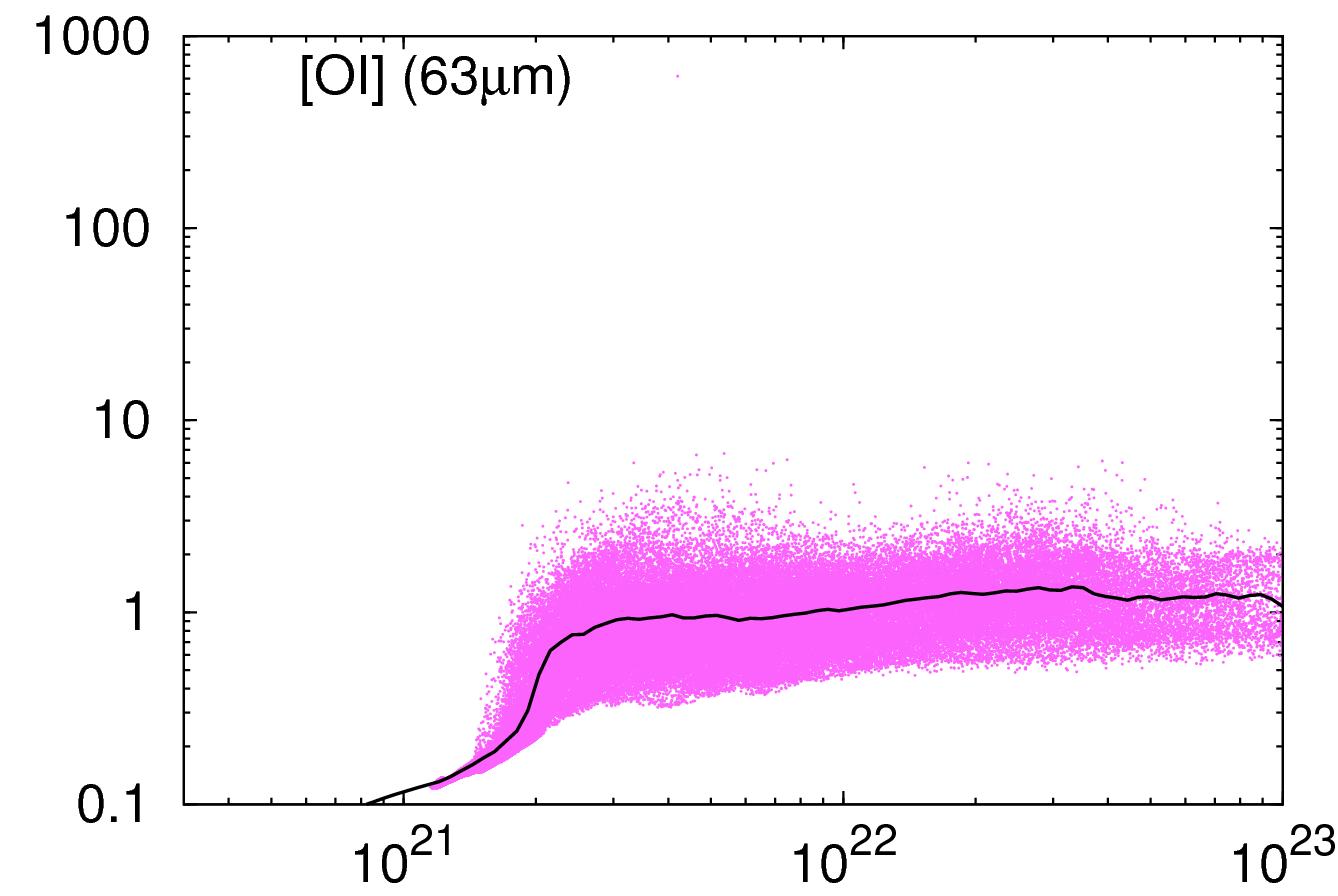}
\includegraphics[height=0.23\linewidth]{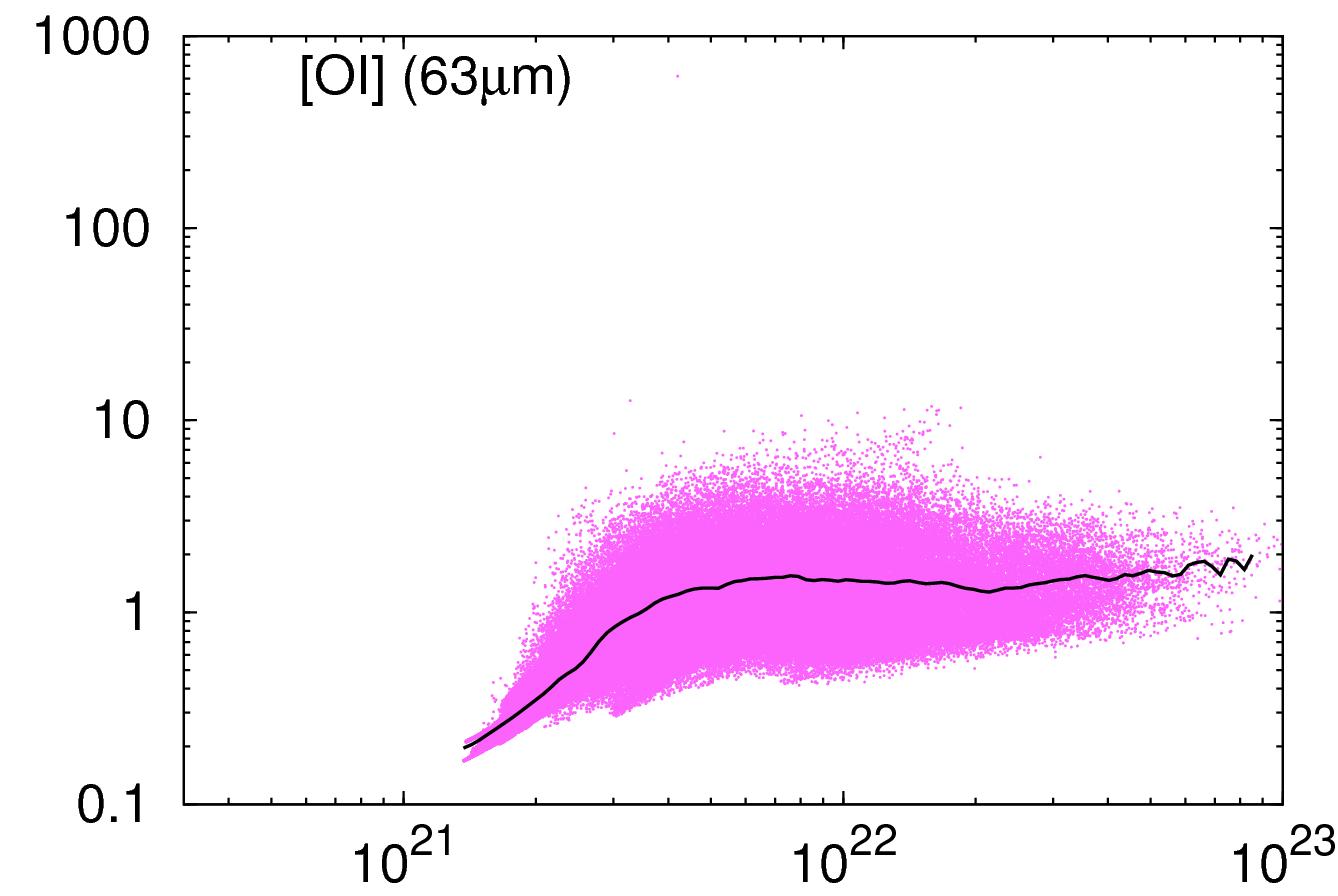}
}
\centerline{
\includegraphics[height=0.255\linewidth]{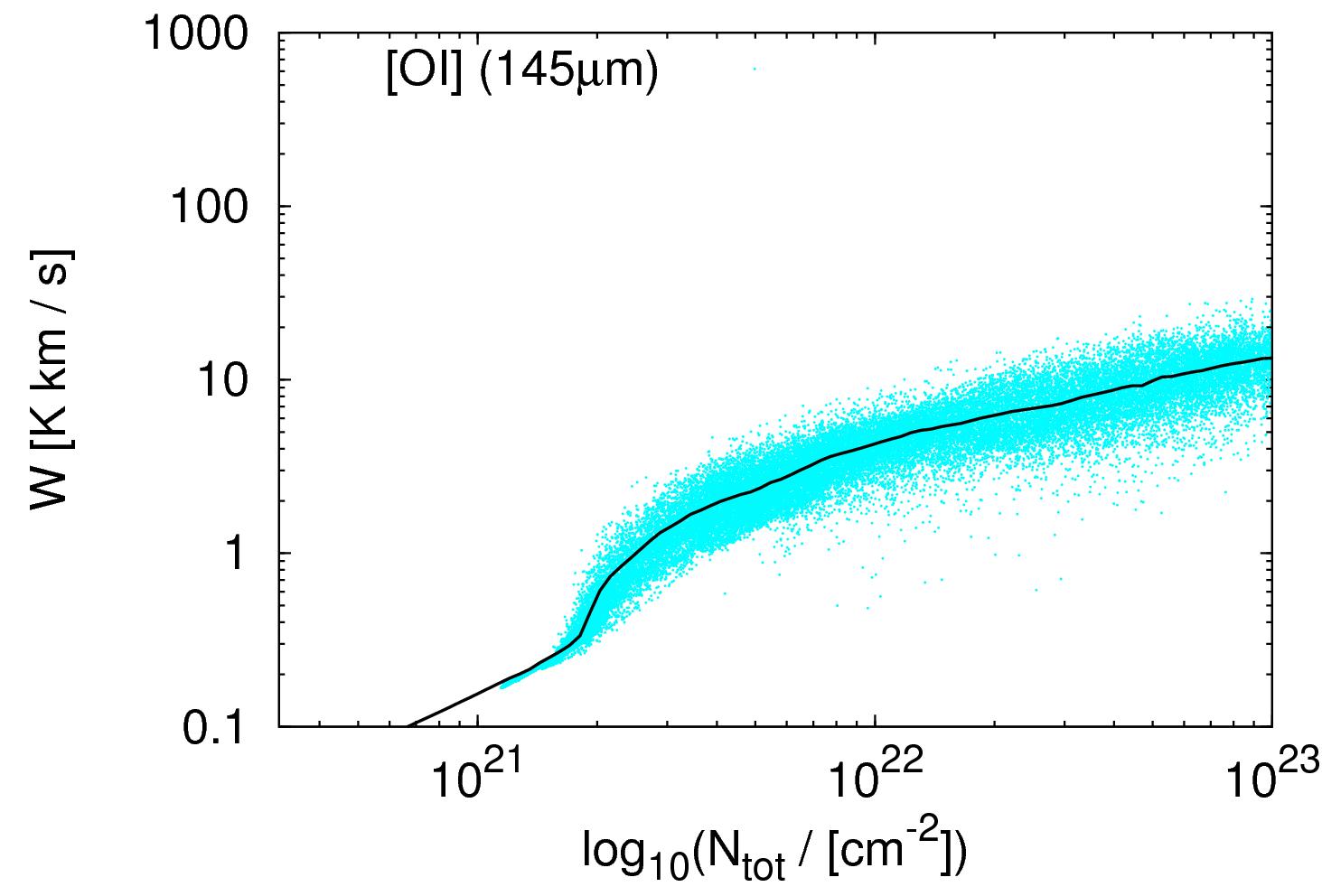}
\includegraphics[height=0.255\linewidth]{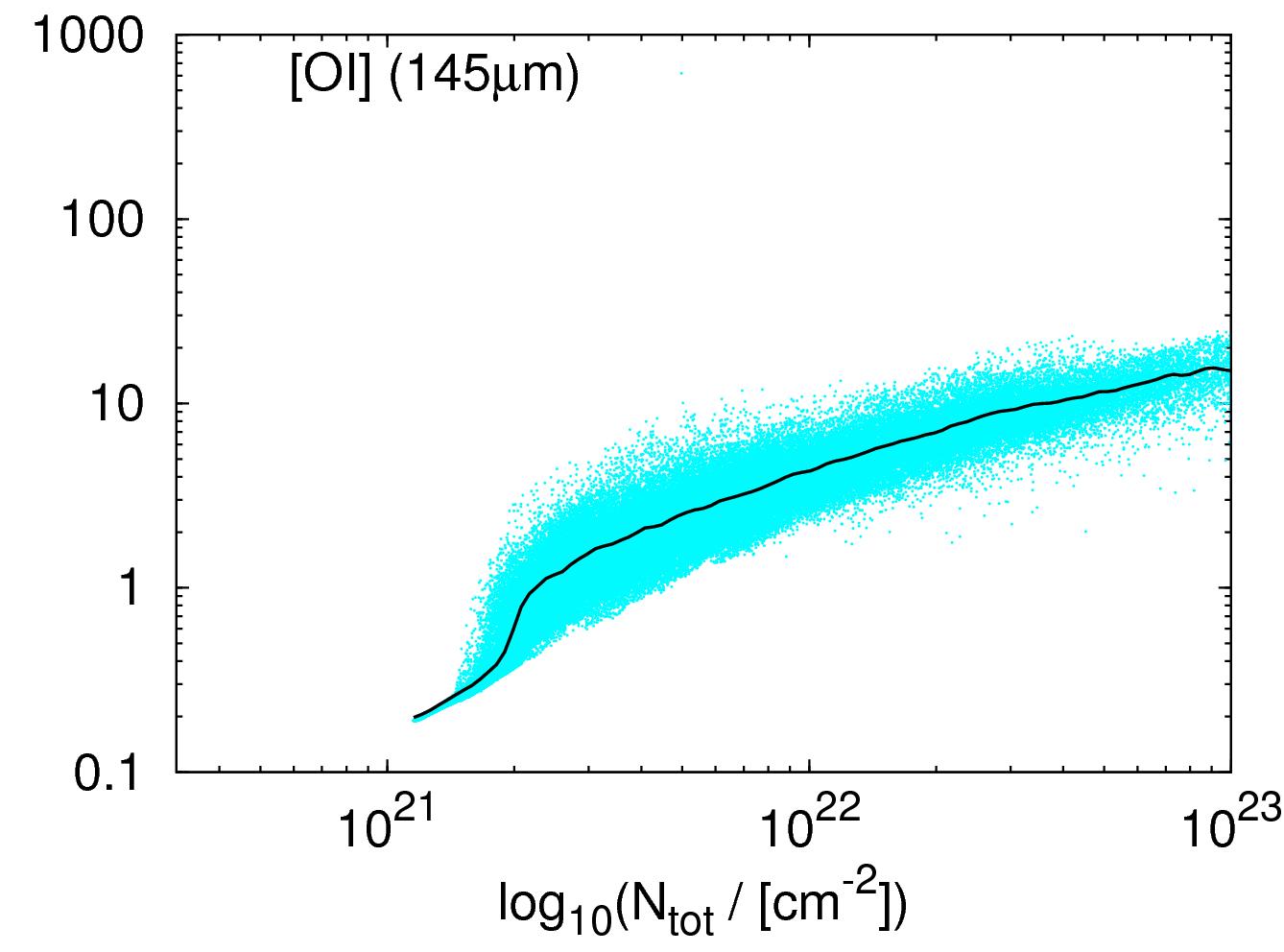}
\includegraphics[height=0.255\linewidth]{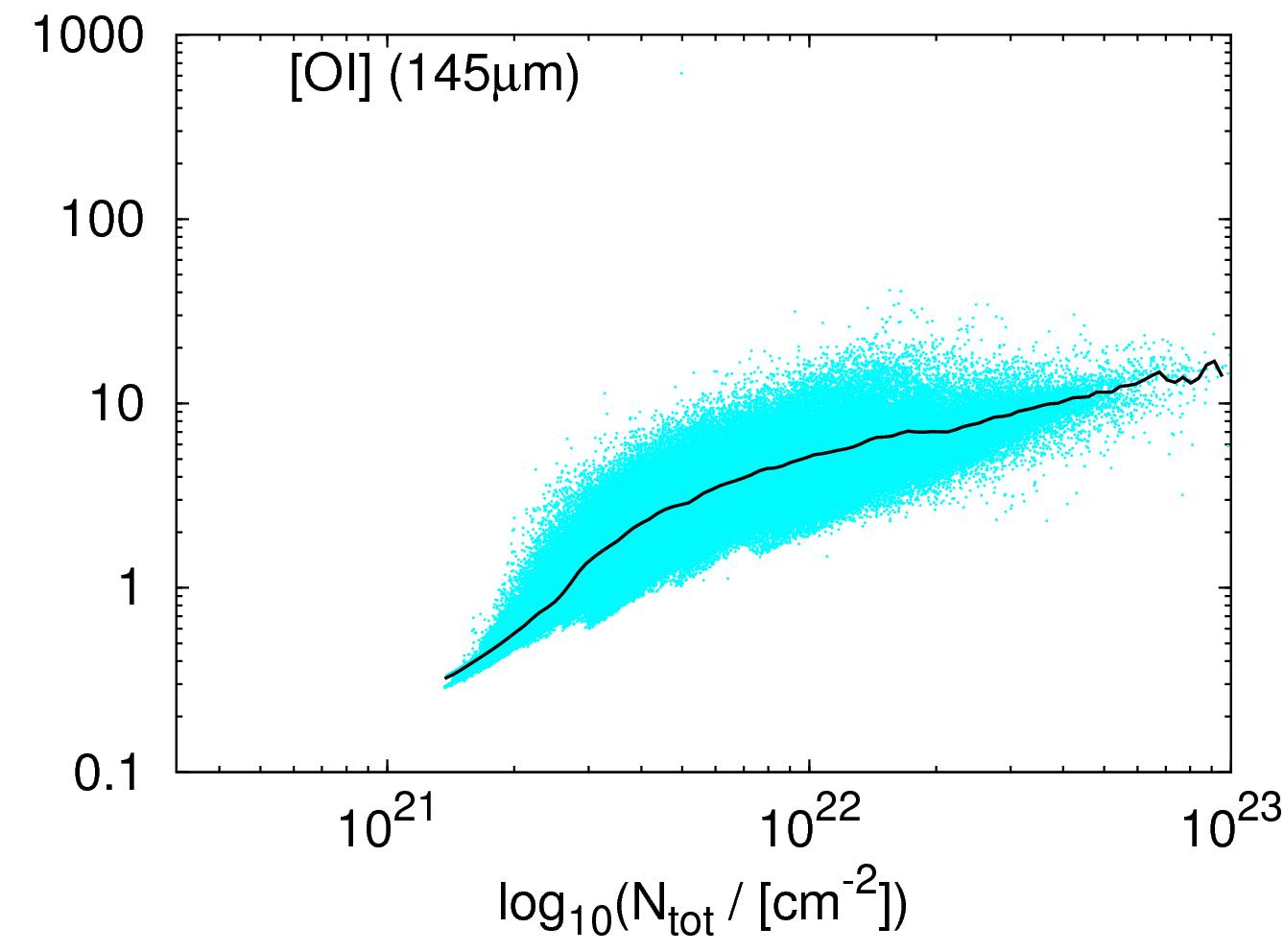}
}
\caption{Log-log plot showing velocity-integrated intensities against total column density for different virial parameters: $\alpha = 0.5$ (left column), 2.0 (middle column) and 8.0 (right column). Shown are all our tracers presented in Table \ref{tab:lines}. In each plot, the mean value is indicated by a black solid line.}
\label{fig:denseregions_tot}
\end{figure*}

\begin{figure*}
\centerline{
\includegraphics[width=0.27\linewidth]{images/alpha05_plot.jpg}
\includegraphics[width=0.27\linewidth]{images/alpha2_plot.jpg}
\includegraphics[width=0.27\linewidth]{images/alpha8_plot.jpg}
}
\centerline{
\includegraphics[height=0.23\linewidth]{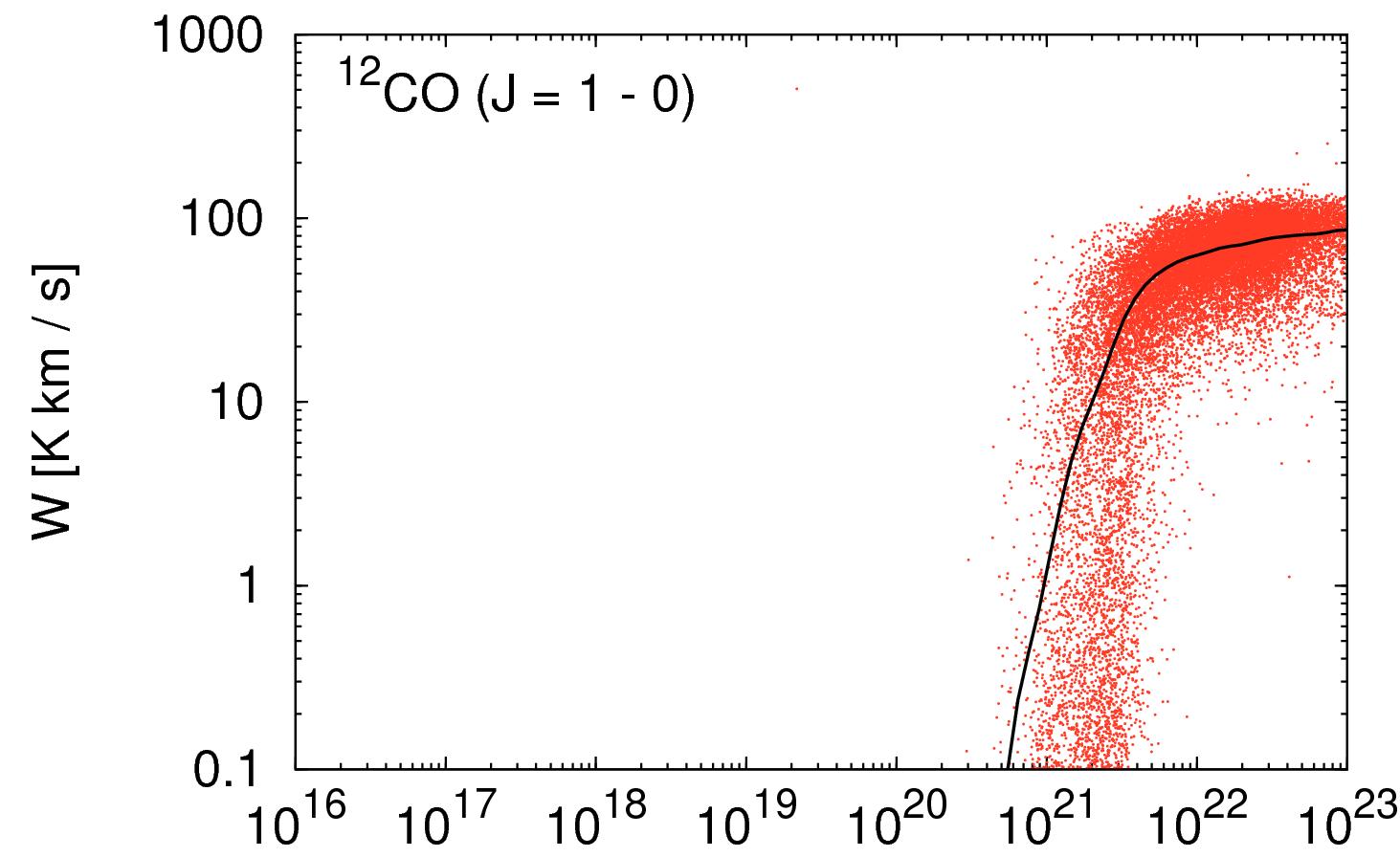}
\includegraphics[height=0.23\linewidth]{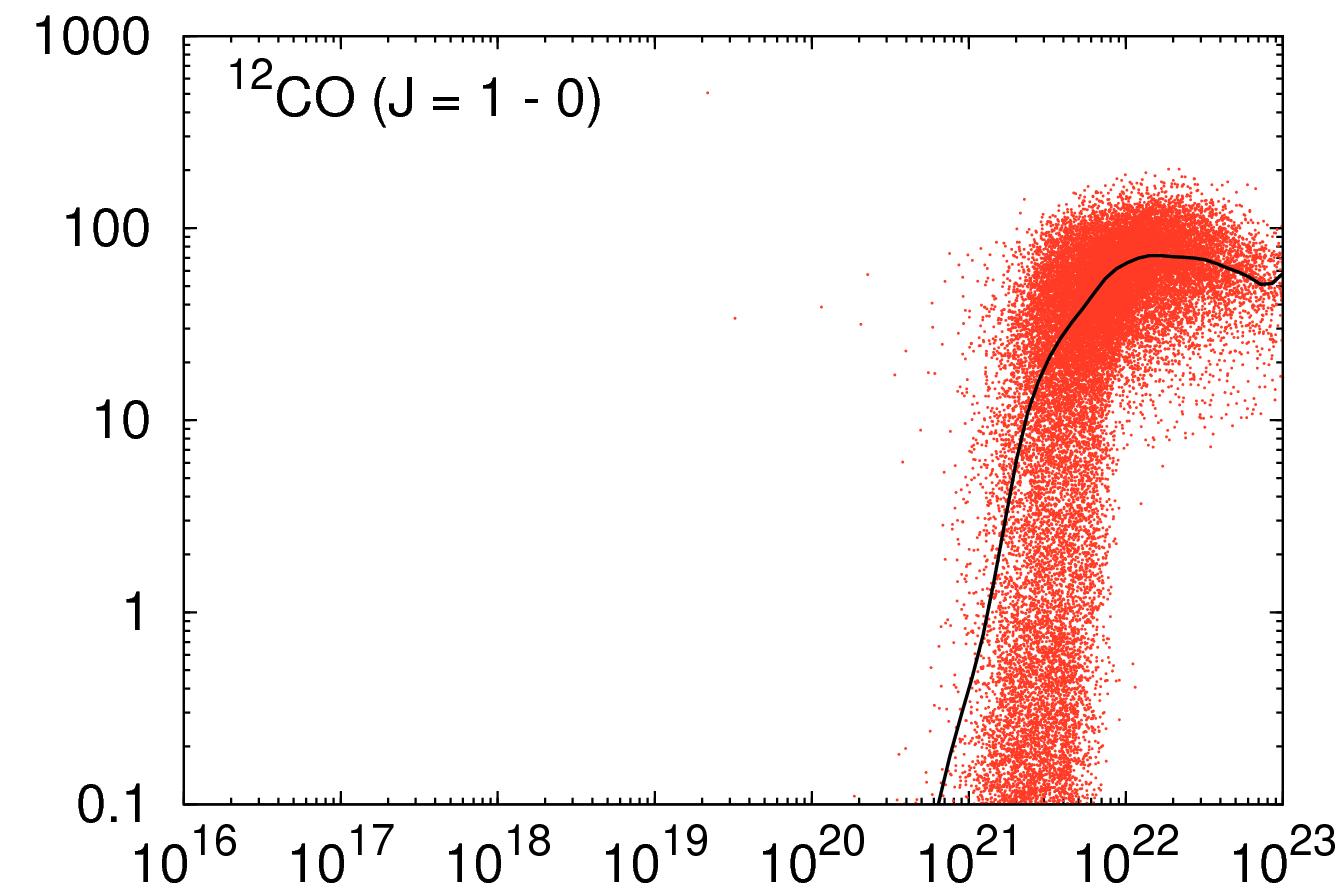}
\includegraphics[height=0.23\linewidth]{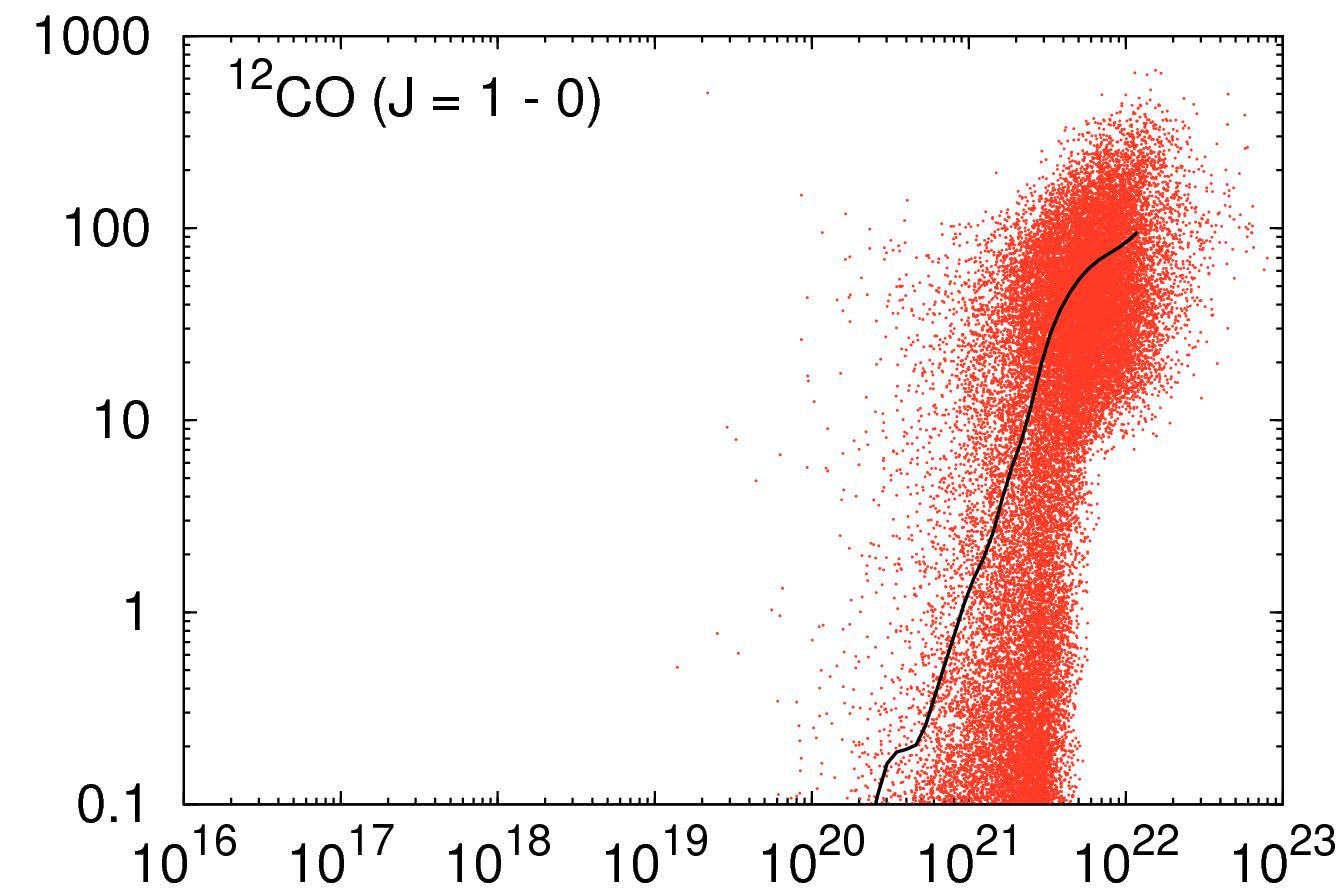}
}
\centerline{
\includegraphics[height=0.23\linewidth]{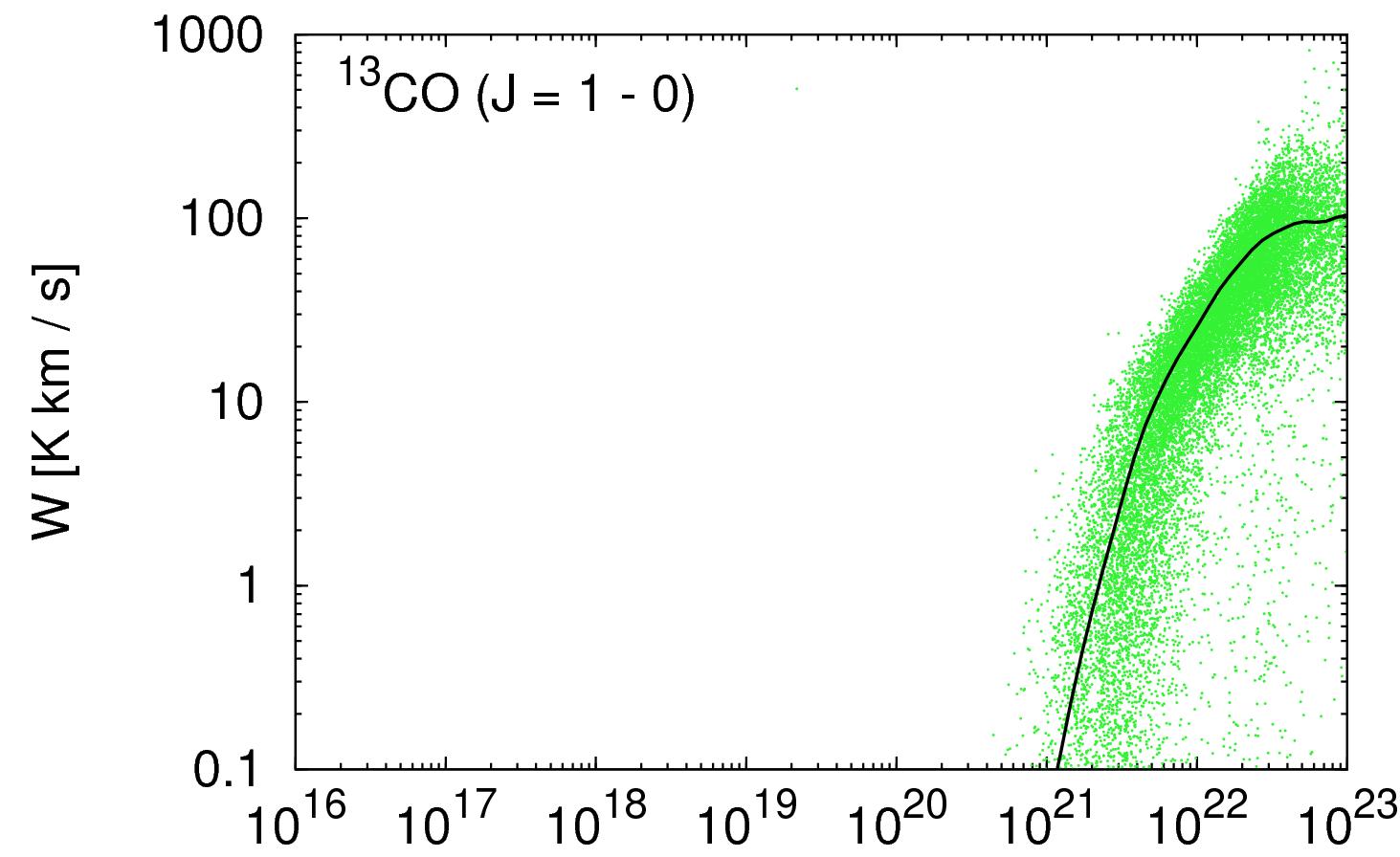}
\includegraphics[height=0.23\linewidth]{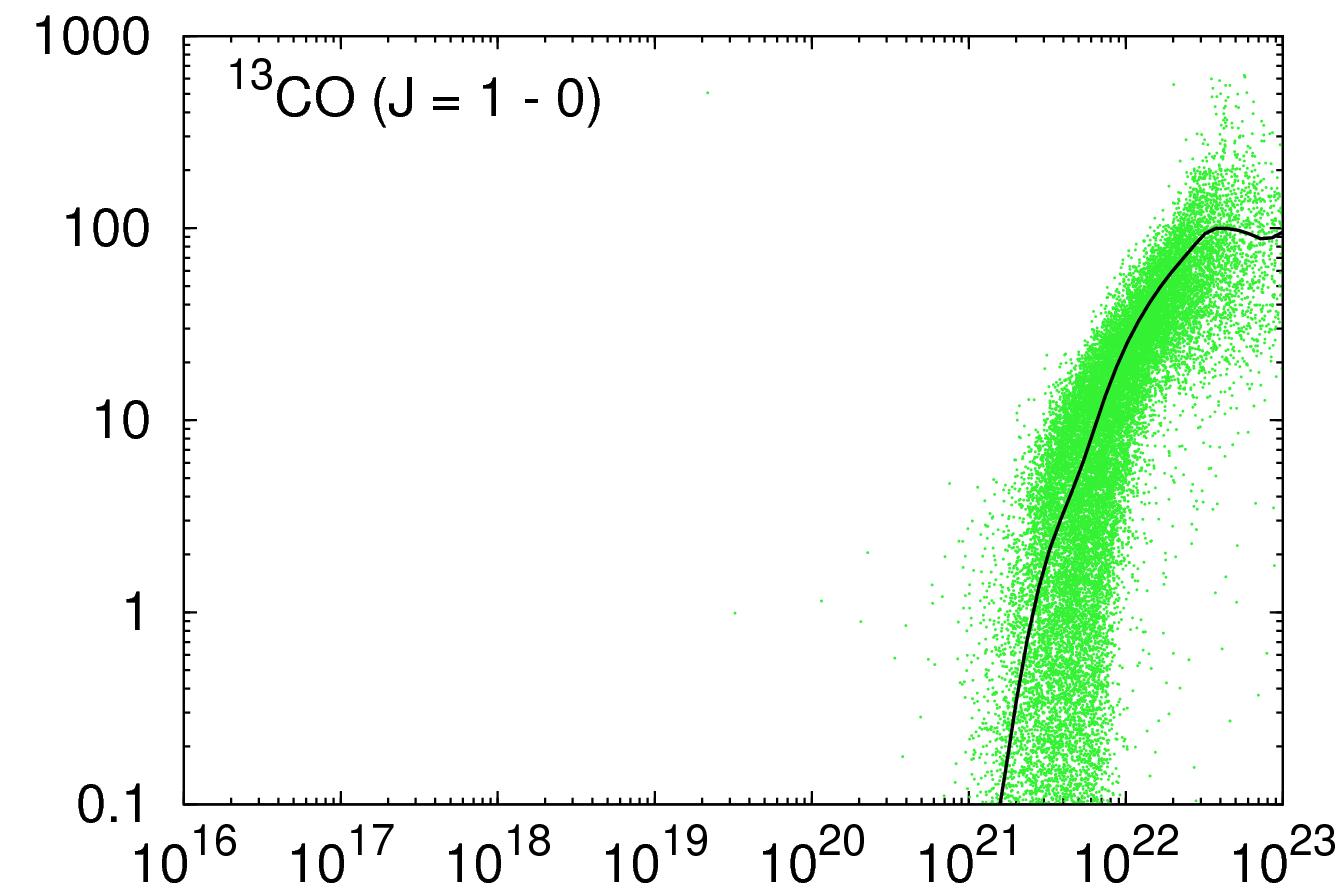}
\includegraphics[height=0.23\linewidth]{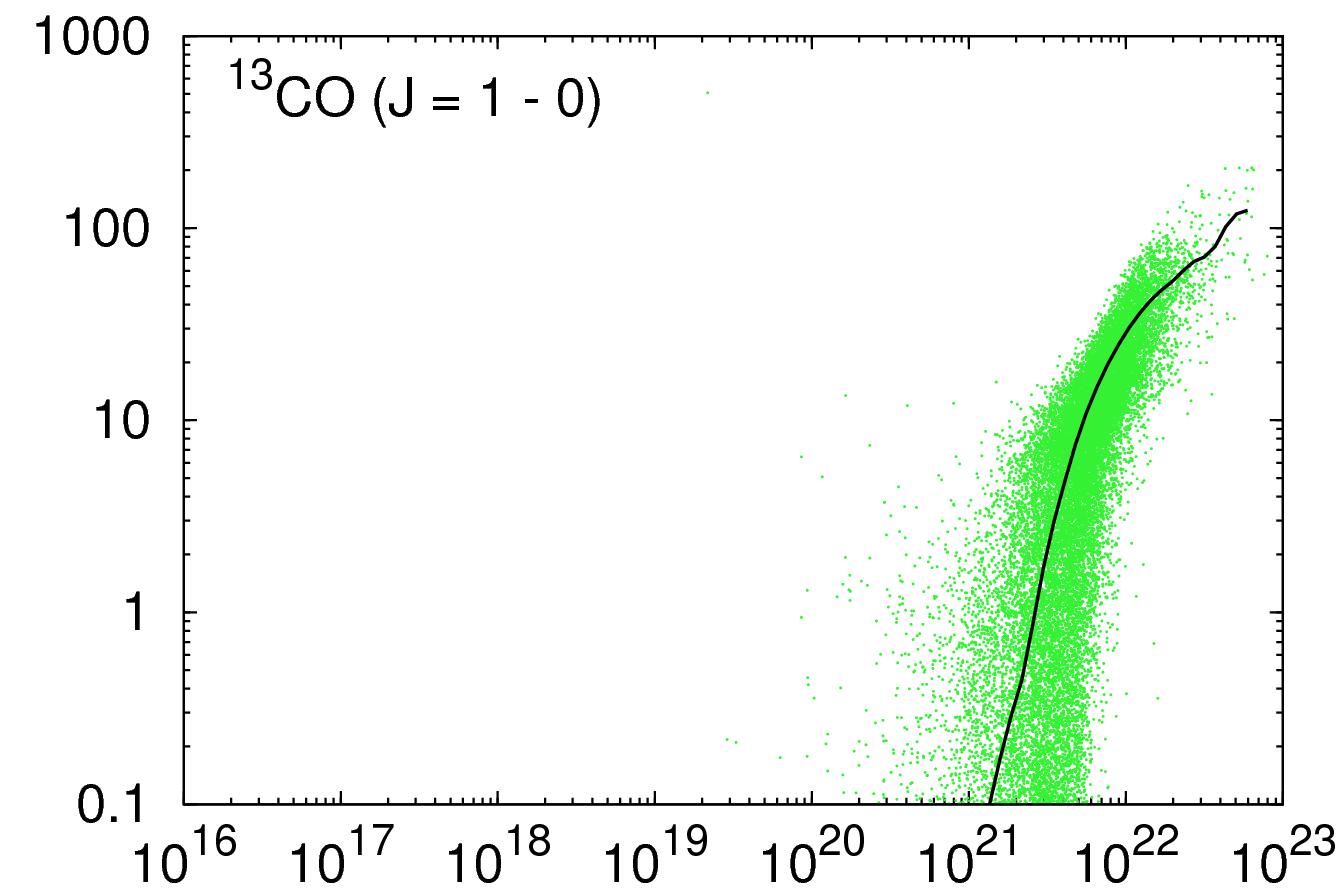}
}
\centerline{
\includegraphics[height=0.23\linewidth]{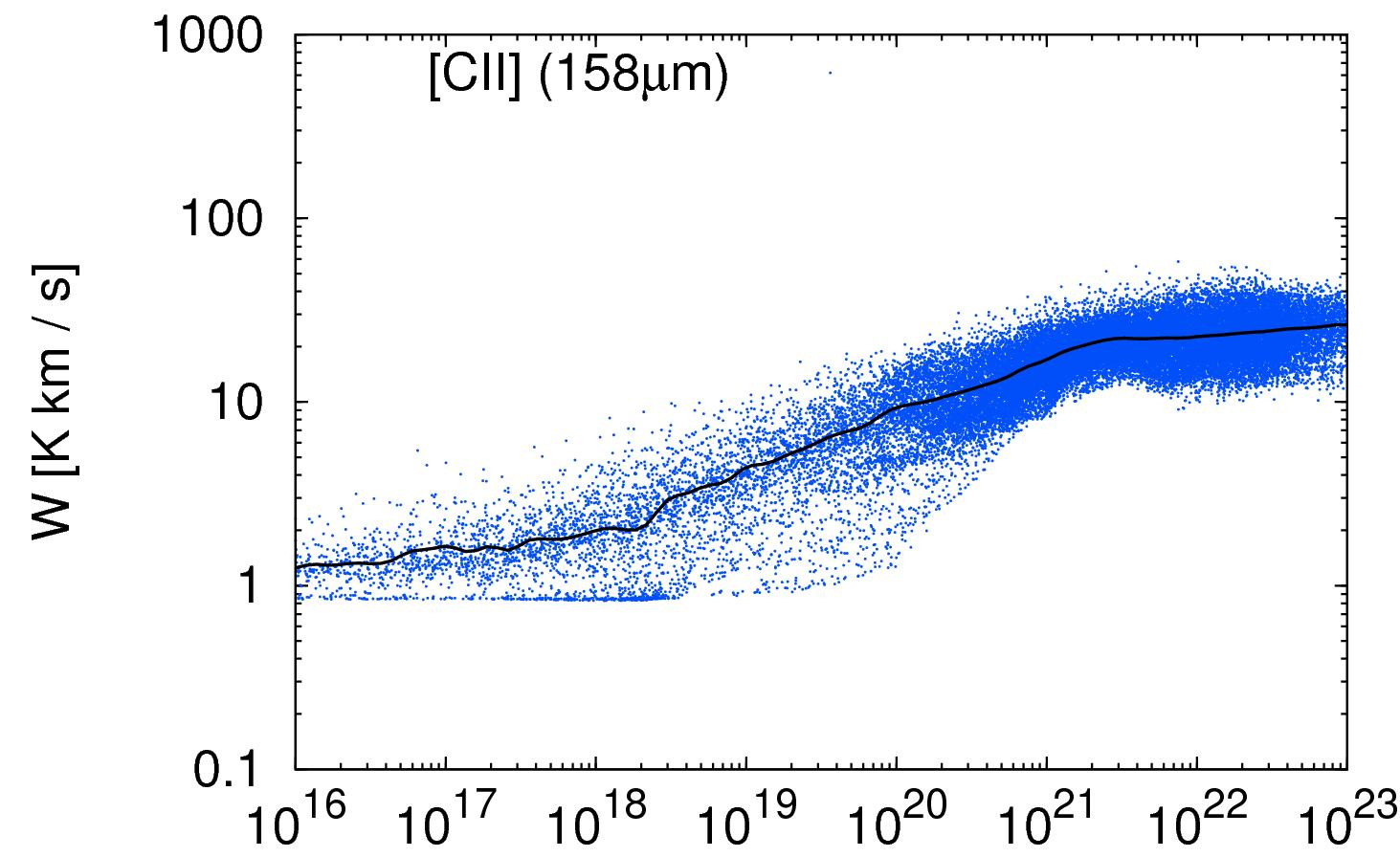}
\includegraphics[height=0.23\linewidth]{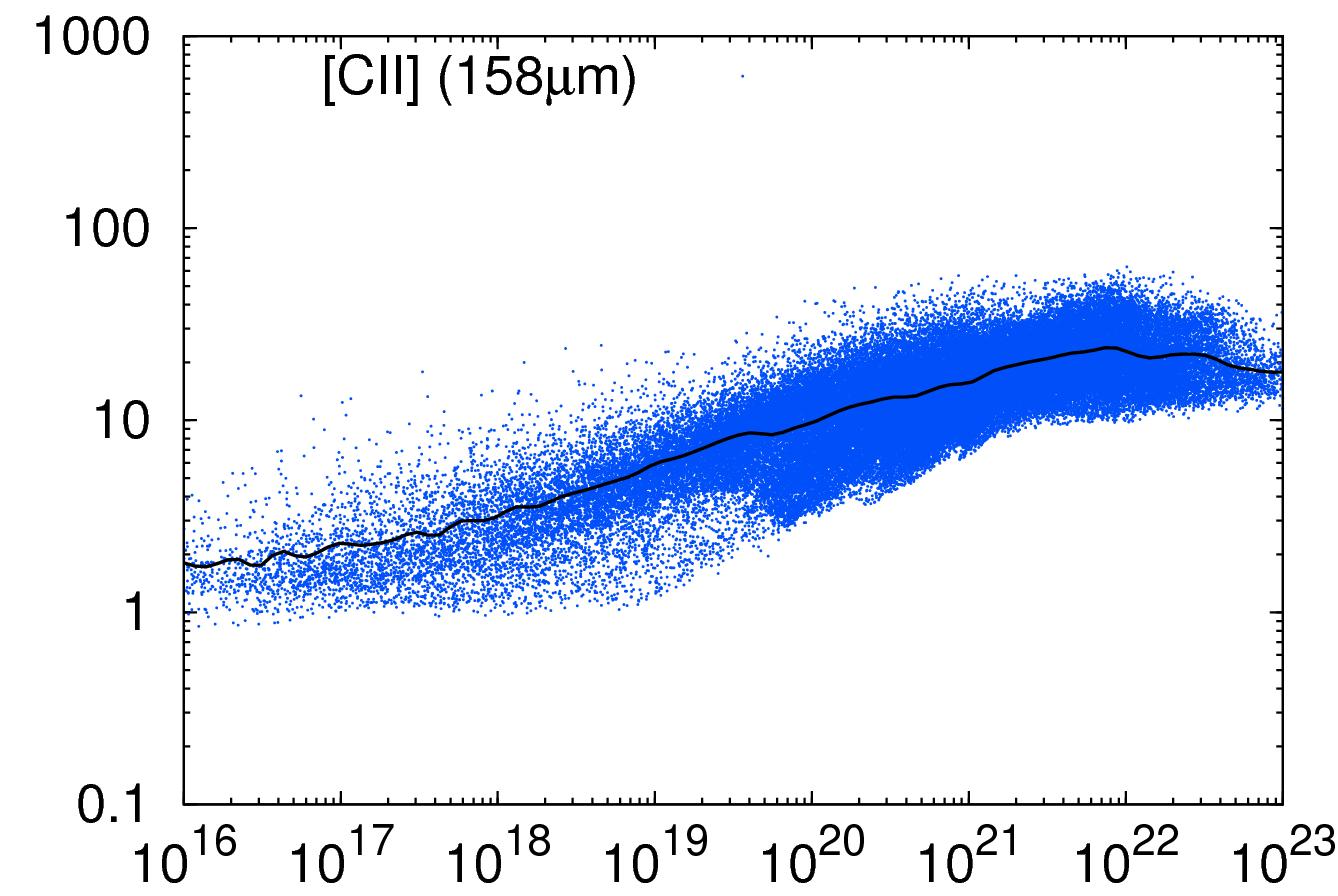}
\includegraphics[height=0.23\linewidth]{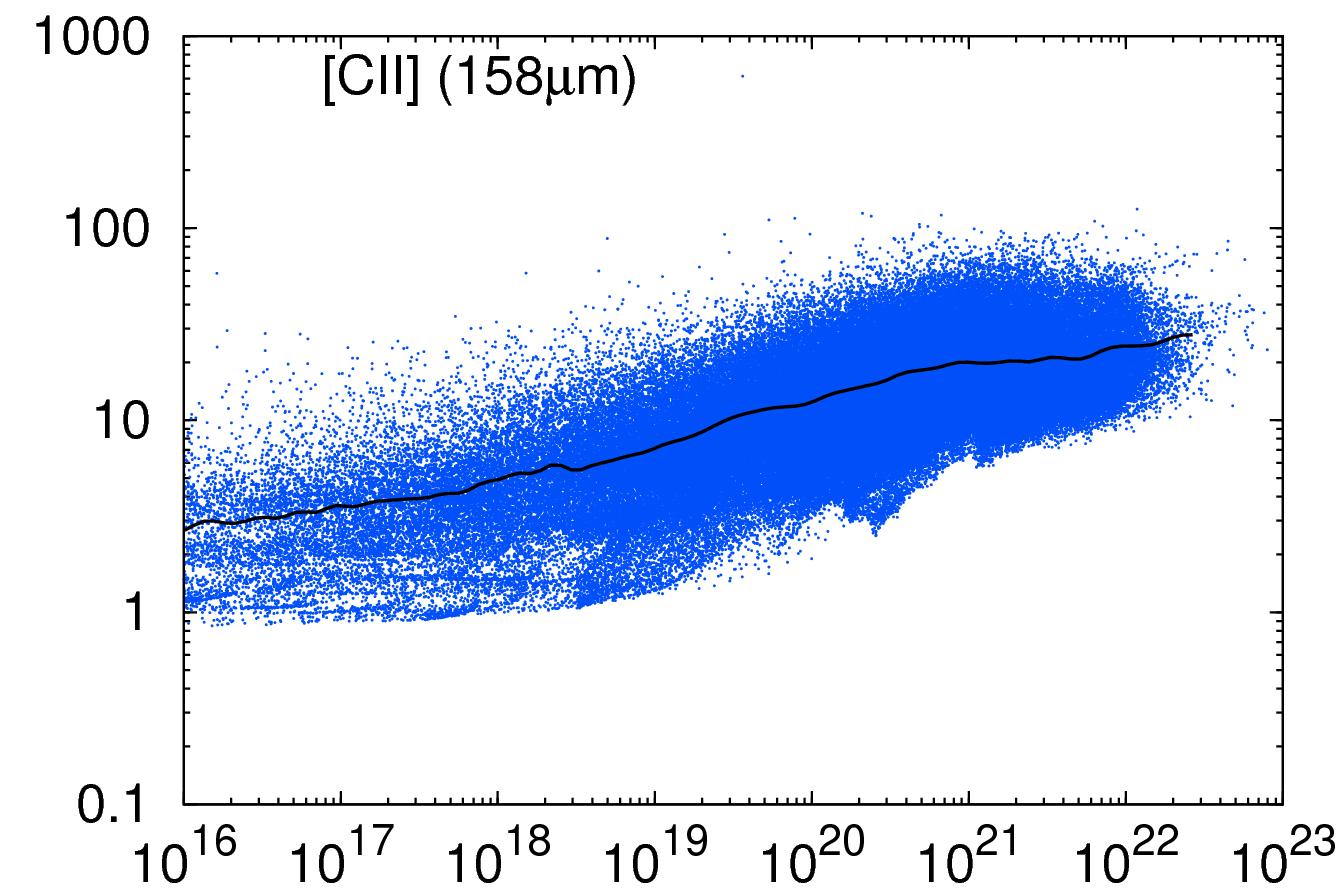}
}
\centerline{
\includegraphics[height=0.23\linewidth]{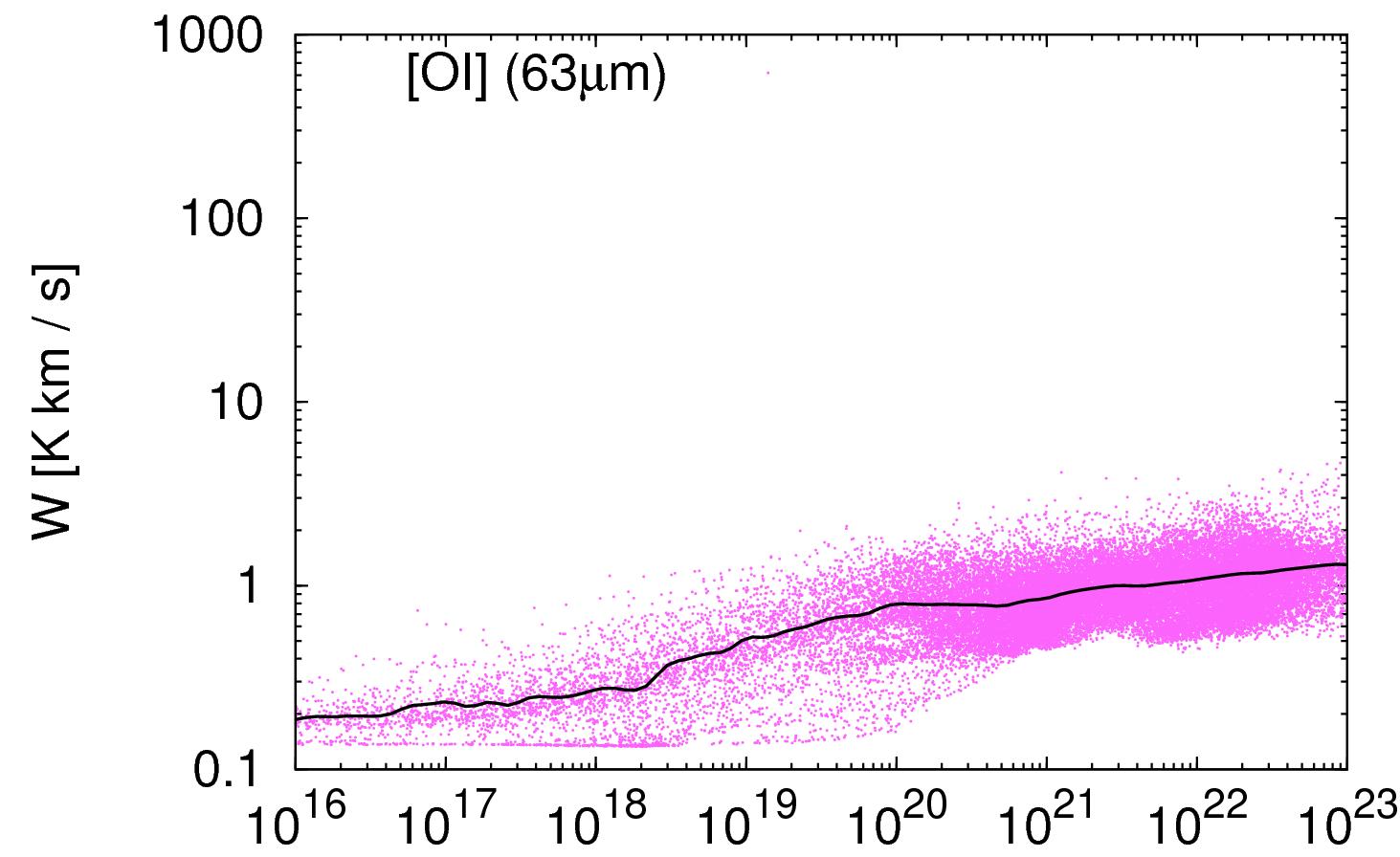}
\includegraphics[height=0.23\linewidth]{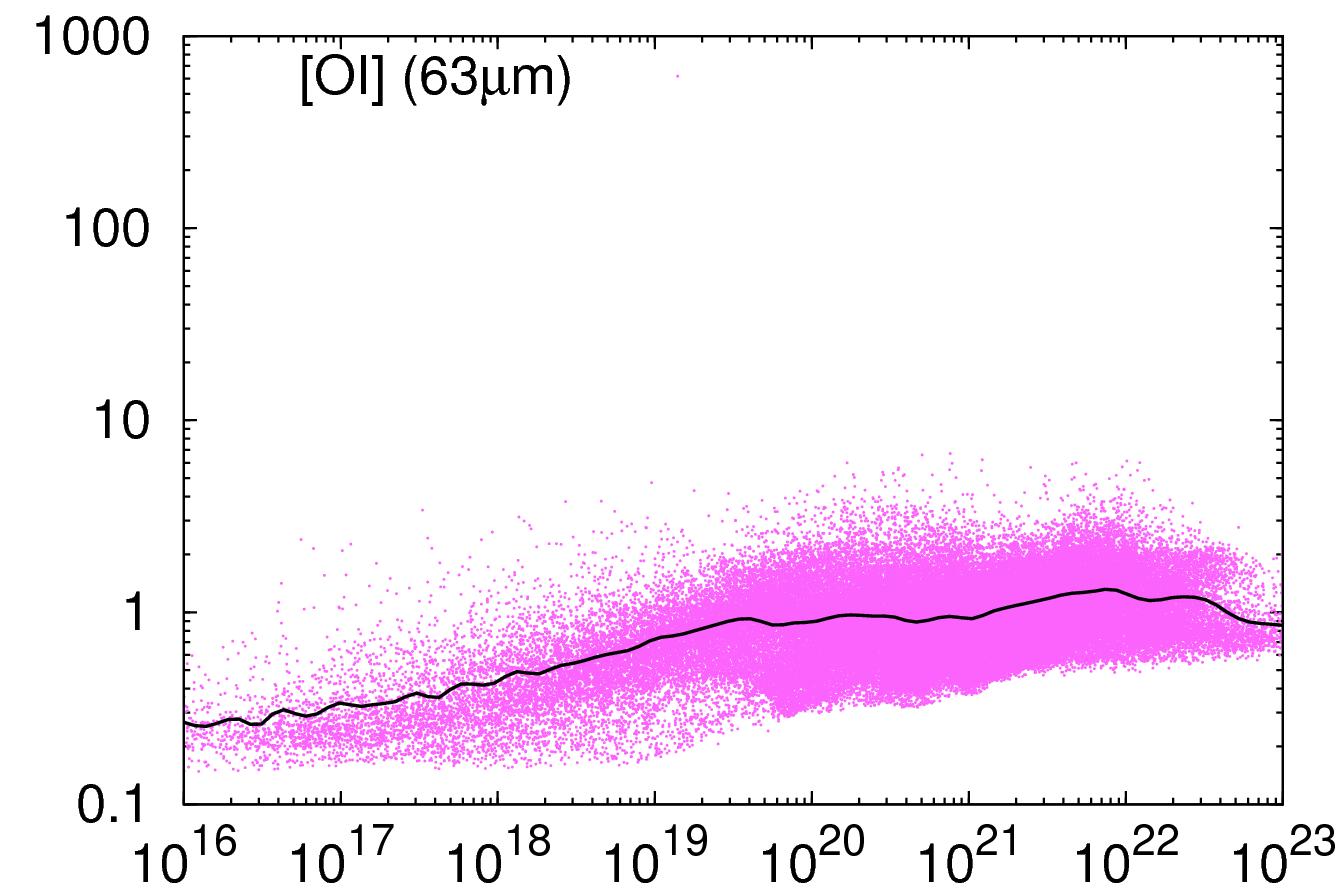}
\includegraphics[height=0.23\linewidth]{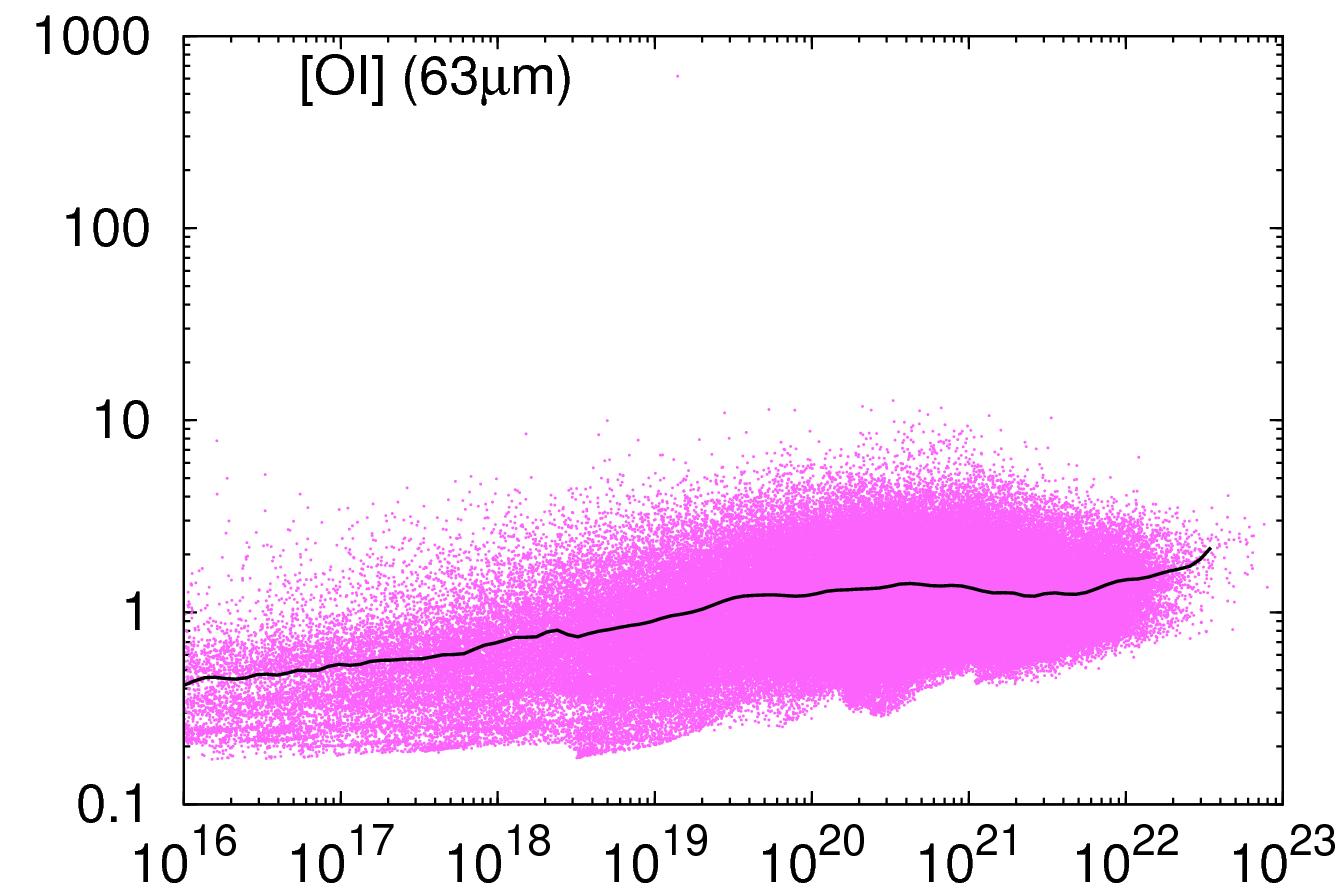}
}
\centerline{
\includegraphics[height=0.255\linewidth]{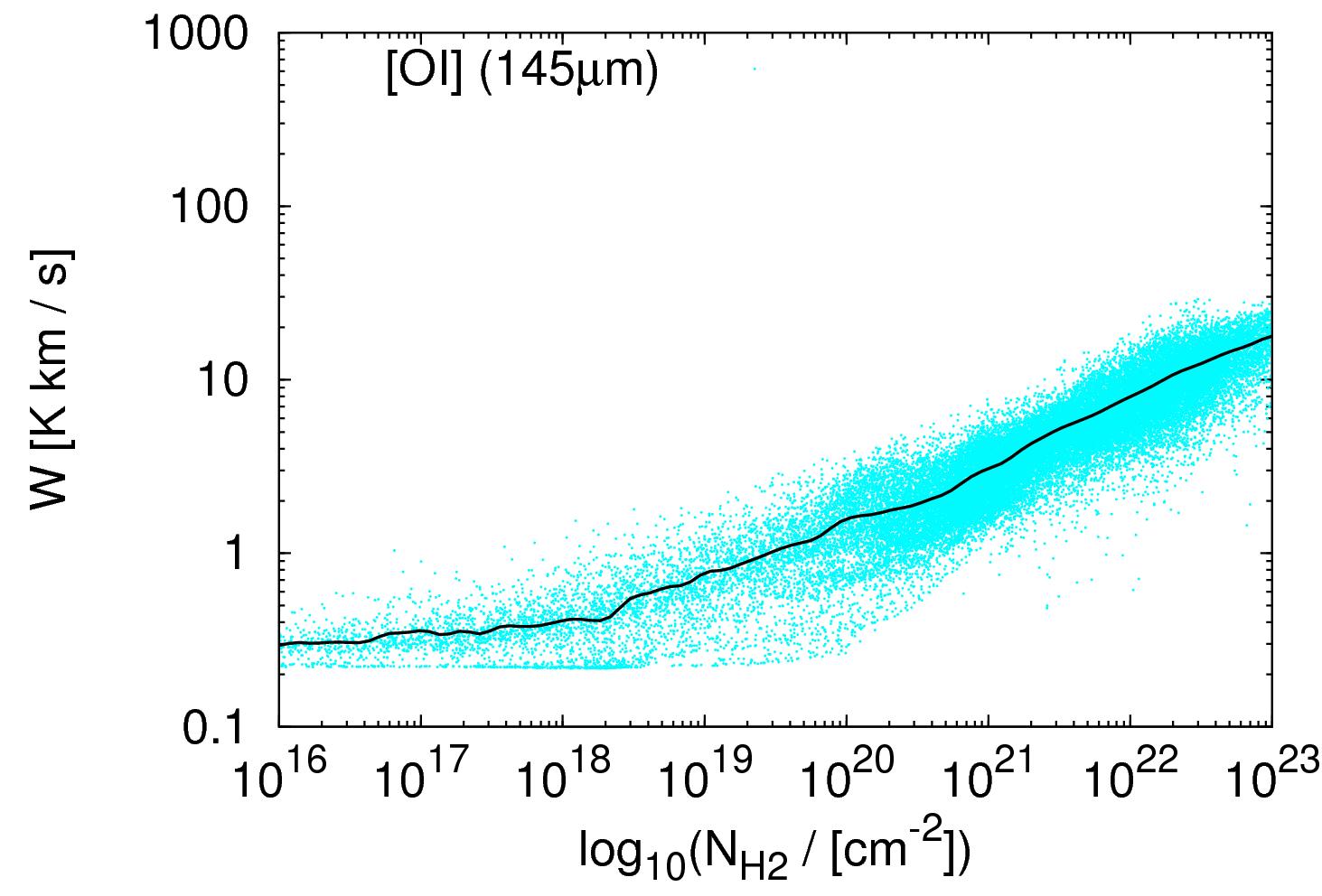}
\includegraphics[height=0.255\linewidth]{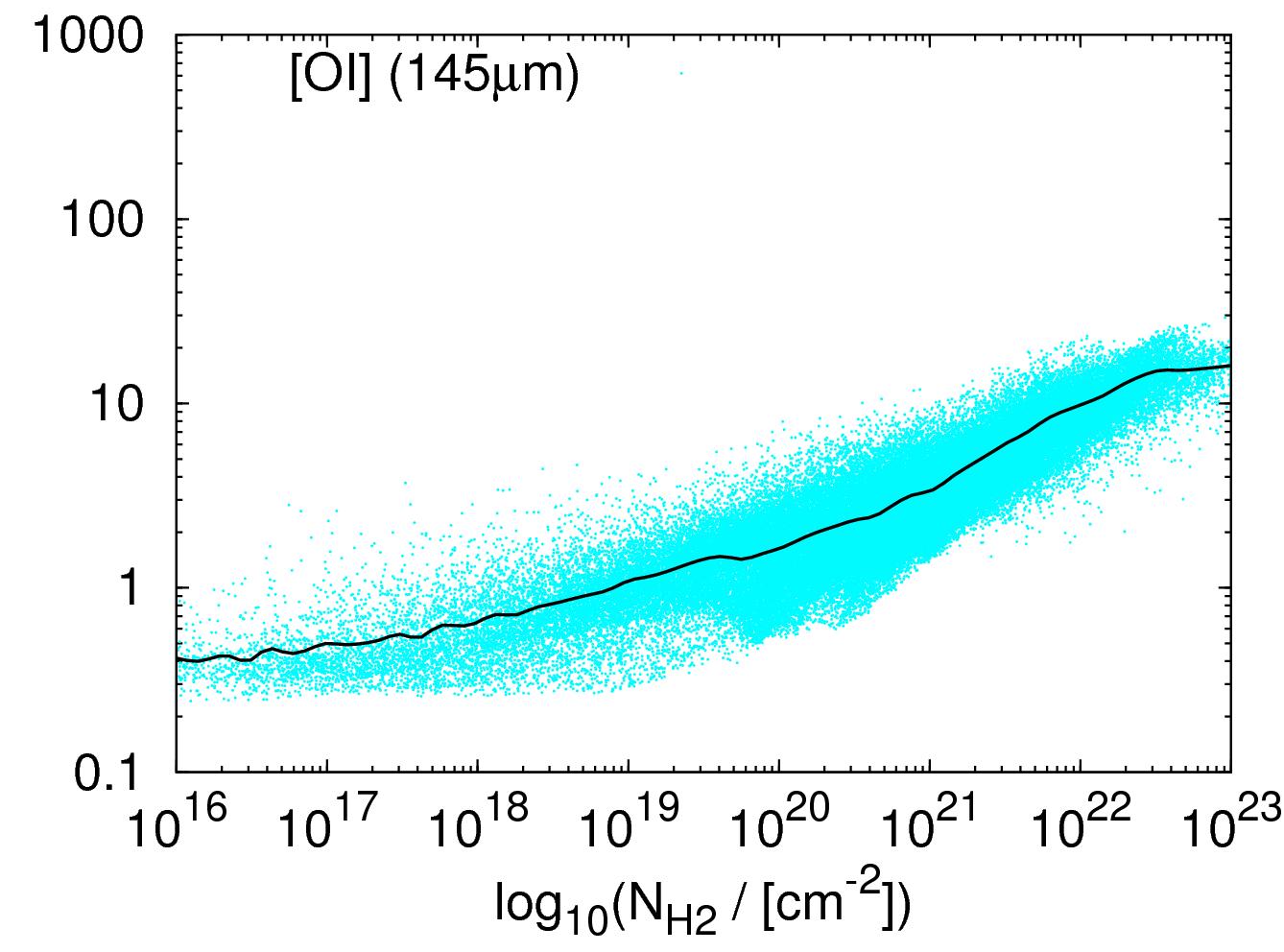}
\includegraphics[height=0.255\linewidth]{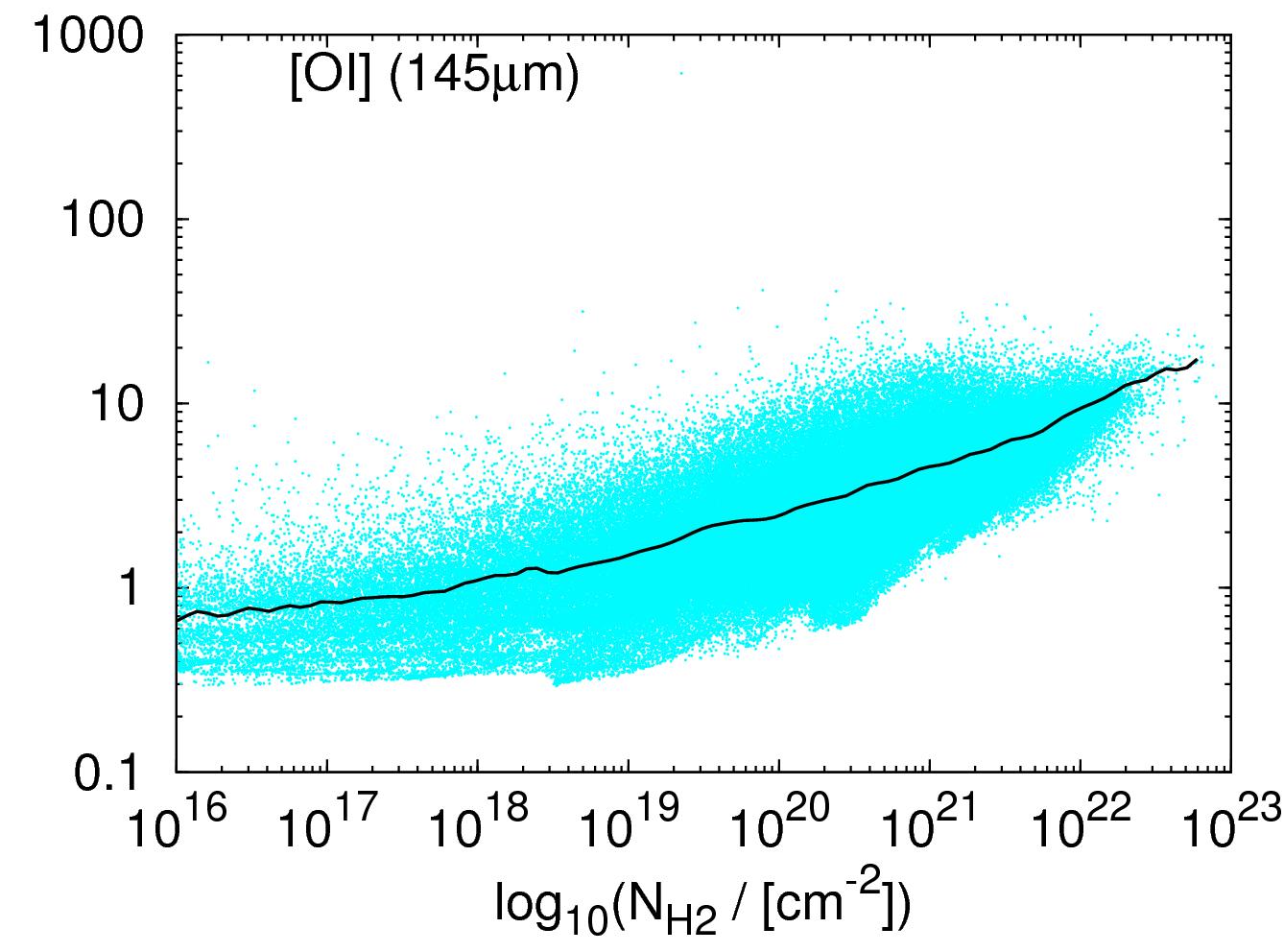}
}
\caption{Same as Fig. \ref{fig:denseregions_tot}, but with the column density of H$_2$.}
\label{fig:denseregions_h2}
\end{figure*}

Now we turn to the densest parts of the cloud, because this is where stars are formed, and we study how well atomic fine-structure lines can be used to trace this regime. Fig. \ref{fig:denseregions_tot} and \ref{fig:denseregions_h2} show velocity-integrated intensities plotted against the total column density and H$_2$ column density for the different virial parameters and all tracers presented in Table \ref{tab:lines}. In these plots, we also show the the mean value at each column density, which is indicated by a black solid line. The small fluctuations at high column densities in each plot are caused by the small number of cells available to compute an average value of $W$.

We find clear differences for the various chemical components. For the total density in Fig. \ref{fig:denseregions_tot}, we observe that the $^{12}$CO and $^{13}$CO tracers fall off sharply with decreasing column density below $\sim10^{22}\,$cm$^{-2}$, which is due to photodissociation of the gas in the diffuse cloud regions, owing to the strong external radiation field. On the other hand, at high column densities, we find a saturation of the $^{12}$CO emission due to the effect of the line opacity, while the $^{13}$CO emission is a significantly better tracer of the column density than $^{12}$CO. This is because $^{13}$CO is optically thin and consequently can better trace compact regions within the cloud. However, even in a small regime where $N_{\text{tot}}$ and $W_{\text{CO}}$ show some degree of correlation, we find a substantial amount of scatter around the mean value for both $^{12}$CO and $^{13}$CO. Furthermore, we also show the scatter plots for the different atomic species in Fig. \ref{fig:denseregions_tot}. As expected, we see that all atomic components extend to significantly lower column densities of $\sim10^{21}\,$cm$^{-2}$ compared to carbon monoxide. Hence, both atomic carbon and oxygen are much better tracers of the low column density material of the total gas in the cloud. However, both the [C{\sc ii}] 158\,$\mu$m and the [O{\sc i}] 63\,$\mu$m line saturate at higher column densities, owing to the effect of the line opacity and thus either $^{13}$CO or the [O{\sc i}] 145\,$\mu$m line is to be preferred as tracers for the dense cloud material.

For the H$_2$ column density in Fig. \ref{fig:denseregions_h2}, we find a different result. Firstly, we see that both $^{12}$CO and $^{13}$CO extend to significantly lower H$_2$ column densities of $\sim5 \times 10^{20}-10^{21}\,$cm$^{-2}$, reflecting the fact that CO is a better tracer of molecular mass than of total mass. At higher column densities, we find a qualitatively similar behavior as for the total gas in Fig. \ref{fig:denseregions_tot}, again owing to the effect of the line opacities. Moreover, we also show the scatter plots for the different atomic species. We note that both the atomic oxygen and carbon extend over a much larger range in H$_2$ column density than in total gas column density, showing that those components can also be used in order to accurately trace a large fraction of the molecular hydrogen mass. However, the correlation between the emission of [C{\sc ii}] 158\,$\mu$m, [O{\sc i}] 63\,$\mu$m and $N_{\text{H}_2}$ breaks down at H$_2$ column densities above $\sim10^{20}-10^{21}\,$cm$^{-2}$, owing to the optical depth of these lines in this density regime. This means that one has to rely on other probes in order to infer information about the compact gas regions. On the contrary, the correlation between the [O{\sc i}] 145\,$\mu$m emission and $N_{\text{H}_2}$ seems to be provide a better way to study the dense H$_2$ clumps in the cloud.

We also analyze the impact that our over-estimation of the CO abundance may have on the relationship between the integrated intensity of the atomic tracers and the column density (see Appendix \ref{sec:noCO}). However, the influence of this turns out to be small.

\subsection{Mass fraction traced by the emission}
\label{subsec:massfraction}

\begin{figure*}
\centerline{
\includegraphics[width=0.27\linewidth]{images/alpha05_plot.jpg}
\includegraphics[width=0.27\linewidth]{images/alpha2_plot.jpg}
\includegraphics[width=0.27\linewidth]{images/alpha8_plot.jpg}
}
\centerline{
\includegraphics[height=0.26\linewidth]{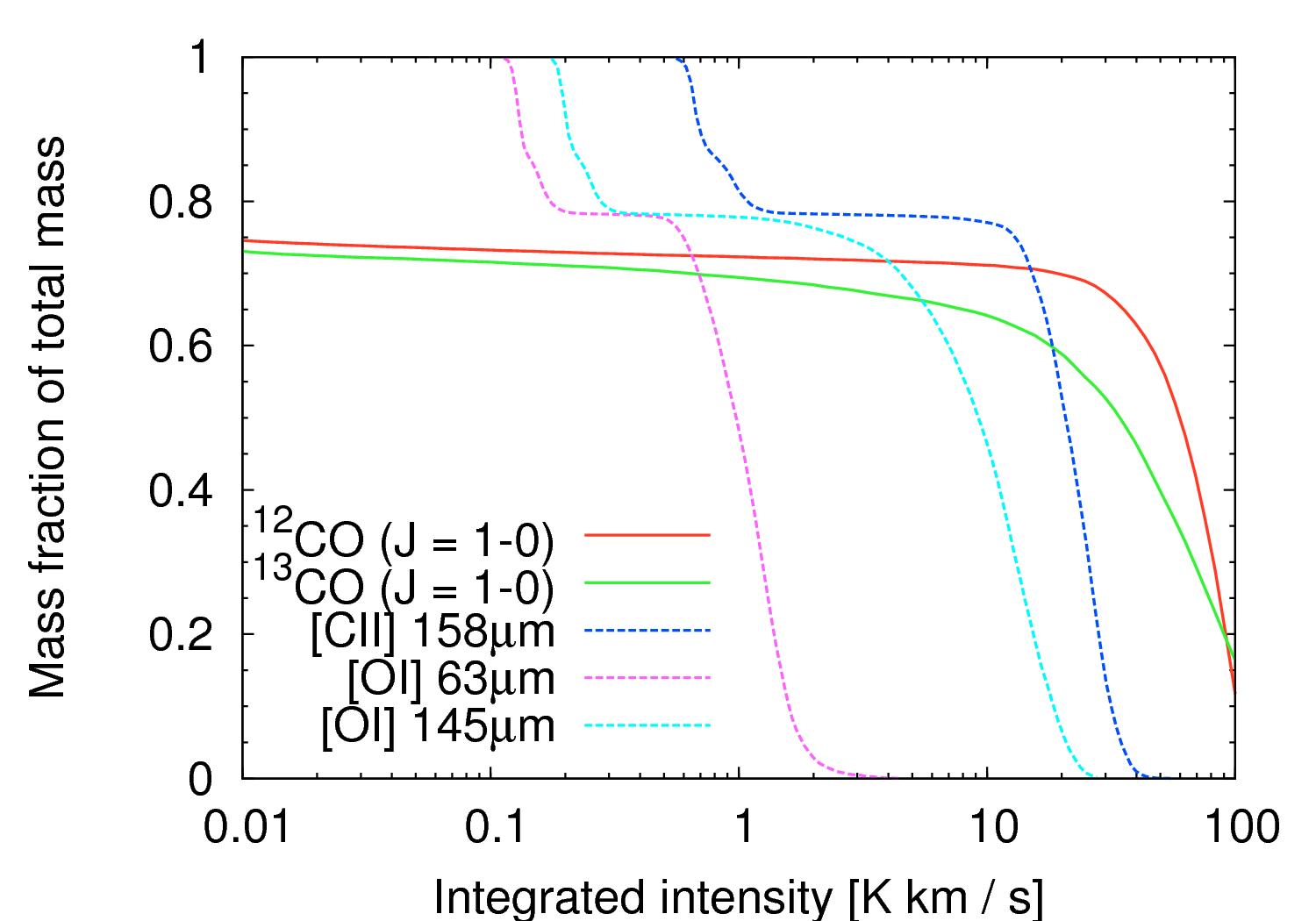}
\includegraphics[height=0.26\linewidth]{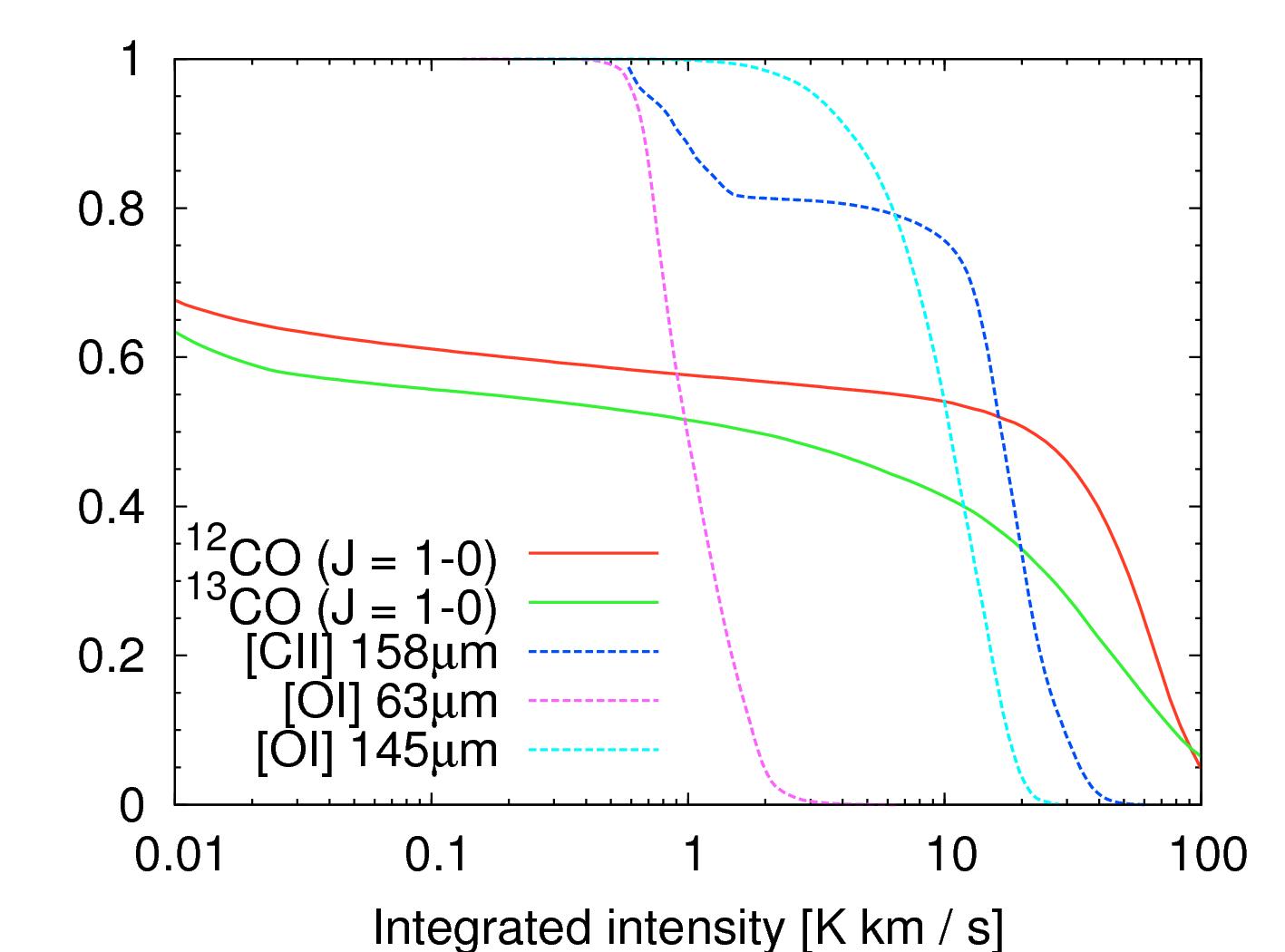}
\includegraphics[height=0.26\linewidth]{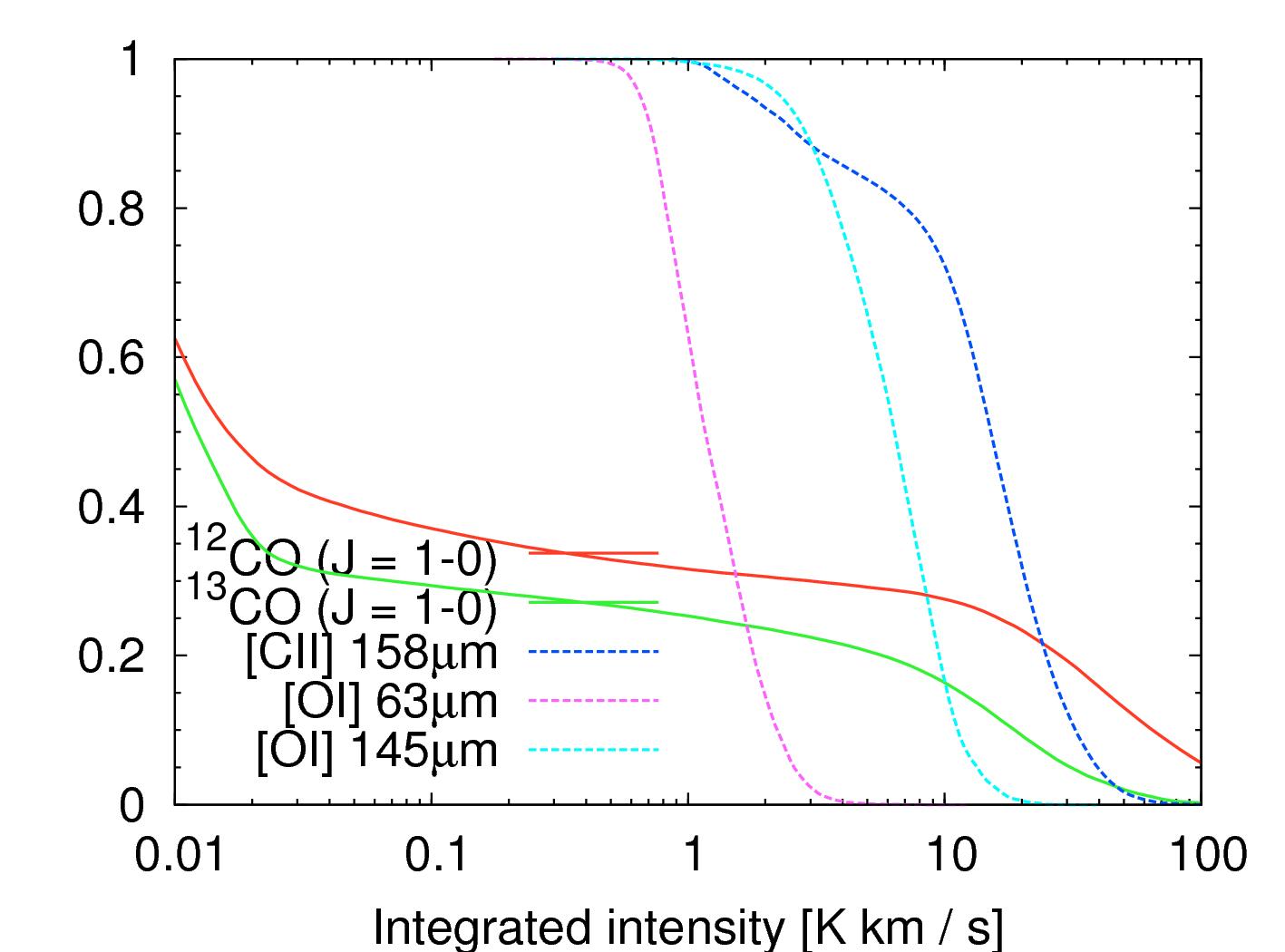}
}
\centerline{
\includegraphics[height=0.26\linewidth]{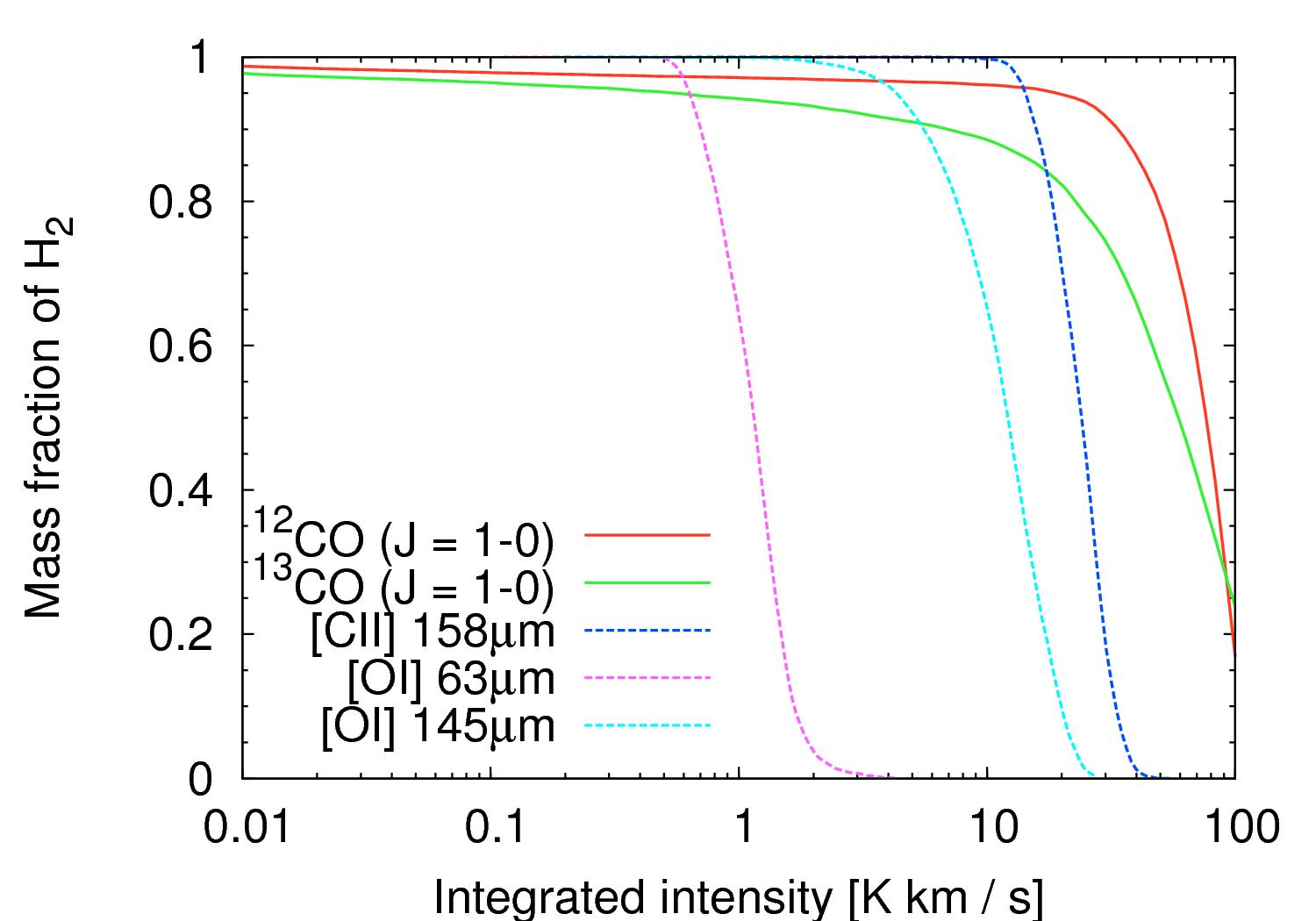}
\includegraphics[height=0.26\linewidth]{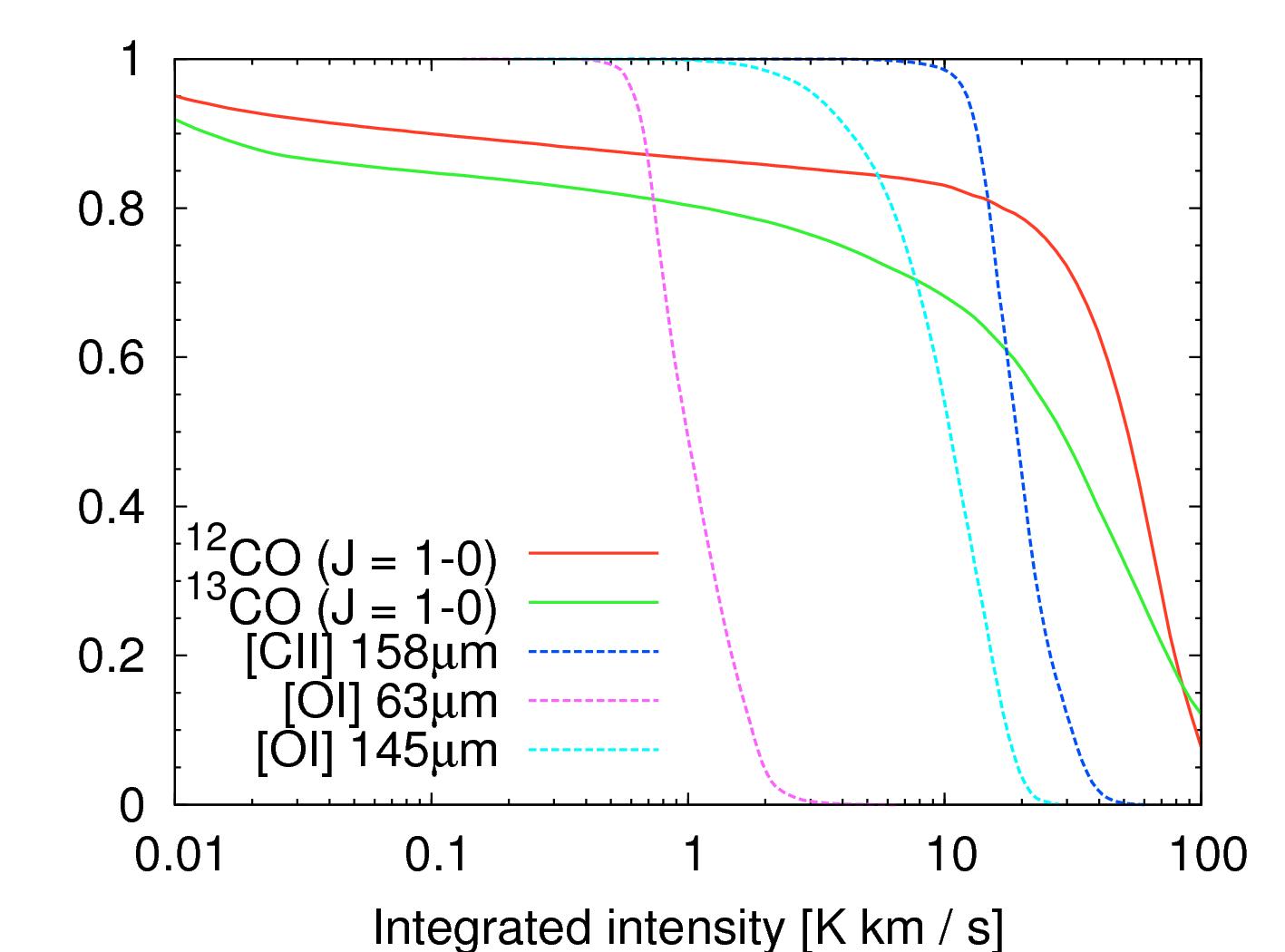}
\includegraphics[height=0.26\linewidth]{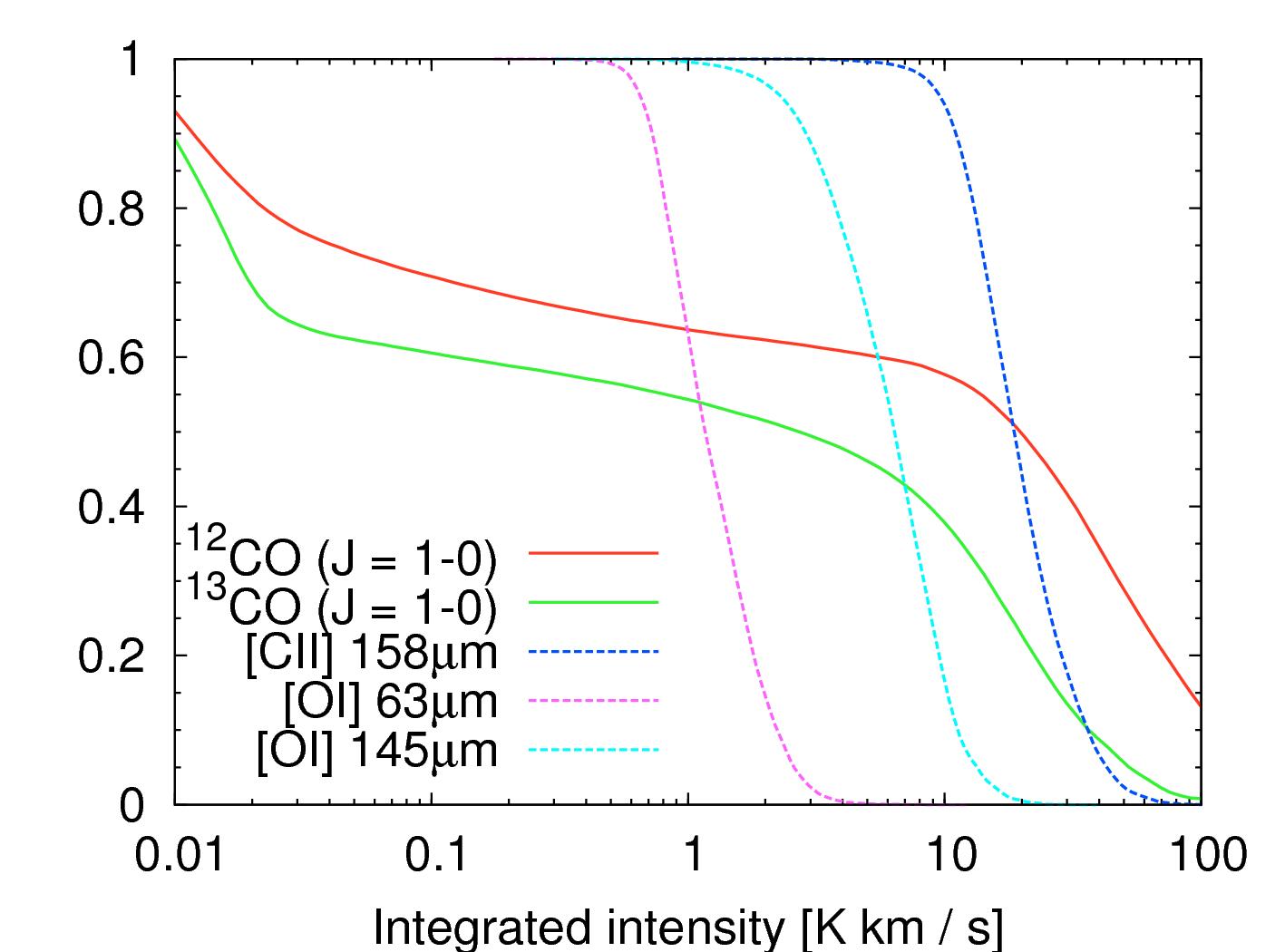}
}
\caption{Cumulative fraction of the total (top row) and H$_2$ (bottom row) mass in the cloud for different virial parameters $\alpha = 0.5$ (left column), 2.0 (middle column) and 8.0 (right column), traced by our various chemical components presented in Table \ref{tab:lines}, as a function of the minimum velocity-integrated intensity in the line.}
\label{fig:massfraction}
\end{figure*}

In Section \ref{subsec:shape} we speculated that a significant fraction of the total gas mass may not be traced by $^{12}$CO or $^{13}$CO. To quantify this finding, Fig. \ref{fig:massfraction} shows the fraction of total and H$_2$ gas mass observed along the LoS having velocity-integrated intensities greater than a minimum value $W_{\text{min}}$ for our various tracers. Furthermore, we analyze how this mass fraction changes as we vary $W_{\text{min}}$. In this context, Fig. \ref{fig:massfraction} defines a theoretical limit for detecting gas above a given sensitivity threshold, e.g. determined by the telescope or the detector.

We see that all lines are a strong function of the minimum velocity-integrated intensity and of the virial parameter. In case of the mass fraction of the total gas, we find that our different atomic tracers recover almost all of the total mass if the minimum integrated intensity is $W_{\text{min}} \approx 0.1\,$K\,km\,s$^{-1}$ in the $\alpha = 0.5$ model and $W_{\text{min}} \approx 1.0\,$K\,km\,s$^{-1}$ in the $\alpha = 2.0$ and $\alpha = 8.0$ models. Furthermore, as already indicated in Section \ref{subsec:shape}, we see that in the case of very small threshold values $W_{\text{min}}$ in the $\alpha = 0.5$ model, both $^{12}$CO and $^{13}$CO only trace about 75\% of the total mass or less. This fraction decreases even further for higher $\alpha$ values. For example, if we adopt $W_{\text{min}} \approx 0.1\,$K\,km\,s$^{-1}$ in the cloud with $\alpha = 8.0$, we find that only $\sim40\%$ of the total mass is traced by CO, while all atomic tracers recover almost 100\% of the total gas mass. Furthermore, we observe narrow emission ranges for some of the atomic components, which trace almost 100\% of the total mass. For example, regarding our intermediate $\alpha = 2.0$ model and the [O{\sc i}] 63\,$\mu$m tracer, almost 100\% of the total mass lies in the narrow emission range between $0.5 < W_{\text{min}} < 3.0$. In the case of the other $\alpha$ models, we find a similar behavior. This is because there are almost no lines-of-sight having higher values of integrated intensity (see also Fig. \ref{fig:denseregions_tot} and \ref{fig:denseregions_h2}), which may be different in even denser clouds.

Regarding the H$_2$ mass, we see that both $^{12}$CO and $^{13}$CO trace a higher H$_2$ mass fraction compared to the total mass of the cloud. This is because the molecular hydrogen can only exist in dense regions of the cloud, owing to the ability to self-shield from the external radiation field. This is similar to our finding for the carbon monoxide. Furthermore, we observe that the atomic components recover almost 100\% of the H$_2$ mass if we set $W_{\text{min}} \approx 0.5\,$K\,km\,s$^{-1}$. Hence, we find that those components again trace a significantly higher mass fraction of the gas compared to the carbon monoxide. Regarding both $^{12}$CO and $^{13}$CO, we see that the amount of mass recovered by those molecular tracers decreases with higher virial parameter for a fixed minimum value of the integrated intensity. For example, if we adopt a value of $W_{\text{min}} \approx 0.1\,$K\,km\,s$^{-1}$, almost 98\% of the H$_2$ mass is traced by $^{12}$CO for our $\alpha = 0.5$ model, while $\sim90\%$ is traced in our $\alpha = 2.0$ and only $\sim70\%$ can be recovered in the $\alpha = 8.0$ model. Moreover, similar to the mass fraction of the total gas mass, we also find a narrow emission range for the various atomic lines in all $\alpha$ models, which trace about 100\% of the H$_2$ mass. In contrast, the emission of the $^{12}$CO and $^{13}$CO molecules is extended over a much wider range in integrated intensities. Hence, a much larger emission range is needed in order to trace the whole mass of molecular hydrogen in the cloud.

In summary, we conclude that CO is only a good tracer for the molecular content of the cloud, but can miss a significant fraction of the total gas mass, particularly for unbound clouds.

\subsection{Estimating the CO-to-H$_2$ conversion factor}
\label{subsec:Xfactor}

\begin{figure}
\centerline{
\includegraphics[height=0.65\linewidth]{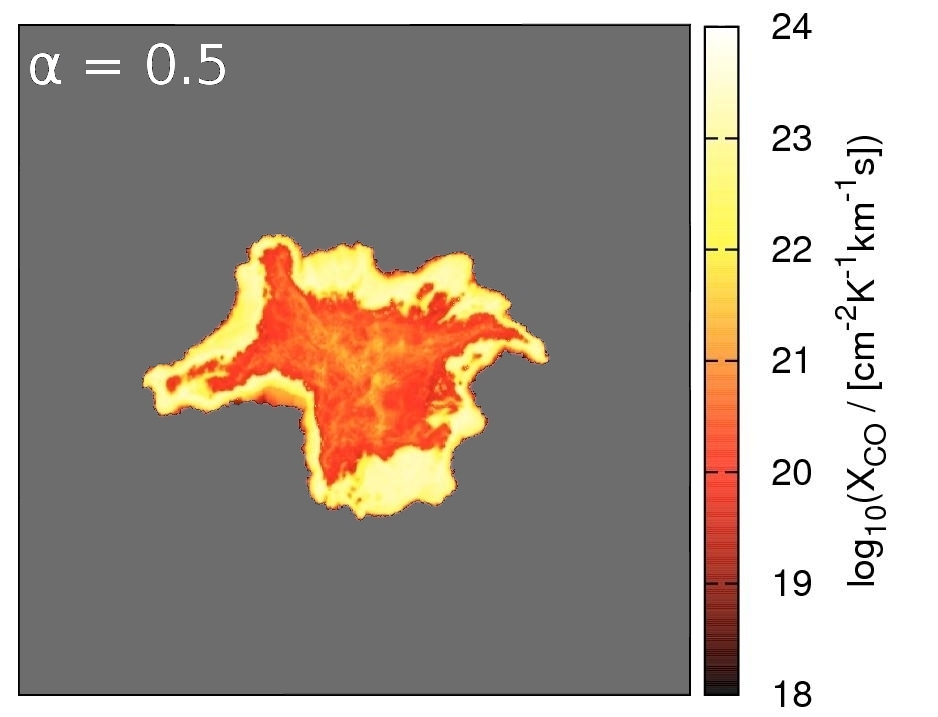}
}
\centerline{
\includegraphics[height=0.65\linewidth]{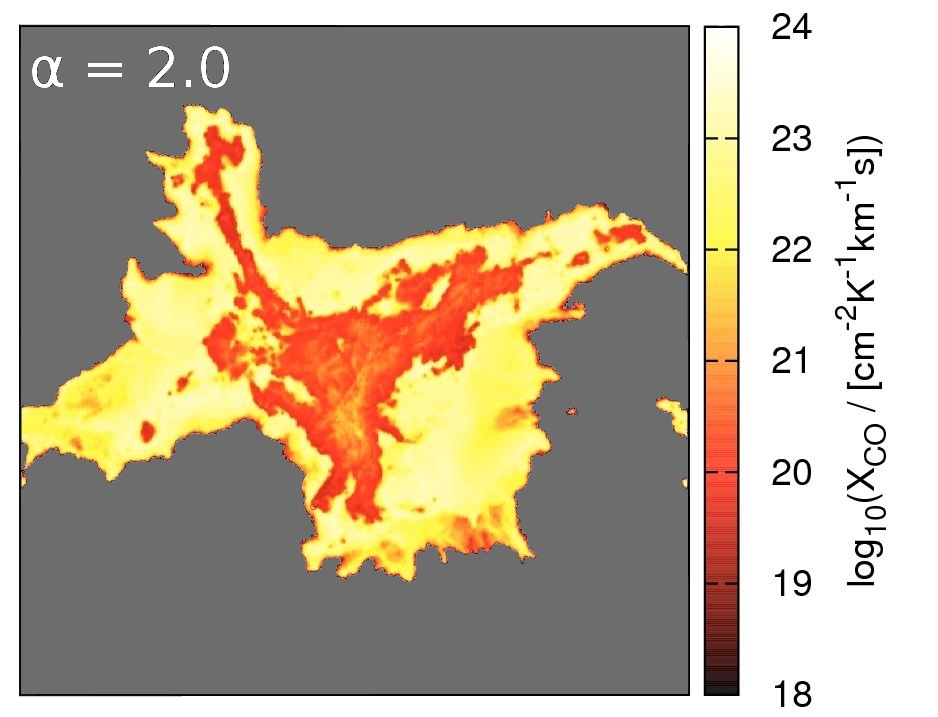}
}
\centerline{
\includegraphics[height=0.65\linewidth]{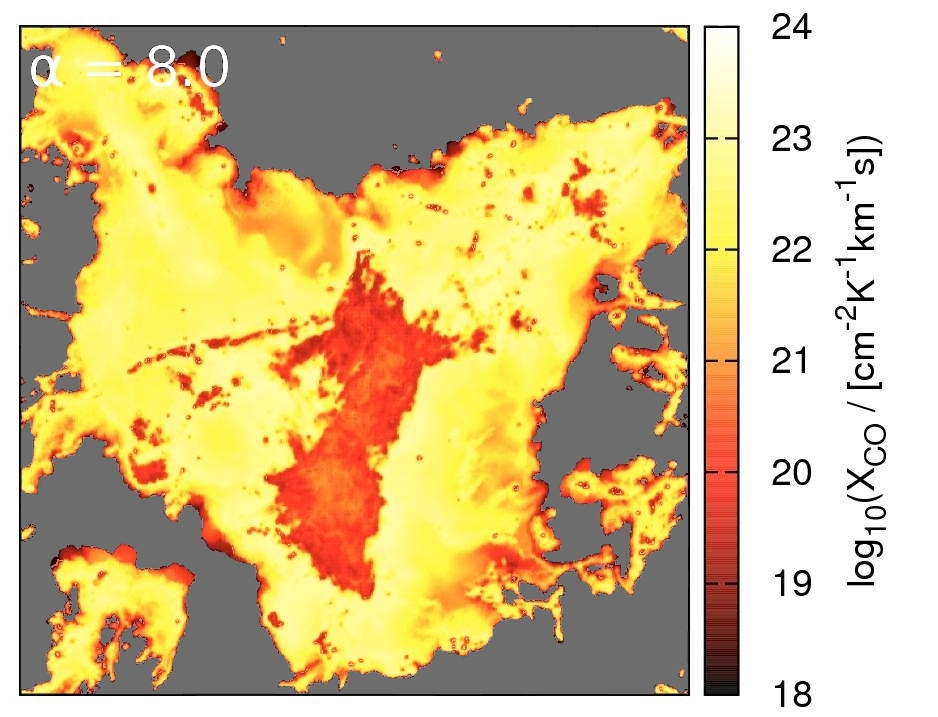}
}
\caption{Logarithmic maps of the CO-to-H$_2$ conversion factor, $X_{\text{CO}} = N_{\text{H}_2} / W_{\text{CO}}$, for the three models: $\alpha = 0.5$ (top), 2.0 (middle) and 8.0 (bottom). The average $X_{\text{CO}}$-factor values are given in Table \ref{tab:Xfactor}. We see that only the inner parts of the cloud yield the canonical $X_{\text{CO}}$-factor value on average, while the outer cloud regions exhibit significantly larger $X_{\text{CO}}$ values. The gray background denotes regions where the CO emission is zero. Hence, it is impossible to compute a value for $X_{\text{CO}}$ there.}
\label{fig:Xfactor}
\end{figure}

Based on our maps of column densities for H$_2$ (Fig. \ref{fig:totdensh2}) and velocity-integrated intensities for $^{12}$CO (Fig. \ref{fig:tracers}), we also estimate the value of the CO-to-H$_2$ conversion factor, $X_{\text{CO}}$, in our various models. $X_{\text{CO}}$ is a widely used quantity in particular in extragalactic astronomy to derive H$_2$ column densities from CO observations \citep{BolattoEtAl2013}. It is defined via
\begin{equation}
\label{eq:Xfactor}
X_{\text{CO}} = \frac{N_{\text{H}_2}}{W_{\text{CO,1-0}}},
\end{equation}
where $N_{\text{H}_2}$ denotes the H$_2$ column density and $W_{\text{CO,1-0}}$ the velocity-integrated brightness-temperature. Previous studies tried to estimate the $X_{\text{CO}}$-factor in both observations and numerical simulations \citep[see, e.g.][]{SolomonEtAl1987,PolkEtAl1988,YoungAndScoville1991,DameEtAl2001,LisztEtAl2010,LeroyEtAl2011,GloverAndMacLow2011,GloverAndClark2012,BolattoEtAl2013,NarayananAndHopkins2013,ClarkAndGlover2015}. For example, \citeauthor{ShettyEtAl2011b}~(2011a,b) investigated the $X_{\text{CO}}$-factor in numerical simulations of turbulent clouds, varying environmental properties such as the initial number density, the metallicity and the external UV field, finding cloud average values $X_{\text{CO}} \approx 2-4 \times 10^{20}\,$cm$^{-2}$\,K$^{-1}$\,km$^{-1}$\,s for solar-metallicity models. Moreover, \citet{BolattoEtAl2013} review the efforts to measure the CO-to-H$_2$ conversion factor for molecular clouds in the Milky Way disk, recommending a similar factor $X_{\text{CO}} \sim 2 \times 10^{20}\,$cm$^{-2}$\,K$^{-1}$\,km$^{-1}$\,s with $\pm30\%$ uncertainty. However, all these studies arrived at the conclusion that the $X_{\text{CO}}$-factor in the Galaxy is remarkably constant, having a value of $X_{\text{CO}} \approx 2-4 \times 10^{20}\,$cm$^{-2}$\,K$^{-1}$\,km$^{-1}$\,s.

Nevertheless, in the last few years concerns have been raised that the $X_{\text{CO}}$-factor may be significantly different from the canonical value in MCs located in the galactic center. For example, \citet{NarayananEtAl2012} have shown that large gas temperatures and velocity dispersions can increase the CO intensity, thus decreasing $X_{\text{CO}}$. On the other hand, the strong ISRF and CRF can also destroy the carbon monoxide, increasing $X_{\text{CO}}$ \citep{ClarkAndGlover2015}. In addition, observational measurements of $X_{\text{CO}}$ at the center of other nearby spiral galaxies typically find values that are lower than the canonical Galactic value \citep{SandstromEtAl2013}. Moreover, conditions in the Galactic Center are somewhat similar to those found in ULIRGs, and there is considerable observational evidence that $X_{\text{CO}}$ is smaller in ULIRGs than in normal spiral galaxies \citep[see e.g. the detailed discussion in][]{BolattoEtAl2013}. So far, we are still missing a reliable picture of the $X_{\text{CO}}$-factor in different (and extreme) physical environments.

\begin{table}
\begin{tabular}{l|c}
\hline\hline
Model & $X_{\text{CO}}$ \\
\hline
 & [cm$^{-2}$\,K$^{-1}$\,km$^{-1}$\,s] \\
\hline
GC-0.5-1000 & $3.9 \times 10^{20}$ \\
GC-2.0-1000 & $2.6 \times 10^{20}$ \\
GC-8.0-1000 & $1.3 \times 10^{20}$ \\
\hline
\end{tabular}
\caption{Values of the CO-to-H$_2$ conversion factor, relating the amount of $^{12}$CO ($J=1 \rightarrow 0$) emission $W_{\text{CO}}$ to the H$_2$ column density $N_{\text{H}_2}$ for all our $\alpha$ runs. For the computation of the $X_{\text{CO}}$-factor, we only include pixels where the velocity-integrated intensity for carbon monoxide is larger than the corresponding threshold given in Table \ref{tab:thresh}. Thus, $X_{\text{CO}}$ only accounts for regions with a significant amount of CO emission in the cloud.}
\label{tab:Xfactor}
\end{table}

In the following, we evaluate the $X_{\text{CO}}$-factor for our three $\alpha$ models and list our results in Table \ref{tab:Xfactor}. In general, the $X_{\text{CO}}$-factors given in Table \ref{tab:Xfactor} are computed by taking the ratio of the mean values of the H$_2$ column density and the velocity-integrated intensity of CO. We only include pixels where the velocity-integrated intensity for carbon monoxide is larger than the corresponding threshold given in Table \ref{tab:thresh}. These are 1.0, 0.3 and 0.1\,K\,km\,s$^{-1}$ for the models with $\alpha = 0.5, 2.0$ and $8.0$. Thus, $X_{\text{CO}}$ only accounts for regions with a significant available amount of CO emission in the cloud, as illustrated in Fig. \ref{fig:tracers}.

We find a range of values for the CO-to-H$_2$ conversion factor in Table \ref{tab:Xfactor}, reaching from $\sim 1-4 \times 10^{20}\,$cm$^{-2}$\,K$^{-1}$\,km$^{-1}$\,s. Our estimates are in good agreement to estimates of the canonical $X_{\text{CO}}$-factor in the Milky Way. However, we again emphasize that our \citet{NelsonAndLanger1997} network overestimates the rate at which CO forms (see also Section \ref{subsec:hydromodel}) and hence we expect that our $X_{\text{CO}}$-factors listed in Table \ref{tab:Xfactor} are lower limits.

Furthermore, we show logarithmic maps of the $X_{\text{CO}}$-factor for all our $\alpha$ models in Fig. \ref{fig:Xfactor}. Here, the $X_{\text{CO}}$-values are evaluated by computing the ratio of the H$_2$ column density and the velocity-integrated intensity of CO for each individual pixel. The gray background denotes regions where the CO emission is zero. Hence, it is impossible to compute a value for $X_{\text{CO}}$ there. We find significantly different values for the inner and the outer parts of the clouds. While the inner cloud regions reproduce the canonical $X_{\text{CO}}$-factor value on average, we find that the outer cloud regions exhibit significantly larger $X_{\text{CO}}$ values compared to the canonical one. This is because the carbon monoxide is photodissociated at the edges of our clouds by the strong ISRF, leading to an increase in $X_{\text{CO}}$. In contrast, the molecular CO is better able to self-shield in the dense parts of the cloud and hence we observe that those regions fairly reproduce the canonical value.

\section{Summary and Conclusions}
\label{sec:summary}

In this paper we have analyzed synthetic images of MCs in a CMZ-like environment. For this purpose, we have performed numerical simulations of model clouds with the moving mesh code A{\sc repo} \citep{Springel2010} using environmental properties comparable to those experienced by typical CMZ clouds. We adopted values for the interstellar radiation field (ISRF) and the cosmic-ray flux (CRF) that are a factor of $\sim1000$ larger than the values measured in the solar neighbourhood \citep{ClarkEtAl2013}. We simulated clouds with an initial number density of $n_0 = 10^3\,$cm$^{-3}$ and studied the impact of different virial $\alpha$ parameters of $\alpha = 0.5, 2.0$ and $8.0$. The total cloud mass was set to a constant value of $M_{\text{tot}} = 1.3 \times 10^5\,$M$_{\odot}$. Furthermore, we used the radiative transfer code R{\sc admc}-3{\sc d} \citep{Dullemond2012} to compute synthetic maps of our clouds in important diagnostic lines, i.e. we model the emission of the cloud in {[C{\sc ii}]} (158\,$\mu$m), [O{\sc i}] (145\,$\mu$m), [O{\sc i}] (63\,$\mu$m), $^{12}$CO (2600\,$\mu$m) and $^{13}$CO (2720\,$\mu$m). We report the following findings:
\begin{itemize}
\item Atomic carbon and oxygen is found in both cold and warm regions in the clouds, but CO only traces cold regions (see Section \ref{subsec:state}).
\item If the cloud is virialized ($\alpha = 0.5$), the H$_2$ gas is much better able to self-shield from the high external ISRF compared to other runs with a larger $\alpha$ value. This is because the gravitational pressure compresses the gas to much higher densities because the amount of turbulent kinetic energy in the box is smaller in the low $\alpha$ models than in the high virial parameter models. In the latter case, radiation can penetrate deeper into the MC, leading to photodissociation of the molecular hydrogen (see Section \ref{subsec:shape}).
\item We show that the atomic components trace the shape of the total cloud very well even in diffuse cloud regions (see Section \ref{subsec:shape}).
\item On the other hand, the molecular gas (H$_2$ as well as CO) is photodissociated at the edges of the MC by the strong external ISRF.
\item As a consequence, the photodissociation has a strong impact on measurements of the effective MC radius using a molecular tracer, which tend to significantly underestimate the radius of the total cloud in environments where the external radiation field and the turbulent kinetic energy is high. However, we find that the atomic tracers recover a significantly larger fraction of the total cloud mass (see Section \ref{subsec:cloudradii}).
\item We find a large dynamical range for the [O{\sc i}] 145\,$\mu$m line in column density space, which traces both the H$_2$ and the total gas up to the most dense regions in the cloud. In contrast, the [O{\sc i}] 63\,$\mu$m and {[C{\sc ii}]} 158\,$\mu$m lines already saturate at significantly lower column densities. Hence, we would not expect that those components trace the compact MC regions very well (see Section \ref{subsec:denseregions_tot}).
\item We find that all atomic components trace almost 100\% of the H$_2$ and the total gas mass above a sensitivity threshold of $W_{\text{min}} \approx 0.1\,$K\,km\,s$^{-1}$ for the velocity-integrated intensities. However, we also find that CO only traces a significantly lower mass fraction of H$_2$ and the total density compared to the atomic components (see Section \ref{subsec:massfraction}).
\item We compute values of the CO-to-H$_2$ conversion factor $X_{\text{CO}} = N_{\text{H}_2} / W_{\text{CO}}$, which relates the H$_2$ column density $N_{\text{H}_2}$ to the amount of $^{12}$CO ($J=1 \rightarrow 0$) emission $W_{\text{CO}}$. We find values in the range $X_{\text{CO}} \approx 1-4 \times 10^{20}\,$cm$^{-2}$\,K$^{-1}$\,km$^{-1}$\,s, in good agreement to the canonical $X_{\text{CO}}$-factor values obtained in observations of MCs in the Milky Way (see Section \ref{subsec:Xfactor}). However, we caution that simplifications in our chemical model mean that these values are lower limits.
\end{itemize}

\section*{Acknowledgements}

We thank the referee for an insightful and very constructive report. We also thank Stella Offner and Diane Cormier for informative discussions about the project. EB, SCOG and RSK acknowledge support from the Deutsche Forschungsgemeinschaft (DFG) via the SFB 881 (sub-projects B1, B2 and B8) ``The Milky Way System'', and the SPP (priority program) 1573, ``Physics of the ISM''. Furthermore, EB acknowledges financial support from the Konrad-Adenauer-Stiftung (KAS) via their ``Promotionsf\"orderung''. The simulations presented in this paper were performed on the Milkyway supercomputer at the J\"ulich Forschungszentrum, funded via SFB 881. The radiative transfer post-processing was performed on the \textit{kolob} cluster at the University of Heidelberg, which is funded in part by the DFG via Emmy-Noether grant BA 3706, and via a Frontier grant of Heidelberg University, sponsored by the German Excellence Initiative as well as the Baden-W\"urttemberg Foundation. RSK acknowledges support from the European Research Council under the European Community's Seventh Framework Programme (FP7/2007-2013) via the ERC Advanced Grant "STARLIGHT: Formation of the First Stars" (project number 339177). SER acknowledges support from VIALACTEA, a Collaborative Project under Framework Programme 7 of the European Union, funded under Contract \#607380.

\bibliographystyle{mn2e}
\bibliography{lit/literature}

\begin{appendix}

\section{Radiative transfer post-processing: influence of grid resolution}
\label{sec:resolution}

\begin{figure}
\centerline{
\includegraphics[height=0.54\linewidth]{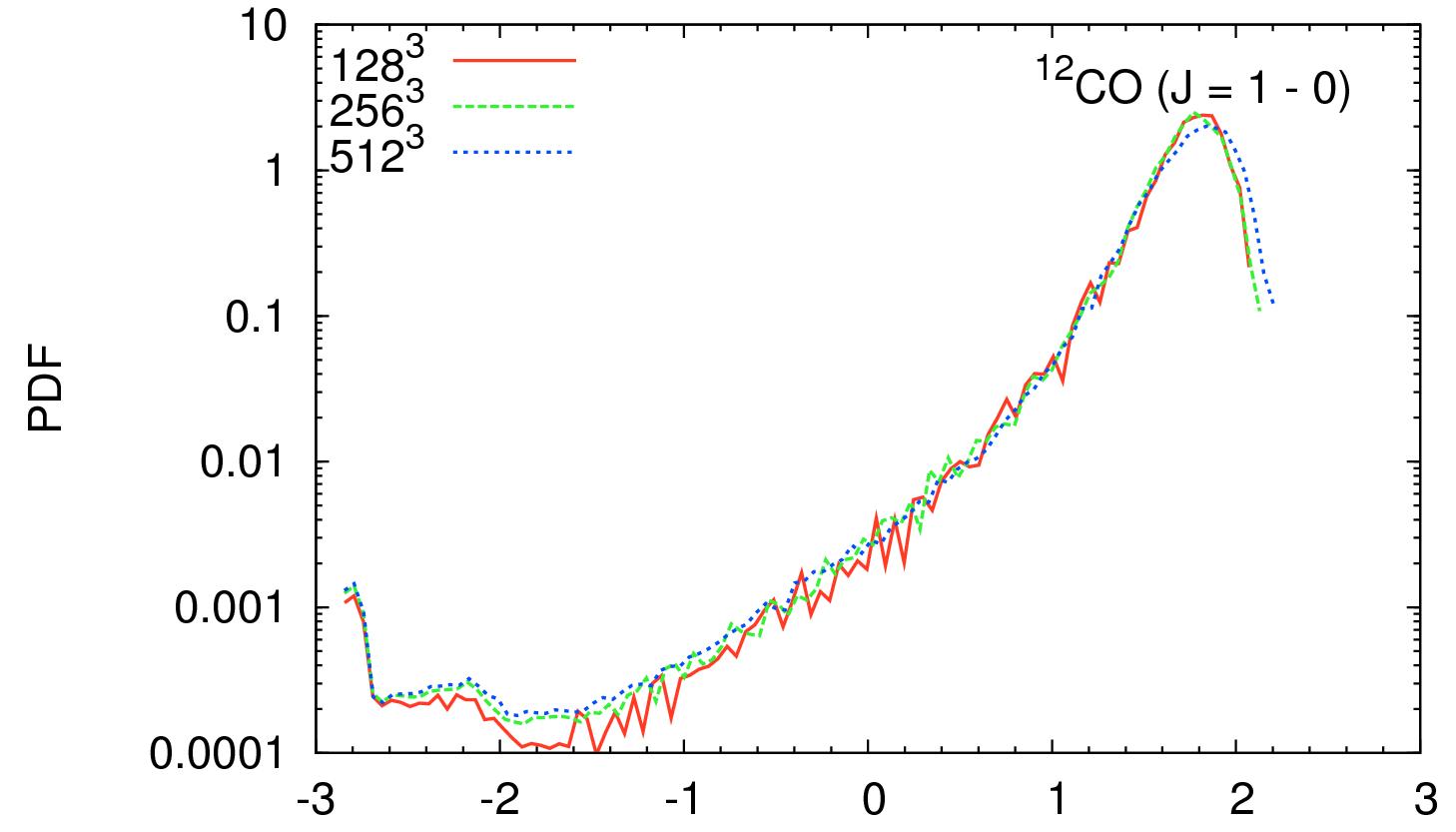}
}
\centerline{
\includegraphics[height=0.54\linewidth]{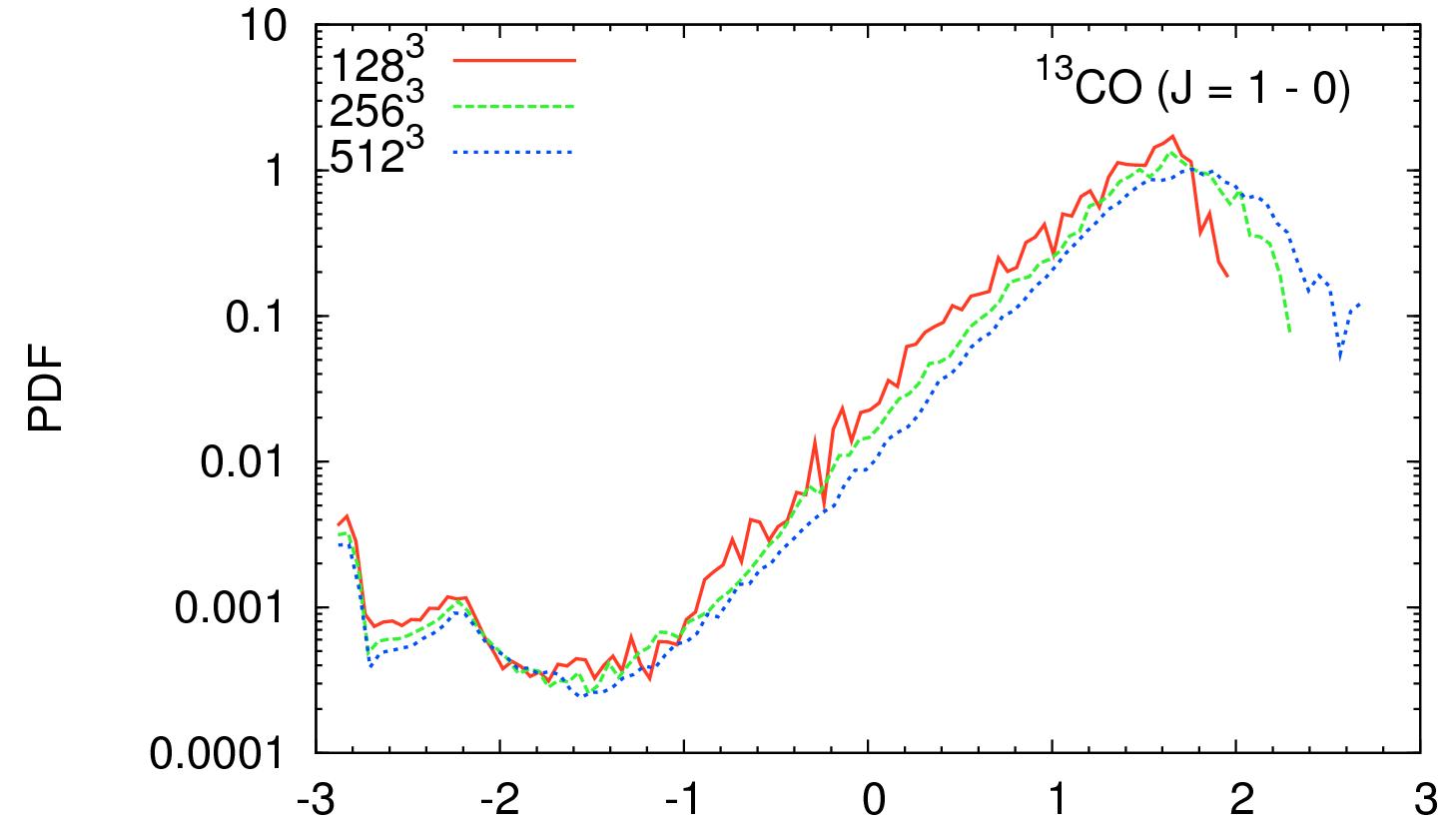}
}
\centerline{
\includegraphics[height=0.52\linewidth]{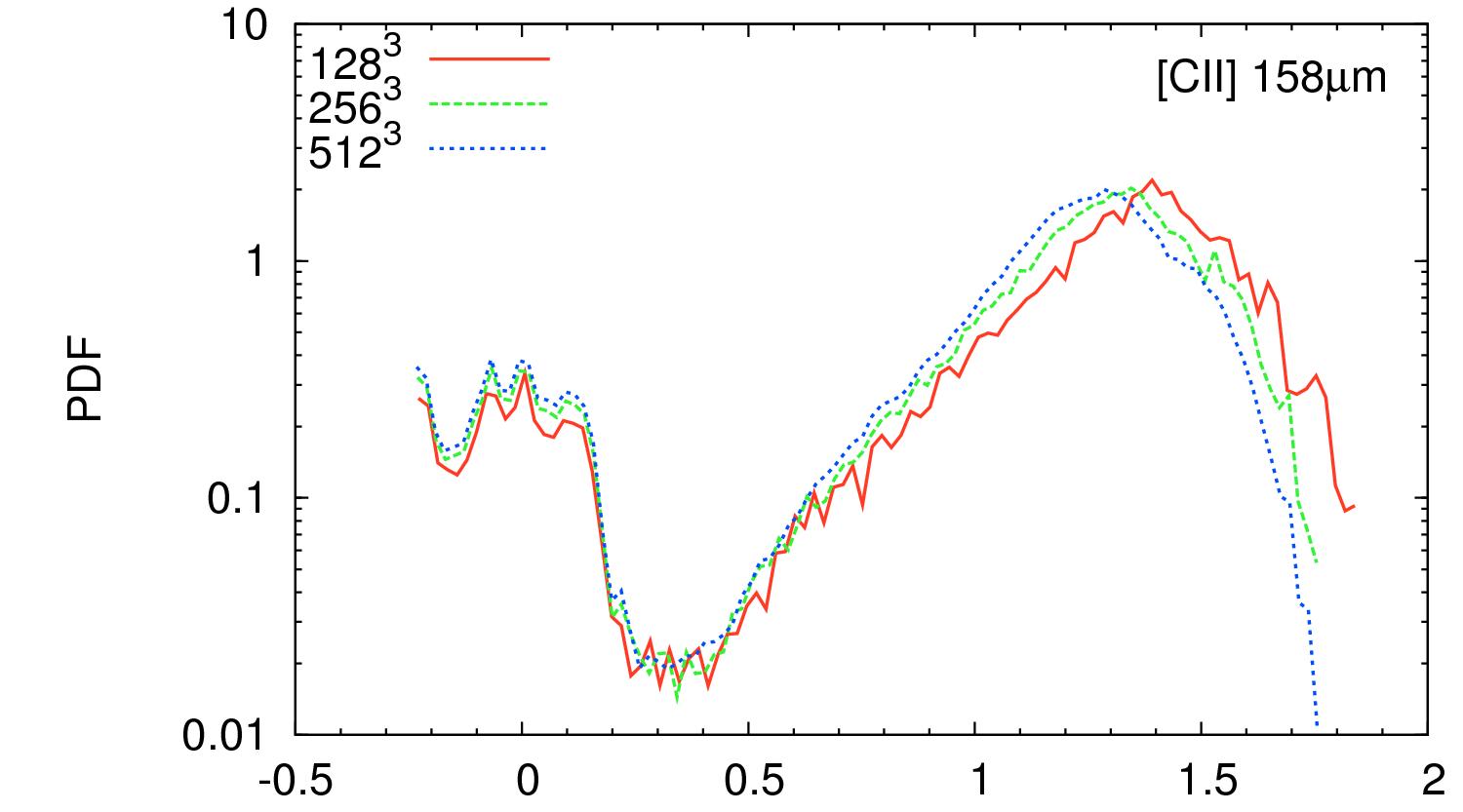}
}
\centerline{
\includegraphics[height=0.53\linewidth]{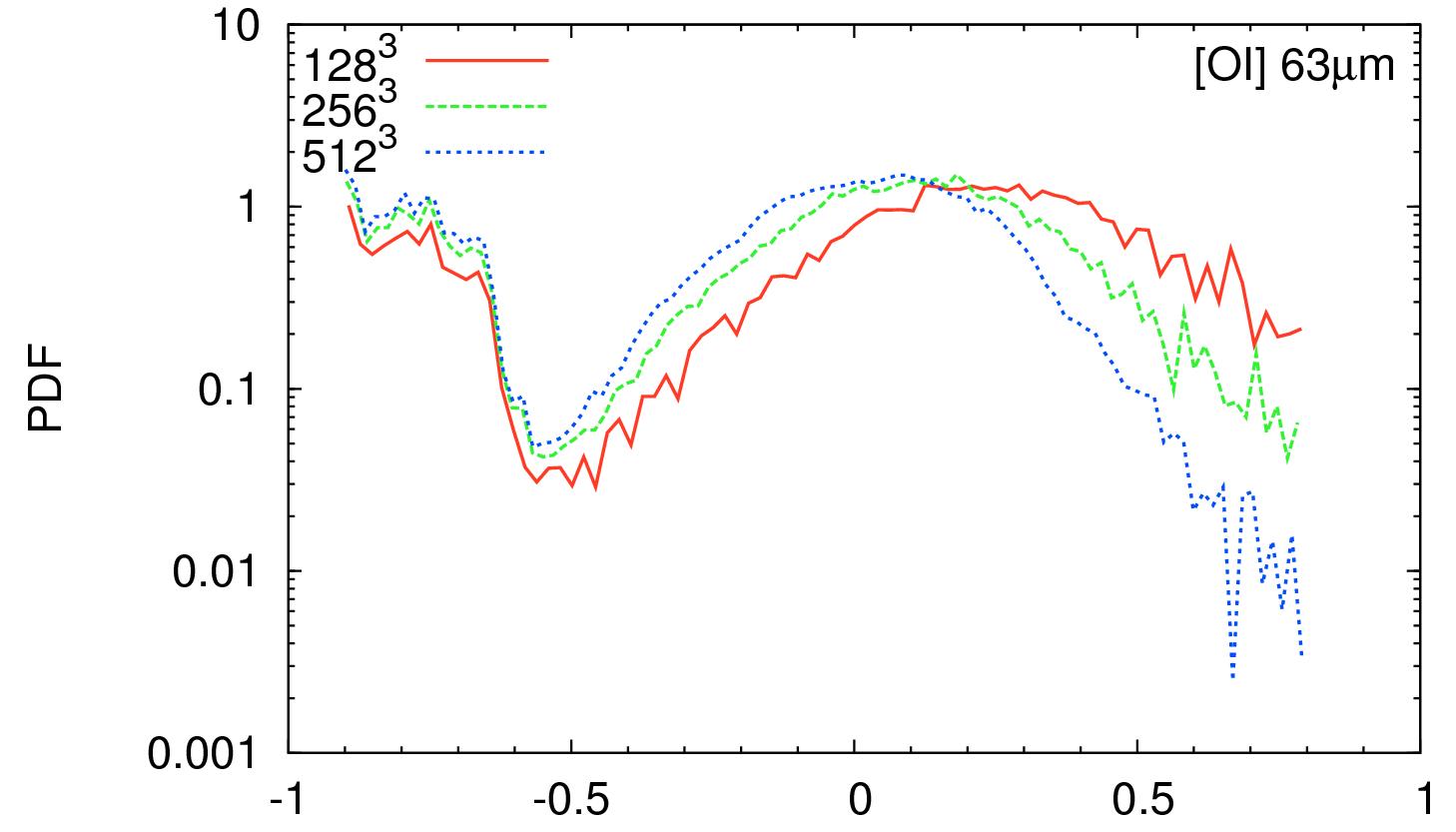}
}
\centerline{
\includegraphics[height=0.58\linewidth]{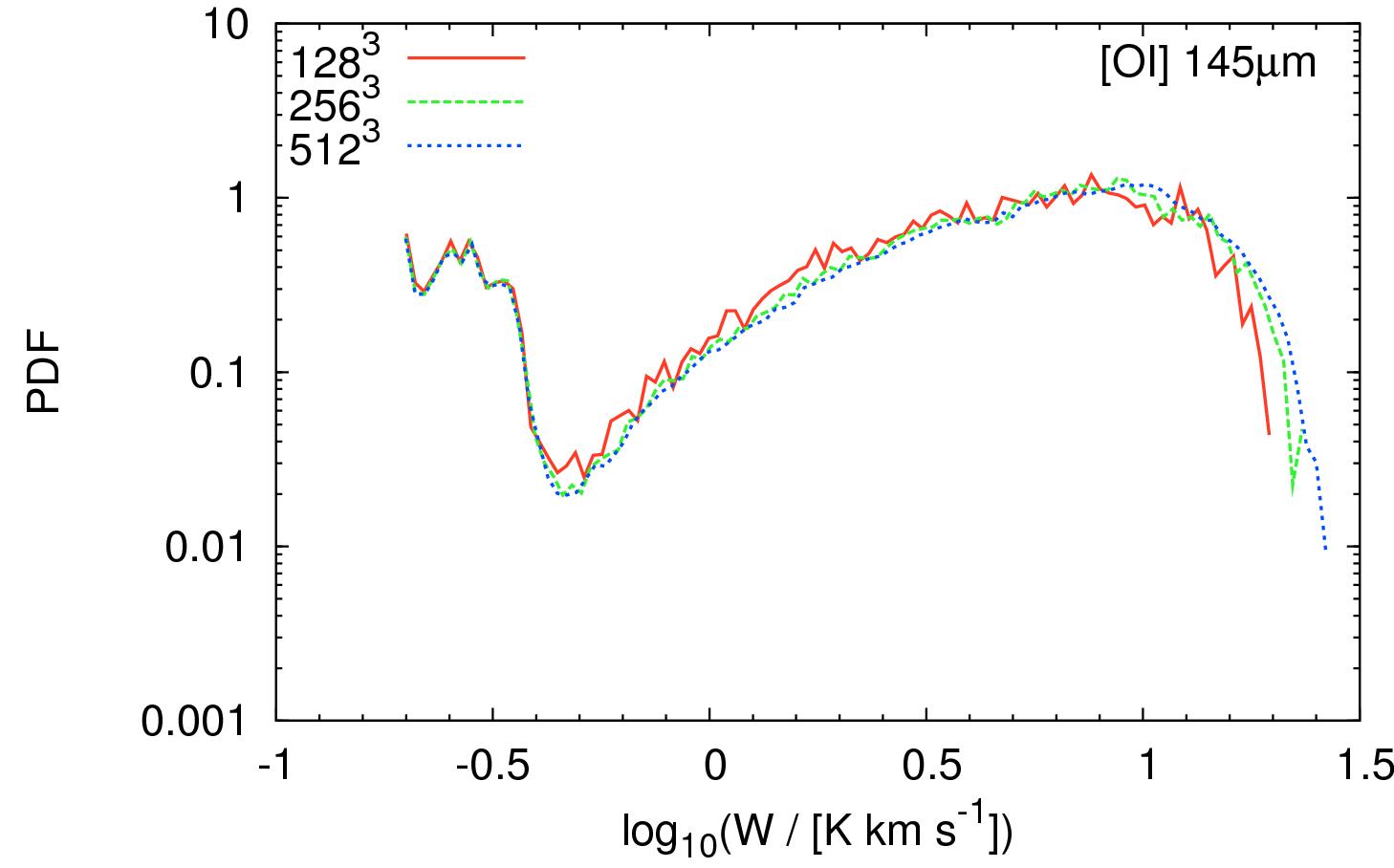}
}
\caption{Emission PDFs for our various tracers for different resolution models with $128^3$, $256^3$ and $512^3$ grid cells for clouds with $\alpha = 2.0$.}
\label{fig:emissionPDFs}
\end{figure}

Since R{\sc admc}-3{\sc d} cannot deal with A{\sc repo} data directly, we have to map the simulation output onto a cubic grid. The results presented in this paper are all based on a grid resolution of 512$^3$ cells. However, we examine the sensitivity of our results to the choice of the number of grid cells in each dimension. Therefore, we run a similar radiative transfer post-processing with R{\sc admc}-3{\sc d}, using resolutions of 256$^3$ and 128$^3$ grid zones. In Fig. \ref{fig:emissionPDFs} we compare the probability density functions (PDF) of the velocity-integrated intensities for our various tracers for the different grid resolutions, while using the intermediate $\alpha = 2.0$ cloud models as an example. We find that a resolution of $512^3$ grid cells is enough in order to properly recover the emission of the cloud for all of the tracers apart from the [O{\sc i}] 63$\mu$m line. In the case of [O{\sc i}] 63$\mu$m, the maps are well-converged for integrated intensities of below $1.5 \, {\rm K \, km \, s^{-1}}$, but at higher integrated intensities, increasing the resolution depresses the PDF. However, as we have already seen in Section \ref{subsec:denseregions_tot},  integrated intensities of this magnitude are only recovered for [O{\sc i}] 63$\mu$m along lines of sight where the line is already optically thick. The precise values of the integrated intensity that we recover in this regime therefore do not significantly affect our conclusions.

For all of the tracers, we find that the differences between the PDFs we recover from the $256^3$ and $512^3$ runs are always significantly smaller than the differences between the $128^3$ and $256^3$ models. Thus, we do not expect to find larger differences in the PDFs for models with even higher resolution, e.g. between $512^3$ and $1024^3$. Hence, due to the significant increase of the computational costs for the radiative transfer post-processing for even higher resolution runs, we focus on models with a resolution of $512^3$ grid cells in this paper, which is enough to demonstrate our basic conclusions.

\section{Dependence on viewing angle}
\label{sec:xyzPDFs}

\begin{figure}
\centerline{
\includegraphics[height=0.65\linewidth]{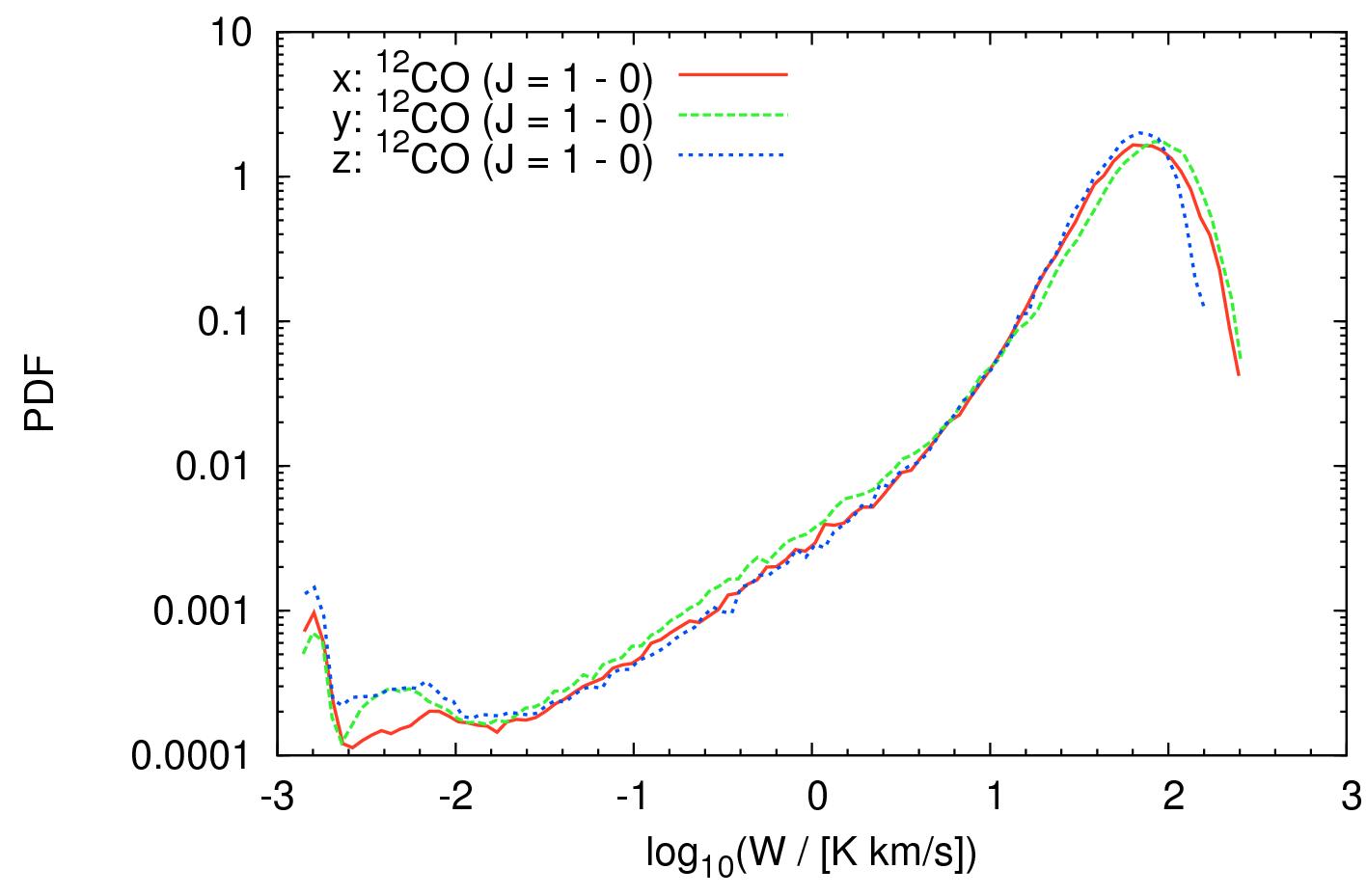}
}
\caption{Example of emission PDFs for the $^{12}$CO tracer for the resolution model with $512^3$ grid cells. We choose the intermediate cloud model with $\alpha = 2.0$. The above figure shows the emission PDF for the $x$-, $y$- and $z$-direction. Although there are slight variations in the different PDFs, the bulk emission is the same for all directions.}
\label{fig:xyzPDFs}
\end{figure}

When constructing our synthetic emission maps, we choose to focus on a line of sight (LoS) parallel to the $z$-axis of the simulation volume. However, as the turbulence in our simulations is isotropic, we expect our results to be insensitive to this choice. To test this, we have also made maps of $^{12}$CO for LoS parallel to the $x$ and $y$ axes, using our standard grid resolution of 512$^{3}$ zones, for the run with $\alpha = 2.0$. The resulting integrated intensity PDFs are shown in Figure \ref{fig:xyzPDFs}. Although there are slight variations in the different PDFs for $x$, $y$ and $z$, the bulk emission is the same for all directions. We find a similar behavior also for the other atomic tracers. Hence, we conclude that it is enough to focus on one specific LoS, for example the $z$-direction, in order to get an idea about the underlying physical parameters of the cloud. We note, however, that for magnetized clouds this is not necessarily true, as in this case the turbulence will  no longer be isotropic, unless the field is very weak.

\section{Influence of the CO abundance on the line emission of [C{\sc ii}] and [O{\sc i}]}
\label{sec:noCO}

\begin{figure}
\centerline{
\includegraphics[height=0.515\linewidth]{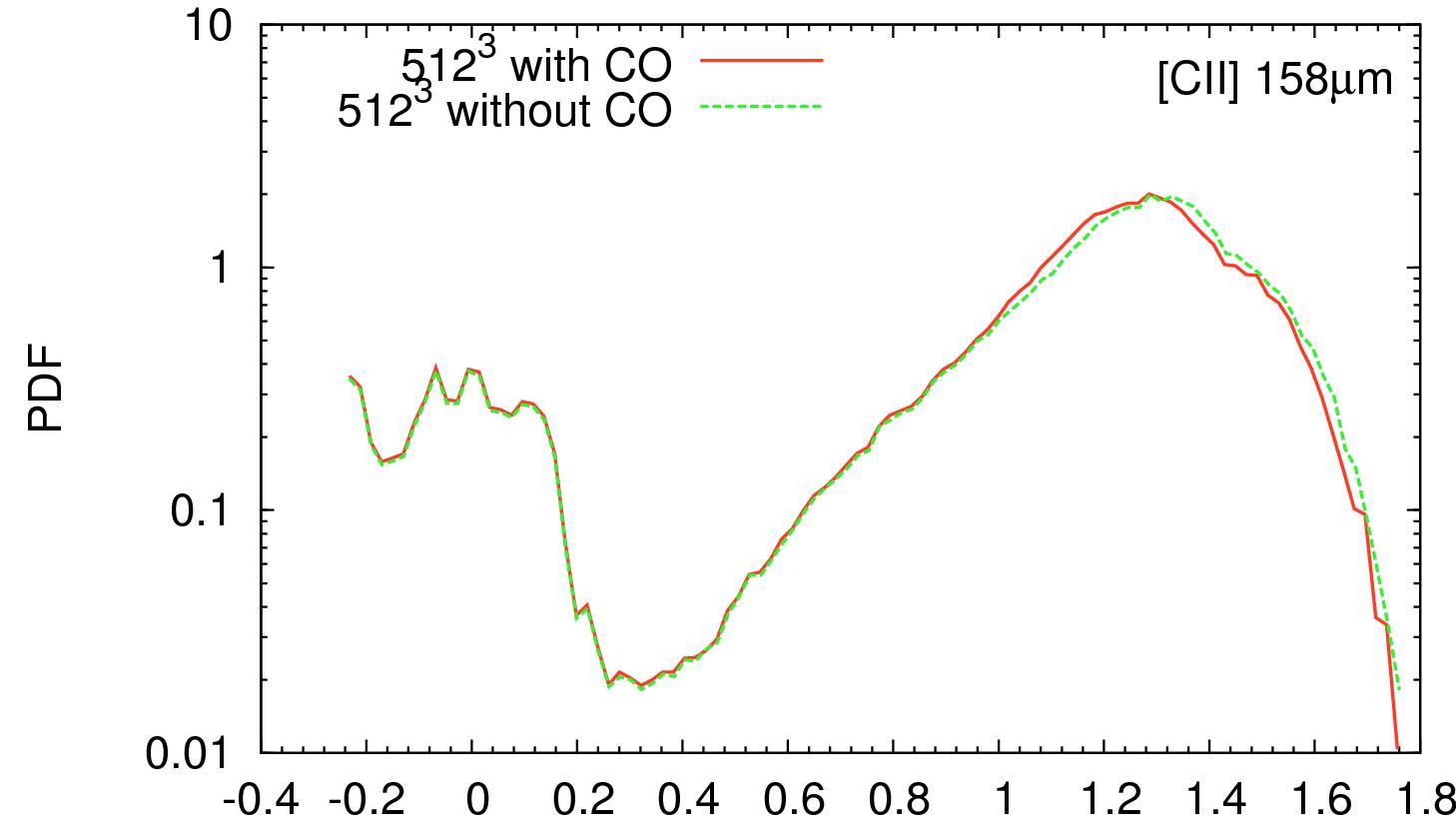}
}
\centerline{
\includegraphics[height=0.53\linewidth]{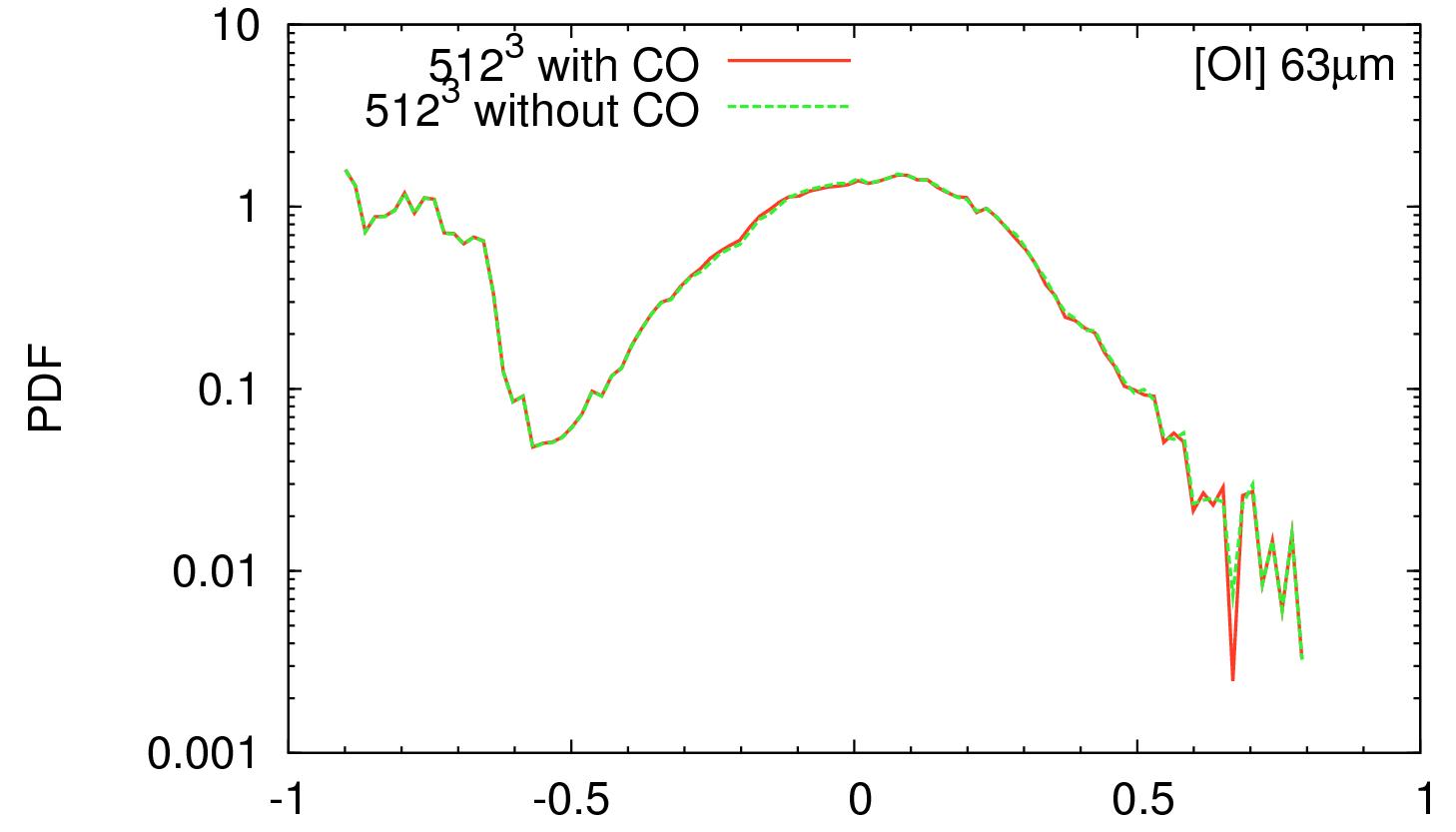}
}
\centerline{
\includegraphics[height=0.58\linewidth]{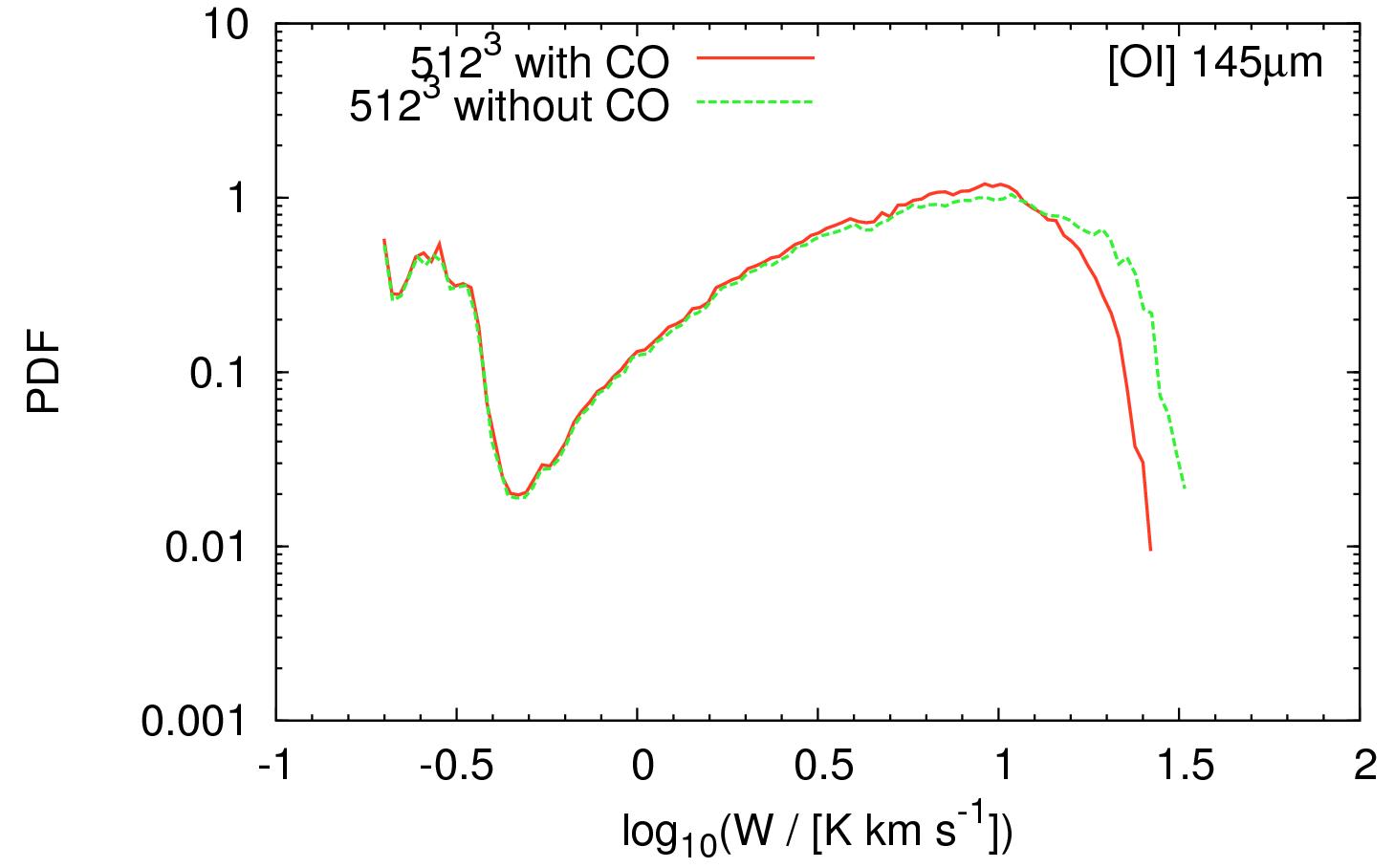}
}
\caption{Emission PDFs for our various atomic tracers for a fixed resolution of $512^3$ grid cells for clouds with $\alpha = 2.0$. Shown are the PDFs given in Fig. \ref{fig:emissionPDFs} (with CO) together with PDFs computed from additional runs of radiative transfer post-processing with updated abundances of carbon and oxygen assuming zero CO abundances in our simulations (without CO). We do not find significant changes in the different emission PDFs of the atomic tracers.}
\label{fig:noCO}
\end{figure}

In Section \ref{subsec:hydromodel} we emphasize that the simplified \citet{NelsonAndLanger1997} network that we use in our study is known to somewhat overestimate the rate at which CO forms in our simulations. Thus, we may expect that this has a consequence on the distribution of the line emission of [C{\sc ii}] and [O{\sc i}]. In order to quantify this, we perform additional runs of radiative transfer post-processing with updated abundances of carbon and oxygen for comparison by assuming zero CO abundances in our simulations. As an example, we plot the different line emission PDFs for our various atomic tracers for a fixed resolution of $512^3$ grid cells for clouds with $\alpha = 2.0$. The result is shown in Fig. \ref{fig:noCO}. We find that the {[C{\sc ii}]} 158\,$\mu$m and {[O{\sc i}]} 63\,$\mu$m line emission are unaffected by the changed abundances. This suggests that in the region where the CO abundance is uncertain, these two lines are already optically thick and hence are in sensitive to this uncertainty. Support for this interpretation comes from the behavior of the [O{\sc i}] 145\,$\mu$m line. As we have already seen, this has a much smaller optical depth, and hence is more strongly altered by the changed abundances. However, even in this case, the effect is relatively small: the brightest regions become around 0.1\,dex brighter, but the majority of the PDF remains unaffected. We can therefore conclude that the inaccuracy introduced into our predicted [C{\sc ii}] and [O{\sc i}] maps by our simplified chemical model is unimportant.

\end{appendix}

\end{document}